\documentclass[10pt,prd,aps,twocolumn,notitlepage,nofootinbib,nobalancelastpage,superscriptaddress]{revtex4-2}

\usepackage{color}

\usepackage{mciteplus}
\usepackage{amsmath}
\usepackage{amssymb}
\usepackage{enumerate}
\usepackage{tikz}
\usepackage{verbatim}
\usetikzlibrary{arrows.meta}
\usetikzlibrary{decorations.pathreplacing}

\usepackage{soul}
\usepackage[normalem]{ulem}

\usepackage{bm}
\usepackage[colorlinks=true,
            urlcolor=blue,
            citecolor=blue,
            linkcolor=red,
            menucolor=blue,
            linktocpage=true]{hyperref}

\usepackage{graphicx}
\makeatletter
\let\old@makecaption=\@makecaption
\usepackage[]{subcaption} % this has the unwanted behaviiour of centering captions so revert to prev caption
\let\@makecaption=\old@makecaption
\makeatother

\usepackage[varg]{txfonts}

\newcommand\mnras{Mon.\ Not.\ Roy.\ Astron.\ Soc.\ }
\newcommand\aap{Astron.\ Astrophys.\ }
\def\mbf#1{\mathbf{#1}}
\def\mrm#1{\mathrm{#1}}
\def\be{\begin{equation}}\def\ee{\end{equation}}
\def\bea#1\eea{\begin{align}#1\end{align}}

\def\ssp{\hspace{0.09em}}
\def\sp{\hspace{0.06em}}
\def\Mid{\,|\,}

\def\rme{\mathrm{e}}
\def\im{\mathrm{i}}
\def\x{\mathbf{x}}
\def\y{\mathbf{y}}
\def\s{\mathbf{s}}
\def\r{\mathbf{r}}

\def\k{\mathbf{k}}

\def\v{\mathbf{v}}
\def\u{\mathbf{u}}
\def\dif{\mathrm{d}}
\def\T{\mathrm{T}}
\def\delD{\delta_\mathrm{D}}
\def\n{\hat{\mbf{n}}}
\def\calH{\mathcal{H}}

\makeatletter
\DeclareFontFamily{OMX}{MnSymbolE}{}
\DeclareSymbolFont{MnLargeSymbols}{OMX}{MnSymbolE}{m}{n}
\SetSymbolFont{MnLargeSymbols}{bold}{OMX}{MnSymbolE}{b}{n}
\DeclareFontShape{OMX}{MnSymbolE}{m}{n}{
    <-6>  MnSymbolE5
   <6-7>  MnSymbolE6
   <7-8>  MnSymbolE7
   <8-9>  MnSymbolE8
   <9-10> MnSymbolE9
  <10-12> MnSymbolE10
  <12->   MnSymbolE12
}{}
\DeclareFontShape{OMX}{MnSymbolE}{b}{n}{
    <-6>  MnSymbolE-Bold5
   <6-7>  MnSymbolE-Bold6
   <7-8>  MnSymbolE-Bold7
   <8-9>  MnSymbolE-Bold8
   <9-10> MnSymbolE-Bold9
  <10-12> MnSymbolE-Bold10
  <12->   MnSymbolE-Bold12
}{}

\let\llangle\@undefined
\let\rrangle\@undefined
\DeclareMathDelimiter{\llangle}{\mathopen}%
                     {MnLargeSymbols}{'164}{MnLargeSymbols}{'164}
\DeclareMathDelimiter{\rrangle}{\mathclose}%
                     {MnLargeSymbols}{'171}{MnLargeSymbols}{'171}
\makeatother

\newcommand{\Geneve}{D\'{e}partement de Physique Th\'{e}orique,
Universit\'{e} de Gen\`{e}ve,
24 quai Ernest-Ansermet, CH-1211 Gen\`{e}ve 4, Switzerland}

\begin{document}
\title{Gravitational redshift from large-scale structure:
nonlinearities, anti-symmetries, and the dipole}
\author{Lawrence Dam}\affiliation{\Geneve}
\author{Camille Bonvin}\affiliation{\Geneve}
% \affiliation{
% D\'{e}partement de Physique Th\'{e}orique,
% Universit\'{e} de Gen\`{e}ve,
% 24 quai Ernest-Ansermet, CH-1211 Gen\`{e}ve 4, Switzerland}

\date{\today}

\begin{abstract}
Gravitational redshift imprints a slight asymmetry in the observed clustering
of galaxies, producing odd multipoles (e.g.\ the dipole) in
the cross-correlation function.
But there are other sources of asymmetry which must also be considered in
any model which aims to measure gravitational redshift from large-scale
structure.
In this work we develop a nonlinear model of the redshift-space correlation
function complete down to these subleading, asymmetric effects. In 
addition to gravitational redshift and the well-known redshift-space 
distortions, our 
model, given by a compact nonperturbative formula, also accounts for
wide-angle effects (to all orders), lightcone effects, and other
kinematic contributions.
We compare our model with $N$-body simulations and find good agreement;
in particular we find that the observed turnover in the dipole moment
around a separation of $20\,h^{-1}\mathrm{Mpc}$ (a feature absent
in the linear predictions) is well accounted for. By examining
the exchange properties of distinct tracers, we identify the
pairwise potential difference as the key physical ingredient of the
dipole. Several new insights related to the theory of redshift-space
distortions are also given.
\end{abstract}
\maketitle
\tableofcontents

\section{Introduction}

The large-scale structure revealed by galaxy surveys discloses important information about how gravity operates on cosmological scales.
This is possible through three major techniques: redshift-space distortions (RSD)~\cite{Kaiser:1987,Hamilton:1992zz}, baryon acoustic
oscillations~\cite{Eisenstein:1998tu,Blake:2003rh,Seo:2003pu}, and weak gravitational lensing~\cite{1991MNRAS.251..600B,1991ApJ...380....1M,1992ApJ...388..272K}.
Together these probes form a large part of the science case of
galaxy surveys, both past and present~\cite{Peacock:2001gs,SDSS:2005xqv,2dFGRS:2005yhx,
Guzzo:2008ac,Beutler:2011hx,Blake:2011en,
BOSS:2016wmc,eBOSS:2020yzd,DES:2021wwk,Wright:2025xka,DESI:2025zgx}.

In recent years interest has turned to more subtle effects imprinted
on large-scale structure.
One such effect is gravitational redshift, the shift in the frequency
of light 
as perceived by an observer outside the gravitational potential
from which the light was emitted.
Like RSD, gravitational redshift distorts the clustering of galaxies in a coherent way.
But unlike RSD, these distortions are directly sensitive to the time-time component of the metric fluctuation, containing information complementary to that from weak lensing (which is sensitive to a degenerate combination of both time and space components in projection).

Despite the increasingly large numbers of galaxies and redshifts
available to us,
isolating the gravitational redshift and accessing this information
still remains a challenge. 
Compared with the Doppler shift, the
gravitational redshift is typically smaller by about two orders of
magnitude. Nevertheless, upcoming surveys
put a significant detection within reach.

Towards this effort, one general method stands as the ideal
way to isolate this small signal. The basic idea is to use that
the gravitational redshift, unlike the Doppler shift, imprints an
asymmetry in the clustering.
This asymmetry can in principle be seen on a range of scales,
from galaxy clusters to large-scale structure. In the case of clusters,
gravitational redshift can be measured through the redshift difference
between the bright central galaxy and its satellites
\cite{1995A&A...301....6C,Wojtak:2011ia,DiDio:2025bff}.
Since the bright central galaxy tends to sit deeper within the potential well 
relative to other cluster members, the anti-symmetry manifests as a net negative
redshift difference. By contrast, there is no net difference expected from
the {linear} Doppler shift; only a (symmetric) dispersion in the
redshift differences. A first detection of gravitational redshift from clusters was reported in Ref.~\cite{Wojtak:2011ia}, followed by various other analyses~\cite{Sadeh:2014rya, Jimeno:2014xma, eBOSS:2021ofn, Rosselli:2022qoz}.

On cosmological scales, gravitational redshift can be probed
by correlating two distinct populations of galaxies
and searching for the dipole moment in their cross-correlation function. The 
idea behind this is similar to the cluster method: gravitational redshift
induces an anti-symmetry along the line of sight, resulting in odd
multipoles in the two-point function~\cite{McDonald:2009ud,Bonvin:2013ogt}. In the linear regime, the dipole has yet to be detected, but in the near future a $19\sigma$ detection of the dipole at redshift $z=0.25$ 
is expected from the DESI Bright Galaxy Sample~\cite{Bonvin:2023jjq} (among other redshift bins). In the nonlinear regime, a first detection of the dipole in the cross-correlations was
reported at $2.8\sigma$ using the BOSS CMASS sample~\cite{Alam:2017izi}. \emph{Euclid}, by the second data release,
should eventually reach a combined detection significance of $6\sigma$
from four redshift bins~\cite{Euclid:2024azy}. These measurements will enable novel tests of gravity and of dark matter~\cite{Bonvin:2018ckp,Castello:2022uuu,Bonvin:2022tii,Tutusaus:2022cab, Castello:2023zjr, Castello:2024jmq, Saga:2021osv,Inoue:2024ikm}.

To arrive at the correct interpretation of the gravitational redshift signal, it is necessary to have a complete model of the total signal. At the very least, this is important for checking the robustness of more approximate models currently in use by the community. The challenge is that gravitational redshift is not the only source of anti-symmetry. Indeed
there are several other effects which 
also lead to anti-symmetry, most notably kinematic effects. On linear scales, a consistent treatment of all possible contributions to the dipole was first presented in Ref.~\cite{Bonvin:2013ogt}. On the scales of clusters, it became clear  shortly after the first measurement~\cite{Wojtak:2011ia} that several
other effects need to be accounted for in the signal modelling~\cite{Zhao:2012gxk,Kaiser_2013,Cai:2016ors}. These include second-order kinematic effects
which, by virtue of the virial theorem, are of the same order as gravitational redshift
and so cannot be neglected in the modelling.
Towards a more robust interpretation of future measurements from clusters, Ref.~\cite{DiDio:2025bff} recently
showed that all effects can be comprehensively accounted for in a fully relativistic framework.

In a similar way, this work aims to put forward a complete and consistent model of the anti-symmetric correlation function, valid down to mildly nonlinear scales (larger than cluster scales but smaller than the large,
linear scales typically treated). The advantage of targeting nonlinear scales is that in this regime  gravitational
redshift begins to dominate over the
kinematic  effects, increasing its signal-to-noise~\cite{Zhu:2017jfl,Saga:2021jrh,Bonvin:2023jjq,Euclid:2024azy}.
This leads to a turnover in the dipole, a clear departure 
from linear theory and an observational sign of gravitational 
redshift~\cite{Zhu:2017jfl,Saga:2021jrh,Euclid:2024azy}.

We note that
a few works have already developed a model of the asymmetric correlations in the mildly nonlinear regime.
Focussing on the $\calH/k$ corrections (where $\calH$ is the
conformal Hubble parameter and $k$ the wavenumber),
Di Dio and Seljak~\cite{DiDio:2018} calculated the dipole in the power spectrum to
one-loop precision, incorporating both the standard treatment of galaxy bias and
Eulerian perturbation theory at third order.
Follow-up studies~\cite{DiDio:2020jvo,Beutler:2020evf} tested the one-loop dipole model
against simulations, finding good agreement down to very
small scales ($k=0.48h\,\mrm{Mpc}^{-1}$), well beyond the scale where
linear theory was found to break down ($k=0.01h\,\mrm{Mpc}^{-1}$).
However, these one-loop fits depend on the addition of an effective-field-theory-like
nuisance parameter which cannot be estimated a priori.

Other works have modelled the dipole in configuration space.
Saga et al.~\cite{Saga:2020tqb,Taruya:2019xsf} presented a model of the
correlation function based on the resummed approach to
Lagrangian perturbation 
theory~\cite{Taylor:1996ne,Matsubara:2007wj,Couchman_Bond88}. Their model
is built on the predictive success of the Zel'dovich approximation~\cite{Zeldovich:1969sb,White:2014,Vlah:2014nta} and the straightforwardness of extending it to include wide-angle effects
and gravitational redshift.
Their model matches well with simulations down to small separations $s\simeq10\,h^{-1}\mrm{Mpc}$,
provided a nonperturbative correction is added to the
gravitational redshift.
Although the model is challenging to evaluate (due to high-dimensional
integration), it was found to be well approximated by
a simple quasi-linear formula. This formula, adopted in recent
forecasts of the dipole~\cite{Saga:2021jrh,Euclid:2024azy}, makes
clear the role of the nonperturbative correction in accounting
for the dipole's turnover and boosting the gravitational redshift
signal on small scales.
But while this model accounts for wide-angle effects and gravitational redshift, both important for the dipole,
it misses several kinematic contributions which are more difficult to
include in this framework.

Though the model of Di Dio et al.\
achieves impressive fits, it is subject to an unknown nuisance parameter.
The model of Saga et al., on the other hand, suffers less of the same problem
but is more approximate and ad-hoc in its coverage of anti-symmetric effects
and nonlinearities.
There is room still for another model which is more
user-friendly yet physically motivated.

In this work we develop a complete model of the correlation function
beyond the linear regime, on the full sky, including a consistent
treatment of all subleading effects---RSD,
gravitational redshift, wide-angle effects, and other
relevant kinematic effects.
Our model is based on the `streaming model', a
nonperturbative approach to RSD~\cite{Peebles_LSS,Fisher:1994ks,Scoccimarro:2004,Vlah:2018ygt}.
As we showed in previous work~\cite{Dam:2023}, the streaming model is not only able to handle more general types of distortions but it
can also be extended without approximation to the full wide-angle regime.
(By contrast, note that wide-angle effects can only be
treated approximately in
Fourier space due to the loss of translation invariance in this regime~\cite{Szalay:1997cc}.)
The important quantity in this approach is the
map from real to redshift space (see Ref.~\cite{Shaw:2008aa} for an
earlier expression of this idea).
This map, normally determined by the Doppler shift (RSD), 
is extended in this work to include the gravitational redshift. 

Importantly, the streaming model naturally 
accounts for nonperturbative effects missed in more 
straightforward perturbative treatments
where a Jacobian is explicitly computed.
The standard example of this is the Finger-of-God effect,
described by the pairwise velocity dispersion.
In this work we show that the inclusion of
gravitational redshift yields a similar kind of contribution,
with its own one-point function. But whereas the Finger-of-God
effect is associated with a second moment, this new contribution
is associated with a first moment, the \emph{pairwise potential difference},
a key anti-symmetric ingredient of the model.
In this the natural emergence of the one-point function,
which was included as an empirical correction in
the model of Saga et al.~\cite{Saga:2020tqb},
can be traced to the inherent density weighting arising when
correlating biased tracers. Since the
gravitational redshift is probed through tracers which are found at biased
locations (e.g.\ near the bottom of potential wells),
there is a large nonperturbative contribution to
the gravitational redshift from the local small-scale potential,
which is strongly correlated with its immediate environment~\cite{Cai:2016ors}.
We thus show that a realistic estimate of the one-point function
is required to reproduce the dipole's turnover as seen in simulations.

In addition to the gravitational redshift, our model also accounts for two kinematic effects, well established from the linear calculations~\cite{Yoo:2009au,Bonvin:2011,Challinor:2011bk,Jeong:2011as},
which also need to be included in a consistent treatment of the {nonlinear} dipole. Both are related to the lightcone.
The first effect is relatively trivial and can be built into the model by simply promoting the usual spatial mapping
to a spacetime mapping on the lightcone. The second effect is less trivial. As we explain in some
detail, heuristically and then more systematically, it originates from the fact that we generally
probe galaxies in motion. This observation, although obvious, implies a
subtle kinematic correction to the underlying clustering statistics,
leading to another source of anti-symmetry.

\subsection*{Summary and overview of this paper}
This paper can be divided into three parts. The first part
presents the analytic development of the model (Sections~\ref{sec:2pcf} and \ref{sec:complete})
and the second contains the numerical results (Sections~\ref{sec:ingredients} and \ref{sec:results}).
The third part presents a formalism for understanding the model
along with a deeper discussion of the dipole (Sections~\ref{sec:pair} and \ref{sec:origin}).

In detail,
Section~\ref{sec:2pcf} reviews and generalises the
streaming model to include gravitational redshift and
wide-angle effects, emphasizing the key role of
the real-to-redshift map which makes possible our treatment.
Section~\ref{sec:complete} continues the model development
by building in the lightcone and lookback-time effects, the two
remaining order $\calH/k$ effects missed
in the first pass of Section~\ref{sec:2pcf}. (The discussion 
on the lightcone effect is a slight digression from the model development;
readers more interested in the finished products can skip ahead to
Section~\ref{sec:number-density}.) Including these two effects, a
complete model of the streaming model is presented.
In Section~\ref{sec:ingredients} we give an inventory of the (real-space)
ingredients needed to evaluate the model.
Section~\ref{sec:results} tests the model against $N$-body simulations
and presents the numerical results.
In Section~\ref{sec:pair} we analyse the main ingredients of the
streaming model, showing that the displacement statistics decompose
into symmetric and anti-symmetric parts, each having distinct physical interpretations. 
We show that many of the usual intuitions and rules-of-thumb
related to anti-symmetry can be understood in terms of a small set of
pairwise correlation functions, independent of survey considerations
(geometry, line of sight, etc).
From this set we identify the pairwise potential difference as the
single most important quantity for the dipole and compute it using the halo model.
In Section~\ref{sec:origin} we make use of the pairwise formalism to
derive a working perturbative model of the dipole which accounts
for the turnover. The physical origin is shown to be due to an advection-like
effect. We summarize our main findings in Section~\ref{sec:conclusions}.

Technical material and further discussion can be found in a number of
appendices. Of note is Appendix~\ref{app:pot-corrs} in which we discuss
the two-point correlations of the gravitational potential, the (apparent)
infrared divergence, and the impact of the local potential on correlations.

\paragraph*{Notation.}
We work in conformal Newtonian gauge with spatially flat
Friedmann--Lema\^itre--Robertson--Walker
line element $\dif s^2=-a^2(\tau)(1+2\Psi)\sp\dif\tau^2+a^2(\tau)(1-2\Phi)(\dif\chi^2+\chi^2\dif\Omega^2)$, 
where $\tau$ is conformal time, $\chi$ is the comoving radial distance, $a(\tau)$ is the scale factor,
and $\Psi$ and $\Phi$ are the Bardeen potentials.
We work in units where the speed of light $c=1$, so that, e.g.\
$\Psi$ and $\Phi$ are dimensionless, the conformal Hubble parameter
$\calH=aH$ has units inverse length, and $\calH/k$ (with $k$ the wavenumber) is dimensionless.
A summary of the rest of our notation can be found in Table~\ref{tab:notation} below.

\section{Galaxy clustering with gravitational redshift}\label{sec:2pcf}
This section begins the model development. To avoid introducing
too many novelties at once, we will gradually build up the model,
starting in this section with the inclusion of gravitational redshift
and wide-angle effects (in addition to RSD).
We postpone to the next section the inclusion of the
two remaining effects needed to complete the model.

Since our model is aimed at the nonlinear regime, and in particular
the anti-symmetric correlations,
we need only treat the order $\calH/k$ relativistic effects
\cite{Yoo:2009au,Bonvin:2011,Challinor:2011bk,Jeong:2011as}.
On sub-Hubble scales ($k\gg\calH$), these are the leading
corrections to the dominant RSD effect.%
\footnote{The remaining corrections are order $(\calH/k)^2$ so
are further suppressed with respect to those considered in this work.
In practice there are additional survey-specific contributions due to
magnification bias (from the magnitude limit) and evolution bias
(from the non-conservation of tracers). These contributions are also
order $\calH/k$ but will be neglected to keep the discussion simple.
We thus consider the modelling problem of an ideal survey. That said,
there is no serious difficulty in including survey non-idealities
in the current formalism, as we showed in Ref.~\cite{Dam:2023} (see Section III therein).} 
Including these corrections, the linear overdensity $\delta^{(\mrm{s})}_A$
of an arbitrary tracer $A$ in redshift space (denoted with superscript s) is
given by~\cite{Bonvin:2013ogt}
\be\label{eq:delta-lin}
\begin{split}
\delta^{(\mrm{s})}_A
&=\delta_A - \frac1\calH\frac{\partial\sp\v\cdot\n}{\partial\chi} 
+ \frac{1}{\mathcal{H}}\frac{\partial\Psi}{\partial\chi} \\
&\qquad
+ \v\cdot\n 
- \left(\frac{2}{\mathcal{H}\chi}+\frac{\dot{\mathcal{H}}}{\mathcal{H}^2}\right) \v\cdot\n
+ \frac{1}{\mathcal{H}}\ssp \dot\v\cdot\n\,,
\end{split}
\ee
where $\delta_A=b_A\delta$ is the linear overdensity in real space
(with $\delta$ the matter overdensity and $b_A$ the linear bias),
$\v$ is the peculiar velocity, 
$\n$ is the line of sight, and an overdot denotes a partial derivative with
respect to $\tau$.
In this expression the second term gives the well-known RSD effect,
the third term is due to
gravitational redshift,
and the second line of terms are kinematic contributions (whose physical origin
we will discuss).

By the end of the model building we will see that the linear
expression~\eqref{eq:delta-lin} actually follows from a very simple
formula [Eq.~\eqref{eq:deltas-all}].
As in the streaming model, the trick is to keep everything resummed by
working with the integral form of number conservation.
This allows us to avoid having to explicitly compute a Jacobian associated
with the transformation from real to redshift space (and the large number
of terms that result from it).

\subsection{Nonlinear modelling of the redshift-space correlation function in the
wide-angle regime}\label{sec:xi-redshift-space}

The wide-angle streaming model we presented in Ref.~\cite{Dam:2023}
forms the basis of our model. Here we will extend it, doing so in
a way that makes clear the possibility of generalisation beyond RSD.
There are two elements to the model:
(\emph{i}) a map from the undistorted, real-space position
$\x$ to the observed, redshift-space position $\s$ (both
in comoving units); and (\emph{ii}) the requirement that the
number of tracers in going between real and redshift space is conserved.

We will focus here on radial distortions, i.e.\ distortions in the clustering
along the line of sight.
Both the Doppler effect and the gravitational redshift (the target of this work)
are distortions of this type.\footnote{There are two other effects which
can also change the apparent position of a galaxy: the integrated Sachs--Wolfe effect,
a radial distortion which is nominally of order $(\calH/k)^2$;
and gravitational lensing, a transverse distortion which has been shown
to contribute negligibly to the dipole~\cite{Bonvin:2013ogt,Breton:2018wzk}.}
Since the line of sight in redshift space and in real space
coincide, $\hat\s=\hat\x\equiv\n$, we can simply work with the radial map,
$\chi(\x)=\chi'+\delta\chi(\x)$,
obtained by projecting the three-dimensional
map $\s(\x)=\x+\delta\x$ onto $\n$. Here
$\chi'=\x\cdot\n$ and $\chi=\s\cdot\n$ are
the comoving distances in real and redshift space, respectively.
The displacement $\delta\chi=\n\cdot\delta\x$ between these
two distances is due to fluctuations to the redshift from the
Doppler shift and gravitational redshift:%
\bea\label{eq:chi-map}
% &\chi(\x)=\chi'+\delta\chi(\x) \,,\\[3pt]
\delta\chi(\x)=\u(\x)\cdot\n+\psi(\x)-\psi_O \, .
\eea
Here we have defined the scaled quantities
$\u\equiv\calH^{-1}\v$ and $\psi\equiv-\calH^{-1}\Psi$ (both
having units length), with  $\psi_O\equiv\psi(\tau_0,\x=0)$ being the
local potential.
Clearly the usual RSD map is recovered by dropping the
potentials so that $\s=\x+(\u\cdot\n)\sp\n$,
or $\delta\chi=\u\cdot\n$. In general, the map \eqref{eq:chi-map} depends on the difference between quantities at the observer and quantities at the source; here we have removed the contribution from the observer velocity since it is routinely subtracted in the measurement of the redshift. Note that the results in the
rest of this subsection hold for any $\delta\chi(\x)$ and thus any radial map,
not just for Eq.~\eqref{eq:chi-map}.

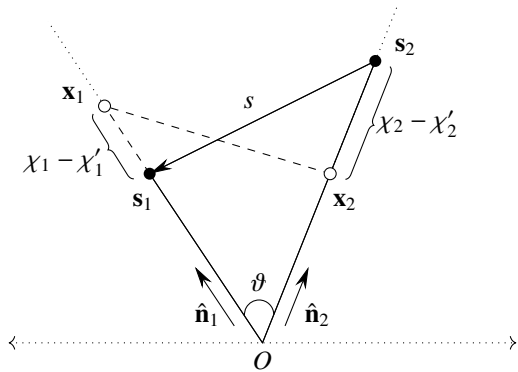
\begin{figure}[!t]
\begin{tikzpicture}[scale=0.75]
\tikzstyle{every node}=[font=\normalsize]
  \coordinate[label=below:$O$] (A) at (0,0);
  \coordinate (B) at (-2,3); % [label=below left:$\s_1$]
  \coordinate (C) at (2,5);
  \coordinate (M) at (0,4);
  \coordinate (B2) at (-3.8,5.7);
  \coordinate (C2) at (2.4,6);
  \draw (A) -- node[left] {} (B) node[circle,fill,inner sep=1.6pt]{}
            -- node[above] {} (C) node[circle,fill,inner sep=1.6pt]{}
            -- node[right] {} cycle;
  % \draw[dashed,->] (0,0) -- (0,5.6) node[above]{}; % $z$
  \draw[dotted,<->] (-4.5,0) -- (4.5,0) node[right]{}; % $x$
  \draw (A) -- node[midway, below left]{} (B);
  \draw (A) -- node[midway, below right]{} (C);
  % \draw[-{Stealth[length=3mm,width=1.5mm]}] (A) -- node[midway, left]{$\mbf{d}$} (M);
  \draw[-{Stealth[length=3mm,width=1.5mm]}] (C) -- node[midway, above left]{$s$} (B);
  \draw (-2.8,4.2) node[circle,fill=white,inner sep=1.6pt,draw](b){}; %x_1 empty O
  \draw (1.2,3) node[circle,fill=white,inner sep=1.6pt,draw](c){}; %x_2 empty O
  \draw[dashed] (b) -- (c);
  \draw[dashed] (B) -- (b);
  \draw[dotted] (b) -- (B2);
  \draw[dotted] (c) -- (C2);
  % \draw (0,4.3) arc (90:206.565:0.3) node[midway, left]{$\theta$};
  % \draw [draw=black, every edge/.append style={draw=black, dashed}]
  %    (C) -- node[sloped] {$\parallel$} (M)
  %    (B) -- node[sloped] {$\parallel$} (M);
  \draw[-{Stealth[length=3mm,width=1.5mm]}] (-0.5,0.3) -- (-1.2,1.35);
  \node[anchor=east] at (-0.6,0.5) {$\n_1$};
  \draw[-{Stealth[length=3mm,width=1.5mm]}] (0.4,0.3) -- (0.82,1.35);
  \node[anchor=west] at (0.6,0.5) {$\n_2$};
  \node[anchor=west] at (-2.45,2.5) {$\s_1$};
  \node[anchor=west] at (2.2,5.25) {$\s_2$};
  \draw [decorate,decoration = {brace}] (-2.28,2.91) -- (-3.0,4.06)
  node[pos=0.5,below=6,left=]{$\chi_1-\chi_1'$};
  \draw [decorate,decoration = {brace}] (2.26,4.9) -- (1.47,2.9)
  node[pos=0.5,below=0.2,right=1.1]{$\chi_2-\chi_2'$};
  \node[anchor=west] at (-3.7,4.4) {$\x_1$};
  \node[anchor=west] at (1.1,2.5) {$\x_2$};
  \draw[black] (0.2,0.5) arc (0:185:0.26) node[midway,above]{$\vartheta$};
\end{tikzpicture}
\caption{Coordinates used to specify a typical configuration
in the wide-angle regime (in which the opening angle $\vartheta$ can
range from $0^\circ$ to the maximum $180^\circ$). Here,
two-point correlations are specified by three independent
variables describing the redshift triangle formed by the two
tracers (solid circles) and the observer $O$. Here we use
two sidelengths $\chi_1=|\s_1|$, $\chi_2=|\s_2|$, and the opening
angle $\vartheta$. The real triangle (dashed lines) is specified by
$\chi_1'=|\x_1|$, $\chi_2'=|\x_2|$, and the same opening angle
$\vartheta$.
The streaming model integrates over the unknown real positions
of the tracers (open circles) which can lie anywhere on their
respective line of sight ($\n_1$ or $\n_2$).}
\label{fig:config}
\end{figure}

Since the anti-symmetric effects can only be revealed in
cross-correlation it will be necessary to consider two distinct tracers, labelled $A$ and $B$.
(We will often treat $A$ as an abstract label for both tracers.)
Denoting by $n_A$ the number density of tracer $A$ and using number conservation of tracers, 
$n^{(\mrm{s})}_A(\s)\ssp\dif^3\s=n_A(\x)\ssp\dif^3\x$, 
the redshift-space number density is 
\bea
n^{(\mrm{s})}_{A}(\s)
&=\int\dif^3\x\,n_A(\x)\ssp\delD\big(\s-\s(\x)\big) \nonumber\\
&=\frac{1}{\chi^2}\int^\infty_0\dif\chi'\chi'^2 n_A(\x)\ssp
\delD\big(\chi-\chi'-\delta\chi(\x)\big)\,,
\label{eq:ns}
\eea
where $\s=\chi\sp\n$,  $\x=\chi'\n$, and $\delD$ is the Dirac
delta function.
This expression relates the redshift-space
density $n^{(\mrm{s})}_A$ to the real-space density
$n_A$ underlying it, through a map (which need not be one-to-one).
The key step above, which makes possible much
of this work, is the change to spherical coordinates, with which
the integration over the angular coordinates
becomes trivial because $\s(\x)$ is purely radial (the $1/\chi^2$
is due to the transformation properties of the delta function).

From Eq.~\eqref{eq:ns} an exact formula for
the wide-angle correlation function can be obtained.
Denoting by $A$ the tracer at $\s_1=\chi_1\n_1$ and
by $B$ the tracer at $\s_2=\chi_2\n_2$,
the redshift-space correlation function
$\xi^{(\mrm{s})}_{AB}\equiv\langle\delta_A^{(\mrm{s})}(\s_1)\sp
\delta_B^{(\mrm{s})}(\s_2)\rangle$ is given by~
(see Ref.~\cite{Dam:2023} for details)
\be\label{eq:gsm}
\begin{split}
&1+\xi^{(\mrm{s})}_{AB}(\chi_1,\chi_2,\cos\vartheta)
=\frac{1}{\chi_1^2\,\chi_2^2}
\int\dif\chi_1'\,\chi_1'^2\int\dif\chi_2'\,\chi_2'^2 \\
&\qquad\quad\times
    \big[1+\xi_{AB}(\chi_1',\chi_2',\cos\vartheta)\big]\,
    p(\bm\chi-\bm\chi',\bm\chi',\cos\vartheta)\,.
\end{split}
\ee
Here $\cos\vartheta=\n_1\cdot\n_2$ is the (cosine) opening
angle between $A$ and $B$;
$\bm\chi=(\chi_1,\chi_2)$ and $\bm\chi'=(\chi_1',\chi_2')$
are two-component vectors consisting of the radial distances to
each tracer; and
$\xi_{AB}$ is the real-space cross-correlation function between
tracer $A$ and $B$.%
\footnote{In Eq.~\eqref{eq:gsm}, usually $\xi_{AB}=\xi_{AB}(r)$
for the real-space correlations, i.e.\ we have dependence on separation
$r$ only (although recall
$r=(\chi_1'^2+\chi_2'^2-2\cos\vartheta\sp\chi_1'\chi_2')^{1/2}$).
But this turns out to be a special case valid when there are no
selection effects.
In general,
$\xi_{AB}$ is a function of all three $\chi_1',\chi_2',\cos\vartheta$ independently. This is the case
when we include the lightcone effect (see
Section~\ref{sec:lightcone}).}
No assumption about galaxy bias needs to be made; Eq.~\eqref{eq:gsm}
simply expresses the relation between pairs of tracers in real space and
in redshift space.

The distribution function $p(\bm\chi-\bm\chi',\bm\chi',\cos\vartheta)$
is at the core of the streaming model~\eqref{eq:gsm}.
It emerges because the map
from real to redshift space is not one-to-one but rather
many-to-one, 
leading to a distribution of
possible pair positions in real space (or distribution in possible
displacements), for a known pair position in redshift space, as depicted in Figure~\ref{fig:config}. 
However, the possible positions of each tracer are not independent
from one another but are instead correlated. This is because 
$\delta\chi$ depends, for instance, on the velocity of the tracer,
which in turn is correlated with the velocity of the other tracer.
Since there are two independent lines of sight the `probability' 
distribution is two-dimensional and can be
formally written as%
\footnote{Despite its form, Eq.~\eqref{eq:gauss} is {not} a proper
probability distribution function in the displacements.
% Due to the scale dependence of the cumulants
% there is no one probability of obtaining a given displacement
% $\delta\bm\chi$ because there is not one mean and covariance;
% these depend on the separation between the pair (or more 
% precisely on $\bm\chi'$ and $\cos\vartheta$).
It can however be understood as a conditional
probability distribution, as we show in Section~\ref{sec:p-interpretation}.}
\bea\label{eq:gauss}
p(\bm\chi-\bm\chi',\bm\chi',\cos\vartheta) 
&=\int\frac{\dif^2\bm\kappa}{(2\pi)^2}\,
    \rme^{-\im\bm\kappa\cdot(\bm\chi-\bm\chi')} \\
&\quad\times\exp\left(
\im\sp\bm\kappa^\T\bm{m}
-\frac12\bm\kappa^\T\mbf{C}\sp\bm\kappa+\cdots\right)\nonumber\,,
\eea
i.e.\ expressed as the Fourier transform of the cumulant
generating function (the second exponential).
The first and second cumulants, the mean $\bm{m}$
and covariance $\mbf{C}$, are given by
\bea
\bm{m}(\bm\chi',\cos\vartheta)
&=\llangle\delta\bm\chi\rrangle\,,\label{eq:mu}\\
\mbf{C}(\bm\chi',\cos\vartheta)
&=\llangle\delta\bm\chi\ssp\delta\bm\chi^\T\rrangle -\bm{m}\sp\bm{m}^\T\,,
\label{eq:mu_cov} 
\eea
where $\delta\bm\chi\equiv(\delta\chi(\x_1),\delta\chi(\x_2))$ is
a two-component vector consisting of the radial displacements given by
Eq.~\eqref{eq:chi-map}.
Explicit expressions for $\bm{m}$ and $\mbf{C}$ are given in Section~\ref{sec:stats} below.
The double bracket in Eqs.~\eqref{eq:mu} and~\eqref{eq:mu_cov}
is a shorthand for the density-weighted ensemble average:
\be\label{eq:dens-weighting}
\llangle F(\x_1,\x_2)\rrangle
\equiv
\frac{\big\langle n_A(\x_1)\sp n_B(\x_2)\sp F(\x_1,\x_2)\big\rangle}
{\big\langle n_A(\x_1)\sp n_B(\x_2)\big\rangle}\,,
\ee
for any pairwise function $F(\x_1,\x_2)$.
These averages naturally arise in the model from our
mapping (or displacing) only those points where there is a galaxy to
be found, i.e.\ where the density is high.
In general each component of the vector
$\bm{m}$ and $2\times2$ matrix $\mbf{C}$ is a
function of $\bm\chi'=(\chi_1',\chi_2')$ and $\cos\vartheta$.
One should keep in mind that $\bm{m}$ and $\mbf{C}$ also depend on
the galaxy bias of each tracer $A$ and $B$ through the
density weighting. For brevity we have suppressed this dependence.

The probability distribution~\eqref{eq:gauss} is in general
determined by an infinite series of cumulants. In our model we will
truncate the series after the second cumulant, keeping the mean and covariance,
and dropping all the non-Gaussian cumulants.
This is the approximation typically used in the streaming model 
(distant-observer limit) and provides an accurate description in the mildly nonlinear regime. We expect a similar level of accuracy in our model.

Finally, by retaining only the mean and covariance in the generating
function, Eq.~\eqref{eq:gauss} integrates to a 
Gaussian, so that the correlation function~\eqref{eq:gsm}
explicitly reads
\bea
&1+\xi^{(\mrm{s})}_{AB}(\bm\chi,\cos\vartheta)
=\frac{1}{\chi_1^2\sp\chi_2^2}
\int^\infty_0\dif\chi_1'\sp\chi_1'^2
\int^\infty_0\dif\chi_2'\ssp\chi_2'^2\sp
    (1+\xi_{AB}) \nonumber\\[-2pt]
&\times
    \frac{1}{2\pi|\mbf{C}|^{1/2}}
    \exp\left(-\frac12
    (\bm\chi-\bm\chi'-\bm{m})^\T
    \mbf{C}^{-1}\sp
    (\bm\chi-\bm\chi'-\bm{m})\right)\,, \label{eq:gsm2}
\eea
where $\xi_{AB}$, $\bm{m}$, and $\mbf{C}$ are all functions
of $\bm\chi'=(\chi_1',\chi_2')$, and $\cos\vartheta$.
This is the wide-angle generalisation of the Gaussian streaming model; it allows us to correlate \emph{any}
two points on the sky, regardless of the size of the opening angle
$\vartheta$ (e.g.\ when one galaxy is in front
of the observer and the other is a full $180^\circ$ behind).
Although Eq.~\eqref{eq:gsm2} is valid for any
radial distortion $\delta\bm\chi$, for the rest of this work we
will specialise to the case of Doppler shift and
gravitational redshift, given by Eq.~\eqref{eq:chi-map}.

It should be noted that although $\chi_1, \chi_2$ and $\vartheta$
are natural coordinates in the wide-angle regime, they
are not appropriate to describe the (line-of-sight) asymmetries
in correlations we are after (or any of the multipoles for that
matter).
For this one needs to make a change of coordinates and express the
natural coordinates in terms of the standard
coordinates $s$ (separation) and $\mu$ (line-of-sight angle),
together with a third coordinate, $d$, describing the length scale
between the tracers and the observer (see Appendix~\ref{app:formulas}
for relations). With these coordinates one can then perform a
multipole decomposition with respect to $\mu$
in the usual way, isolating the anti-symmetric effects
through the odd multipoles.

\subsection{Displacement statistics}\label{sec:stats}
The mean and covariance of the displacement field~\eqref{eq:chi-map}
are two key quantities in the model~\eqref{eq:gsm2}.
Although the treatment in the previous section was kept
largely general, it will be useful to plug in the particular
displacement~\eqref{eq:chi-map} to see the structure
of these quantities. Compared with the standard model in the
distant-observer limit, there are two main differences to be noted.
First, in the wide-angle regime the mean and covariance are vector and
matrix quantities. Second, each consist of contributions due to
RSD and gravitational redshift.

Since the mean~\eqref{eq:chi-map} is
linear in the displacements we can decompose it into two parts:
\be\label{eq:mu-map}
\bm{m}(\bm\chi')
=\bm{m}_\text{RSD}(\bm\chi')+\bm{m}_\text{grav}(\bm\chi')\,,\\[2pt]
\ee
where again we dropped the dependence on $\cos\vartheta$ and
\be
\bm{m}_\text{RSD}(\bm\chi')
=
\begin{pmatrix}
\ssp\llangle u_{\|}(\x_1)\rrangle \\[3pt]
\ssp\llangle u_{\|}(\x_2)\rrangle
\end{pmatrix},
\quad
\bm{m}_\text{grav}(\bm\chi')
=
\begin{pmatrix}
\llangle\psi(\x_1)-\psi_O\rrangle\ssp\\[3pt]
\llangle\psi(\x_2)-\psi_O\rrangle\ssp
\end{pmatrix}.
\label{eq:mrsd-mgrav}
\ee
Here we used the shorthands
$u_{\|}(\x_1)\equiv\u(\x_1)\cdot\n_1$ and $u_{\|}(\x_2)\equiv\u(\x_2)\cdot\n_2$.
Note that the mean of these fields is \emph{not} zero because
of the density weighting (as it would be with volume weighting).
As we will see in Section~\ref{sec:origin}, this has an
important consequence for the dipole in the nonlinear regime.

By size alone, the gravitational redshift is about
two orders of magnitude smaller than the Doppler shift. However, as already mentioned, each effect has a different line-of-sight dependence,
so contributes to different parts of the overall
multipole structure. This can be used to measure these effects separately. For instance, in terms of
the line-of-sight angle $\mu=\hat\s\cdot\n$ we have ${m}_\mrm{RSD}\sim\mu$ while
${m}_\mrm{grav}\sim1$, leading to $\partial\ssp{m}_\mrm{RSD}\sim\mu^2$ while
$\partial\ssp{m}_\mrm{grav}\sim\mu$ (upon perturbatively expanding the
exponential in Eq.~\eqref{eq:gsm2}). In other words, at leading order
RSD contributes to even multipoles while gravitational
redshift contributes to odd multipoles. The separation
of RSD from gravitational redshift is in practice not perfect
(there is contamination of RSD and other effects into
odd multipoles), but this is the basis for isolating gravitational redshift from the dominant RSD. 

We can similarly decompose the covariance~\eqref{eq:mu_cov}:
\be
\mbf{C}(\bm\chi')
=\mbf{C}_\mrm{RSD}(\bm\chi')+\mbf{C}_\mrm{grav}(\bm\chi')
    +\mbf{C}_\mrm{cross}(\bm\chi')\,,
\ee
where now we have a cross term for the covariance between
the Doppler shift and gravitational redshift. Inserting the
particular displacement~\eqref{eq:chi-map} into the general 
formula~\eqref{eq:mu_cov} yields
\begin{widetext}
\begin{subequations}\label{eq:covs}
\bea
\mbf{C}_\text{RSD}(\bm\chi')
&=\begin{pmatrix}
\llangle u_{\|}(\x_1)\sp u_{\|}(\x_1)\rrangle & \llangle u_{\|}(\x_1)\sp u_{\|}(\x_2)\rrangle \\[3pt]
\llangle u_{\|}(\x_1)\sp u_{\|}(\x_2)\rrangle & \llangle u_{\|}(\x_2)\sp u_{\|}(\x_2)\rrangle
\end{pmatrix} 
-\bm{m}_\mrm{RSD}\ssp\bm{m}_\mrm{RSD}^{\T}\,, \label{eq:cov-RSD}\\[5pt]
\mbf{C}_\text{grav}(\bm\chi')
&=\begin{pmatrix}
\llangle[\psi(\x_1)-\psi_O][\psi(\x_1)-\psi_O]\rrangle & \llangle[\psi(\x_1)-\psi_O][\psi(\x_2)-\psi_O]\rrangle \\[3pt]
\llangle[\psi(\x_1)-\psi_O][\psi(\x_2)-\psi_O]\rrangle & \llangle[\psi(\x_2)-\psi_O][\psi(\x_2)-\psi_O]\rrangle
\end{pmatrix}
-\bm{m}_\mrm{grav}\ssp\bm{m}_\mrm{grav}^{\T}\,, \label{eq:cov-gravz}\\[5pt]
\mbf{C}_\text{cross}(\bm\chi')
&=\begin{pmatrix}
2\ssp\llangle u_{\|}(\x_1)\sp[\psi(\x_1)-\psi_O]\rrangle & \!\!\!\llangle u_{\|}(\x_1)\sp[\psi(\x_2)-\psi_O]\rrangle + (1\leftrightarrow2) \\[3pt]
\llangle u_{\|}(\x_1)\sp[\psi(\x_2)-\psi_O]\rrangle + (1\leftrightarrow2) & 2\ssp\llangle u_{\|}(\x_2)\sp[\psi(\x_2)-\psi_O]\rrangle 
\end{pmatrix}
-\bm{m}_\mrm{RSD}\ssp\bm{m}_\mrm{grav}^{\T}
-\bm{m}_\mrm{grav}\ssp\bm{m}_\mrm{RSD}^{\T}\,, \label{eq:cov-cross}
\eea
\end{subequations}
\end{widetext}
where in the last line $(1\leftrightarrow2)$ is an instruction
to exchange the arguments of $u_\|$ and $\psi$ only (the positions
in the density weighting are to be left unchanged).
Again, RSD gives the dominant contribution by magnitude,
with $\mbf{C}_\mrm{grav}
\lesssim\mbf{C}_\mrm{cross} \lesssim\mbf{C}_\mrm{RSD}$.
In particular, the rms dispersion $\langle u_\|^2\rangle^{1/2}$ is about
$300\,\mrm{km}\,\mrm{s}^{-1}$ while $\langle\psi^2\rangle^{1/2}$
is about $10\,\mrm{km}\,\mrm{s}^{-1}$; for the cross term
we have roughly $(300\times 10)^{1/2}\simeq 50\,\mrm{km}\,\mrm{s}^{-1}$.
As with the mean, each covariance has a different
angular dependence so can be separated to some extent.

As will become clear, $\bm{m}$ plays a more important
role than $\mbf{C}$ in determining the odd multipoles.
We can already understand this on a heuristic level
given that $\bm{m}$
is associated with a direction (e.g.\ the line of sight)
and thus an asymmetry along it, while $\mbf{C}$ is driven
by (RSD) auto-correlations which are symmetric in nature.
There is however an asymmetry from $\mbf{C}_\mrm{cross}$, based on
its parity, but this is suppressed relative to $\bm{m}$ (correlations in
$\mbf{C}_\mrm{cross}$ are suppressed by an additional factor $\calH/k$).

\begin{table*}[t!]
    \centering
    \renewcommand{\arraystretch}{0.1}
    \begin{tabular}{|l|l|l|}
        \hline
        {Quantity} & {Symbol} & {Comments} \\[2pt]
        \hline
        Real-space comoving position & $\x_1,\x_2$, etc & $\hat\x_1=\x_1/|\x_1|$, $\hat\x_2=\x_2/|\x_2|$ \\
        Redshift-space comoving position & $\s_1,\s_2$, etc & $\hat\s_1=\s_1/|\s_1|$, $\hat\s_2=\s_2/|\s_2|$ \\
        Real-space separation & $\r=\x_1-\x_2$ & $r=|\r|$, $\hat\r=\r/|\r|$ \\
        Redshift-space separation & $\s=\s_1-\s_2$ & $s=|\s|$, $\hat\s=\s/\s|$ \\
        Distance to pair & $\mbf{d}=(\s_1+\s_2)/2$ & $d=|\mbf{d}|$ \\
        Line of sight & $\n_1=\hat\s_1$, $\n_2=\hat\s_2$, etc & — \\
        Angular separation & $\vartheta=\cos^{-1}(\n_1\cdot\n_2)$ & — \\
        Real-space conformal distance & $\chi_1',\chi_2'$, etc & $\bm\chi'=(\chi_1',\chi_2')$ \\
        Redshift-space conformal distance & $\chi_1,\chi_2$, etc & $\bm\chi=(\chi_1,\chi_2)$ \\
        Wide-angle expansion parameter & $\epsilon=s/d$ & — \\[4pt]
        \hline
        Conformal Hubble parameter & $\calH=aH$ & units inverse length ($c=1$) \\
        Peculiar velocity & $\v(\x)$ & unitless \\
        Newtonian potential & $\Psi(\x)$ & unitless \\
        (Scaled) Peculiar velocity & $\u=\calH^{-1}\v$ & units length \\
        (Scaled) Line-of-sight velocity & $u_{\|}(\x)=\u(\x)\cdot\hat\x$ & — \\
        (Scaled) Newtonian potential & $\psi=-\calH^{-1}\Psi$ & units length \\
        (Scaled) Potential of dark matter halo & $\phi(\x)$ & — \\
        (Scaled) Potential at observer & $\psi_O=\psi(\tau_0,\x=0)$ & — \\
        (Scaled) Velocity difference & $\Delta\u=\u(\x_1)-\u(\x_2)$ & — \\
        (Scaled) Potential difference & $\Delta\psi=\psi(\x_1)-\psi(\x_2)$ & — \\
        (Scaled) Mean velocity & $\bar\u=\frac12[\u(\x_1)+\u(\x_2)]$ & — \\
        (Scaled) Mean potential & $\bar\psi=\frac12[\psi(\x_1)-\psi_O+\psi(\x_2)-\psi_O]$ & — \\[4pt]
        \hline
        Density-weighted ensemble average & $\llangle\cdots\rrangle$ & see Eqs.~\eqref{eq:dens-weighting}, \eqref{eq:lc-weighting} \\
        Pairwise velocity difference & $\llangle\Delta u\rrangle(r)\equiv\llangle\Delta\u\rrangle\cdot\hat\r$ & also known as `mean streaming velocity' \\
        Pairwise potential difference & $\llangle\Delta\psi\rrangle(r)$ & — \\
        Pairwise mean velocity & $\llangle\bar{u}\rrangle(r)\equiv\llangle\bar\u\rrangle\cdot\hat\r$ & — \\
        Pairwise mean potential & $\llangle\bar\psi\rrangle$ & — \\[4pt]
        \hline
        Mean number density & $\bar{n}_A$ & $A,B,\ldots$ tracer labels \\
        Real-space number density & $n_A(\x_1)=\bar{n}_A\sp[1+\delta_A(\x_1)]$ & — \\
        Redshift-space number density & $n^{(\mrm{s})}_A(\s_1)=\bar{n}_A\sp[1+\delta_A^{(\mrm{s})}(\s_1)]$ & — \\
        Lightcone-corrected number density & $\tilde{n}_A(\x_1)$ & — \\
        Real-space correlation function & $\xi_{AB}(r)\equiv\langle\delta_A(\x_1)\sp\delta_B(\x_2)\rangle$ & — \\
        Redshift-space correlation function & $\xi_{AB}^{(\mrm{s})}(\chi_1,\chi_2,\cos\vartheta)
        \equiv\langle\delta_A^{(\mrm{s})}(\s_1)\sp\delta_B^{(\mrm{s})}(\s_2)\rangle$ & — \\[5pt]
        \hline
    \end{tabular}
    \caption{Notation of some important quantities appearing in this work.}
    \label{tab:notation}
\end{table*}

\subsection{A probabilistic interpretation}\label{sec:p-interpretation}
In our treatment above we have taken the view that
$p(\bm\chi-\bm\chi',\bm\chi')$ gives the distribution
of displacements $\delta\bm\chi=\bm\chi-\bm\chi'$, in
a generalisation of the pairwise velocity
distribution of the standard RSD streaming model.
Here we offer an alternative and perhaps more natural way to understand
$p(\bm\chi-\bm\chi',\bm\chi')$, namely,
as the \emph{transition probability} of a galaxy pair jumping
from configuration $\bm\chi'$ to configuration $\bm\chi$ (where real and
redshift space should now be thought of as one and the same). Now, 
attention is placed on the initial and final positions, rather than
the distance between them.
This allows us to understand
$p(\bm\chi\Mid\bm\chi')\equiv p(\bm\chi-\bm\chi',\bm\chi')$ 
as a genuine probability distribution, specifically, the conditional
probability distribution of $\bm\chi$ given $\bm\chi'$,
with (fixed) mean [cf.~Eq.~\eqref{eq:mu}]
\bea
\bm{m}_{\sp|\bm\chi'}
\equiv\llangle\bm\chi'+\delta\bm\chi\rrangle
=\bm\chi'+\bm{m}(\bm\chi')\,
\eea
and a covariance unchanged from before:
\bea
\mbf{C}_{\sp|\bm\chi'}
&\equiv\llangle(\bm\chi'+\delta\bm\chi)(\bm\chi'+\delta\bm\chi)^\T\rrangle 
- \llangle\bm\chi'+\delta\bm\chi\rrangle\llangle\bm\chi'+\delta\bm\chi\rrangle^\T\nonumber\\
&=\llangle\delta\bm\chi\delta\bm\chi^\T\rrangle-\llangle\delta\bm\chi\rrangle\llangle\delta\bm\chi\rrangle^\T
=\mbf{C}(\bm\chi')\,,\nonumber
\eea
where we used that $\llangle\bm\chi'\rrangle=\bm\chi'$.
Thus by absorbing $\bm\chi'$ into the mean, the 
probability distribution in
Eq.~\eqref{eq:gsm2} then reads
\be
p(\bm\chi\Mid\bm\chi') 
=\frac{1}{2\pi|\mbf{C}|^{1/2}}
    \exp\left(-\frac12
    (\bm\chi-\bm{m}_{\sp|\bm\chi'})^\T
    \mbf{C}^{-1}\sp
    (\bm\chi-\bm{m}_{\sp|\bm\chi'})\right),
    \nonumber
\ee
which is now to be understood as the probability of the transition
$\bm\chi'\to\bm\chi$.

A useful way to view this transition is as a two-step sequence
$\bm\chi'\to\bm\chi''\to\bm\chi$, where the first transition
is due to RSD and the second due to gravitational redshift.
By the chain rule we can then write the transition probability as
\be\label{eq:p-chain}
p(\bm\chi\Mid\bm\chi')
=\int\dif^2\bm\chi'' p_\mrm{grav}(\bm\chi\Mid\bm\chi'',\bm\chi')\sp
    p_\mrm{RSD}(\bm\chi''\Mid\bm\chi')\,,
\ee
i.e.\ the transition from the initial position $\bm\chi'$ to the intermediate
position $\bm\chi''$ with probability $p_\mrm{RSD}(\bm\chi''\Mid\bm\chi')$, then
the transition from the intermediate position $\bm\chi''$ to
the final position $\bm\chi$ with probability $p_\mrm{grav}(\bm\chi\Mid\bm\chi'',\bm\chi')$, integrated over all $\bm\chi''$.

Thus the transition probability splits into two parts. However, each part
is in general linked together so cannot be modelled independently, as one might hope to do.
While $p_\mrm{RSD}$ can be taken to be the Gaussian distribution discussed so far
(but with $\bm{m}_\mrm{grav}=\mbf{C}_\mrm{grav}=\mbf{C}_\mrm{cross}=0$),
the probability $p_\mrm{grav}(\bm\chi\Mid\bm\chi'',\bm\chi')$ is more complicated due to
dependence on the entire transition history. In other words, one cannot assume $p_\mrm{RSD}$ and
$p_\mrm{grav}$ are, for instance, both Gaussians with
$p$ given by their convolution.
The obstacle preventing this reduced description of $p$ is that the transition due to RSD
and the transition due to gravitational redshift are correlated (they are generated by the same underlying density field). In the full
model this is indicated by the presence of $\mbf{C}_\mrm{cross}$, 
giving the correlations between the velocity and potential fields.
Neglecting this covariance allows us to treat the two distributions
$p_\mrm{RSD}$ and $p_\mrm{grav}$ as independent Gaussians. This
amounts to assuming that the transitions are a Markov process
(current state depending only on the previous state) so that
$p_\mrm{grav}(\bm\chi\Mid\bm\chi'',\bm\chi')=p_\mrm{grav}(\bm\chi\Mid\bm\chi'')$.
(As we will see in Section~\ref{sec:results}, $\mbf{C}_\mrm{cross}$ does
not play an important role for the dipole.)
Under this assumption Eq.~\eqref{eq:p-chain}, can be used
as the basis for another modelling strategy for 
$p(\bm\chi\Mid\bm\chi')$, one that provides an alternative to
the standard truncation beyond second order. We elaborate on
this in Appendix~\ref{app:transitions}.

In fact, there is another intriguing way to view the model.
Since the correlation function gives the excess probability
(relative to Poisson) of finding a galaxy pair at a given separation (or configuration),
the streaming model~\eqref{eq:gsm} is essentially the marginal integral
$p(\bm\chi)=\int\dif^2\bm\chi'\sp p(\bm\chi\Mid\bm\chi')\sp p(\bm\chi')$,
with $p(\bm\chi)\propto1+\xi^{(\mrm{s})}(\bm\chi)$ and
$p(\bm\chi')\propto1+\xi(\bm\chi')$ (ignoring the non-uniform
radial measure).
This point of view is suggestive and brings to mind the Fokker--Planck equation
in that the truncation of the infinite
series after second cumulant corresponds to the 
Fokker--Planck approximation~\cite{binney_tremaine}, with
the first and second cumulants analogous to the diffusion coefficients.
This hints at a dynamical way of viewing clustering in
redshift space in the sense that we can also view the marginal integral
as an integration over the `initial conditions' given by $\bm\chi'$,
in the same way that in Lagrangian perturbation theory one integrates
over the initial separations~\cite{Couchman_Bond88}.
There are further hints of such a description, as we will see in
Section~\ref{sec:origin} when we analyse the model perturbatively.

\section{Accounting for the lightcone: the complete model}\label{sec:complete}
As it stands, $\xi^{(\mrm{s})}_{AB}$ given by
Eq.~\eqref{eq:gsm2} accounts for RSD,  gravitational redshift, and their
associated wide-angle effects. But there are two other
effects which can lead to asymmetries in the apparent clustering
and which are missed in the modelling of Section~\ref{sec:2pcf}.
These effects arise when relaxing the routine (but unrealistic) assumption that the
mapping of galaxies is done on a fixed-time spatial surface.
To bring out these effects we need to build into the model that
observations are based on light received from distant sources and
therefore necessarily lie on our past lightcone.
Accounting for this leads to two additional effects which,
while small compared to RSD, also contribute terms of order $\calH/k$
in $\delta^{(\mrm{s})}$, i.e.\ formally of the same size as that of
gravitational redshift.  While the existence of these additional
terms are well established in linear perturbation theory,
including them in the current framework requires a conceptually different approach to the streaming model. Starting from
first principles, we aim to give a derivation of
these effects with a minimum of technical detail.

\subsection{Lookback time}
As discussed in Refs.~\cite{Dam:2023,Bonvin:2013ogt},
the first effect is due to different redshifts
corresponding not only to different
positions but also different times.
Thus, given a galaxy with redshift $z$ we assign to it a distance
$\chi(z)$ according to Eq.~\eqref{eq:chi-map}, and now also a time
$\tau(z)$.
Since on the lightcone distance is degenerate with time, with a
larger distance implying an earlier time, $\tau(z)$ is
naturally the lookback time, $\tau(z)=\tau_0-\chi(z)$.
Altogether we have a spacetime map:%
\be\label{eq:spacetime-map}
(\tau_0-\chi',\x)\to(\tau_0-\chi,\s)\,,
\ee
where $\x=\chi'\n$ and $\s=\chi\sp\n$ as before.
This effect is naturally included in the model~\eqref{eq:gsm2} by evaluating all quantities at the time $\tau'=\tau_0-\chi'$,
e.g.\
\be
\delta_g(\tau',\chi'\n)=\delta_g(\tau_0-\chi',\chi'\n)
\equiv\delta_g(\chi')\,.
\ee
This means that the time variable plays an active role in the 
model~\eqref{eq:gsm2}, with the line-of-sight integrals
similar to those of weak lensing.
These integrals generally depend on unequal-time
correlations, bringing an additional
layer of complexity in the evaluation of the model.
With one exception (see Section~\ref{sec:potential-nl}), it is enough to compute the
correlation functions entering the model using linear theory, meaning that
we can deal with this effect by simply scaling the power spectra according to the growth factor.

\begin{figure*}[t]
  \centering
  \includegraphics[scale=0.23]{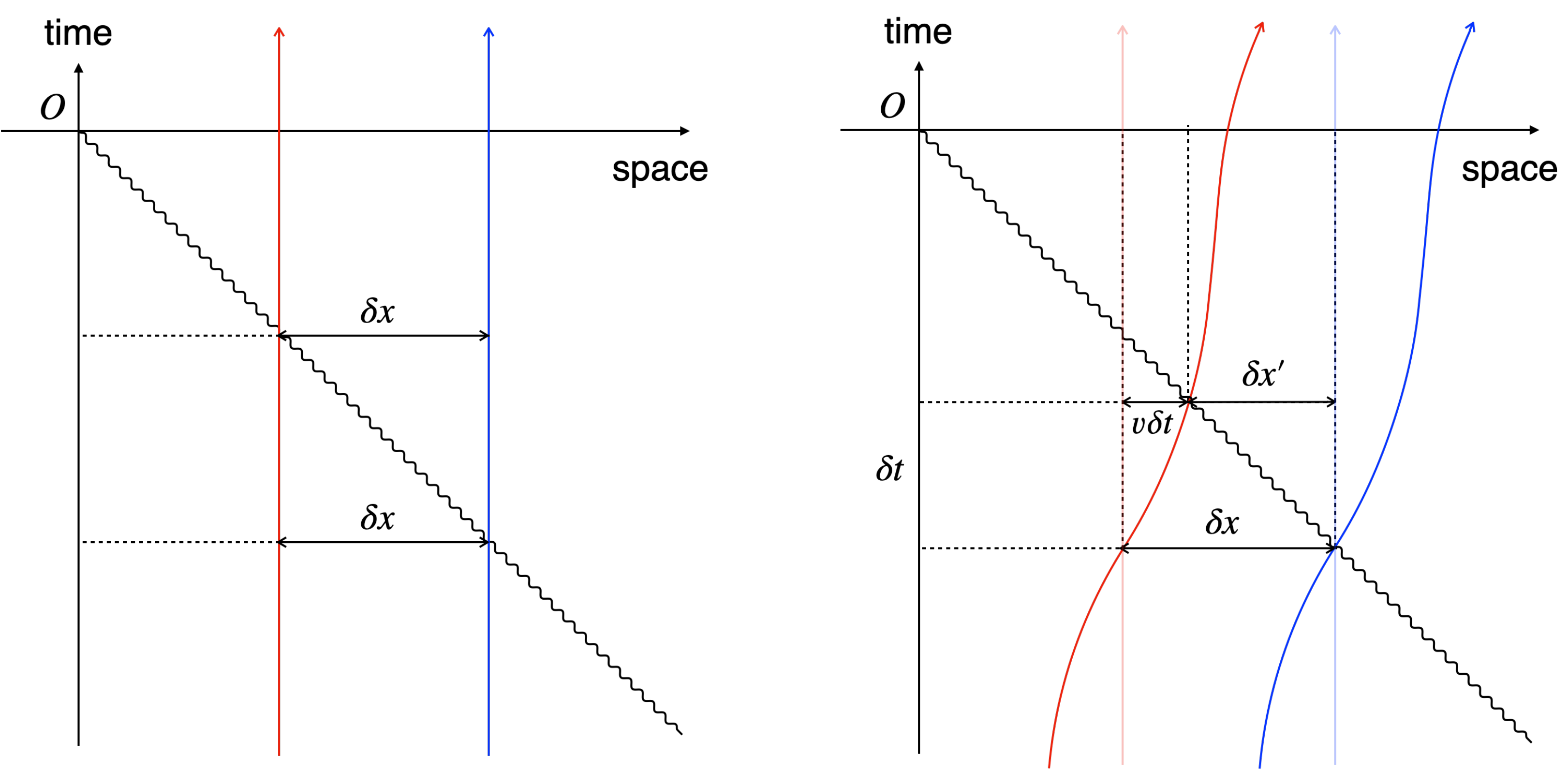}
  \caption{Spacetime diagram for two neighbouring galaxies (red and blue)
  as they cross the lightcone.
  In the left panel the galaxies are at rest with respect to the observer,
  while in the right panel they are moving away from the observer.
  The lightcone effect is that the density of crossings depends on the
  frame of observation, which will generally be boosted relative to the
  rest frame of the galaxies.
  Note that the total \emph{number} of galaxies observed is equal to the
  number of worldlines crossing the past lightcone and is of course 
  conserved in all frames. % (Lorentz invariant).
}
  \label{fig:lightcone_effect}
\end{figure*}

\subsection{Lightcone effect}\label{sec:lightcone}

In addition to the lookback time, there is also the 
so-called lightcone effect~\cite{Kaiser_2013,Bonvin:2013ogt}.
This arises from the use of light as a means to probe the positions
of galaxies, leading to a subtle correction to the number density
if the galaxies are in motion (which they typically are). 
{More precisely, when we measure the distance between two galaxies, what we do in practice is to measure the distance between two photons received at the same time by the observer. If the two galaxies are at different distances from the observer, the two photons were clearly not emitted at the same time. This leads to change in their observed separation if the galaxies are not at rest with respect to the observer.}
The overall
correction is very similar to but distinct from the Doppler shift
(the wave-like properties of light not needed).

The effect can be seen by considering the separation between two
neighbouring galaxies along the same line of sight,
as in Figure~\ref{fig:lightcone_effect}. The effect arises because
the separation $\delta x'$ inferred from light received
and the ideal separation $\delta x$ measured at
fixed time do not coincide \emph{if galaxies are observed in motion}.% 
\footnote{One should not confuse $\delta x$ nor $\delta x'$ with
the proper separation $\delta x_0$ measured at fixed time by an
observer at rest with respect to the galaxies. The
difference between $\delta x$ and $\delta x_0$ is due to
Lorentz contraction, which is not the effect we are describing.}
(Conversely, there is no effect if the galaxy pair
is at rest with respect to the observer, $\delta x'=\delta x$.)
% (when the four-velocities of the observer and galaxies coincide)
This means that the number density of galaxies
reckoned by the observer according to their lightcone is not necessarily
the same as the number density on a surface of fixed time, which is
what is typically assumed.
The effect is a consequence of the travel time of
light between objects (along the line of sight) and is present even
if one could measure the real-space positions of galaxies.

The relation between $\delta x$ and $\delta x'$ can be found by
considering the time it takes for a photon emitted by the blue galaxy
to reach the red galaxy (see Figure~\ref{fig:lightcone_effect}).
If the galaxies are at rest with respect to the observer then
the distance travelled is $\delta x$. However,
if the galaxies are moving away from the observer
with velocity $v>0$,
the red galaxy moves towards the photon (in the observer frame),
so that the photon does not have to cover the full distance $\delta x$
(as depicted in the right panel of Figure~\ref{fig:lightcone_effect}).
In the time $\delta t$ it takes for the photon from the blue galaxy
to reach the red galaxy the red galaxy covers a distance
$v\sp\delta t$, meaning that the photon only has to travel a distance
$\delta x'=\delta x-v\sp\delta t$.
Since the travel time of the photon between the blue
and red galaxies is $\delta t=\delta x'/c$, the inferred
separation is therefore 
\be
\delta x'=\delta x/(1+v/c)\,.
\ee
{This is also the separation seen by the observer, since once the photon emitted by the blue galaxy has reached the red galaxy, the two photons travel together at the same speed and reach the observer at the same time.}

In the primed coordinates the observer then measures a number density
$\dif N/\dif x'=(1+v/c)\,\dif N/\dif x$, that is, the observer sees
the density $\dif N/\dif x$ either enhanced or suppressed,
depending on the
state of motion of galaxies (at the time of photon emission).
In three dimensions this becomes
\be\label{eq:n'}
n(\x)\to \tilde{n}(\x) \equiv n(\x)\sp[1+\v(\x)/c\cdot\hat\x] \,,
\ee
where $\tilde{n}(\x)$ is the apparent number density and, as before,
$n(\x)=\bar{n}[1+\delta(\x)]$ is the underlying number density.
Note that this is a line-of-sight effect so only the radial density
is affected. This leads to a front-back asymmetry just as we have with
the gravitational redshift.
For instance, on the near side of an overdense patch, where
galaxies are infalling away from us, the number density is
biased high, whereas on the far side, where they are streaming
towards us, it is biased low. For typical speeds,
$v\simeq300\,\mrm{km\,s}^{-1}$, the bias (or asymmetry) is at the
$0.1\%$ level, small but not negligible in the context of gravitational
redshift.

As emphasized by Kaiser~\cite{Kaiser_2013}, the lightcone effect is not specific to cosmology but can
be seen in a variety of contexts.
Take for instance a photograph of a particle cloud, where at any given 
instant of time there are on average as many particles moving away from
us as towards us~\cite{Kaiser_2013}. According to the lightcone effect,
the photograph---which captures each particle at slightly different
times and not at fixed time---will show more particles moving away
from us than towards us (which is not to say that
the velocities can be inferred from the photograph itself).
A useful way to understand this
is by following the photons, arriving (simultaneously) from a given
line of sight, back down the lightcone. We can imagine replacing the
received photons by a single messenger travelling backwards in time
from the observer, passing through the particles on the lightcone
without stopping. From this perspective, the messenger
will overtake more particles moving towards them than moving away from
them---in exactly the same way that a trail runner will encounter more hikers
coming towards them than going away from them.

\subsubsection{Covariant formulation}
The lightcone effect, along with all other relativistic effects,
can of course be obtained systematically from a covariant treatment
of the number counts
of galaxies~\cite{Yoo:2009au,Bonvin:2011,Challinor:2011bk,Jeong:2011as}.
Here we will not repeat the full calculation but simply point out the
origin of the lightcone effect in the calculation.
We will hence focus only on the relevant kinematic
part needed to see the effect, ignoring gravitational effects, e.g.\
the metric perturbations which lead to order $(\calH/k)^2$ corrections
in the number counts. In this section we thus assume a
metric with line element $\dif s^2=g_{\mu\nu}\dif x^\mu\dif x^\nu
=a^2(\tau)(-\dif\tau^2+\dif\x^2)$.

In the relativistic context, both the galaxy number density and the
volume element are frame-dependent quantities. Therefore to obtain a
covariant expression of number conservation it is preferable to work with
the covariant galaxy four-current
$J^{\sp\mu}=n_\mrm{p}\sp u^\mu$, where $u^\mu$ is the galaxy
four-velocity and $n_\mrm{p}(x)$ is the physical number
density, as would be measured in the rest frame of the galaxies.
The volume element on an arbitrary three-surface is given by~\cite{1967JMP.....8.1315H}
\be
\dif\Sigma_\mu=\frac{1}{3!}\sqrt{-g}\,
\varepsilon_{\mu\nu\rho\lambda}\,
\dif x^\nu\wedge\dif x^\rho\wedge\dif x^\lambda \,,
\ee
where $g$ is the metric determinant and
$\varepsilon_{\mu\nu\rho\lambda}$ is the Levi-Civita symbol.
Note that $\dif\Sigma_\mu$ is a vector of three-forms whose
direction is specified by the four-vector normal to $\dif\Sigma_\mu$
(or, more accurately, the one-forms dual to the three-forms).
The usual Euclidean volume element 
corresponds to that of a three-surface of constant time:
$\dif\Sigma_\mu=\delta_{\mu0}\sqrt{-g}\ssp\dif^3\x$, with
$\dif^3\x
=\frac1{3!}\sp\varepsilon_{ijk}\,\dif x^i\wedge\dif x^j\wedge\dif x^k
=\dif x^1\wedge\dif x^2\wedge\dif x^3$.

The differential number of galaxies $\dif N$ is
the flux of the galaxy four-current $J^\mu$ across
three-surface $\dif\Sigma_\mu$:
\be
\dif N
={J}^{\sp\mu}\sp \dif\Sigma_\mu
=\frac{1}{3!}\sqrt{-g}\,
n_\mrm{p}\sp u^\mu\sp\varepsilon_{\mu\nu\rho\lambda}\,
\dif x^\nu\wedge\dif x^\rho\wedge\dif x^\lambda\,.
\ee
The total number of galaxies on the surface $\Sigma$
is given by the flux integral $\int_\Sigma {J}^{\sp\mu}\sp \dif\Sigma_\mu$.
To recover the conventional expression {$\dif N=n(x)\ssp\dif^3\x$},
take a three-surface of constant time,
$\dif\Sigma_\mu=\delta_{\mu0}\sqrt{-g}\ssp\dif^3\x$.
Then
$\dif N=n_\mrm{p}(x)\sp u^\mu\sp\delta_{\mu0}\sqrt{-g}\sp\dif^3\x=a^3n_\mrm{p}(x)\sp\dif^3\x=n(x)\sp\dif^3\x$,
where $a^3 n_\mrm{p}(x)= n(x)$ is the comoving number density
and we used that $u^0=a^{-1}$.
But as emphasised in the previous section,
galaxies are {counted on a surface of constant time for the observer, which differs from the} surface of constant time {for the galaxies, if the galaxies are moving with respect to us.}
Since we cannot perform local measurements,
the number counts are obtained {remotely} from light received,
and we thus define $\dif\Sigma_\mu$ with respect to our past lightcone,
such that the null vector $k^\mu\propto(1,-\hat\x)$ is normal to the
surface $\dif\Sigma_\mu$.
In other words, the `volume element' $\dif\Sigma_\mu$,
in which we observe galaxies, lives on a lightlike three-surface,
not a spacelike one.

Now to show that there is a lightcone effect of the form given by
Eq.~\eqref{eq:n'} it is enough to consider a galaxy living on the
$z$-axis at $x\equiv(\tau,\x)=(\tau_0-\chi',0,0,\chi')$ and moving purely
along the $z$-axis at a non-relativistic speed, 
$u^\mu=a^{-1}(1,0,0,v)$ in the observer's frame.
The null vector is $k^\mu\propto(1,0,0,-1)$ and the nonzero entries of
$\dif\Sigma_\mu$ are $\dif\Sigma_0=\dif x^1\dif x^2\dif x^3$ and
$\dif\Sigma_3=-\dif\tau\sp\dif x^1\dif x^2$. The observed number of
galaxies $\dif N =J^\mu \dif\Sigma_\mu$ about the point $x$ on the
lightcone is then
\bea
\dif N
% =n^\mrm{prop}_g \big(\dif\Sigma_0 + v^3\sp \dif\Sigma_3\big)
&=n_\mrm{p}(x)\,a^{-1}\sqrt{-g}\sp\big(\dif x^1\dif x^2\dif x^3
    - v\, \dif\tau\sp\dif x^1\dif x^2 \big) \nonumber\\
% &=n(x) \sp(1+v_z)\,\dif x^1\wedge\dif x^2\wedge\dif x^3 
&=n(x) \sp(1+v)\ssp\dif^3\x\,,\nonumber
\eea
where the second line is obtained by using
that on the past lightcone $\dif\tau=-\dif x^3$. If we now generalise to a
galaxy observed in arbitrary direction $\hat\x$ with velocity $\v=\v(x)$,
we will instead have $\dif N=n(x)(1+\v\cdot\hat\x)\ssp\dif^3\x$, thus
recovering Eq.~\eqref{eq:n'}.

Compared with the analysis of the previous section,
here $1+\v\cdot\hat\x$ emerges as the tilt between the galaxy's
worldline (with tangent $J^\mu$) and the
lightcone (with tangent $k^\mu$)
at the point of intersection $x$.
The $+\v\cdot\hat\x$ is because both light and galaxies are moving
with respect to the observer. If one or both are at rest with
respect to the observer there is no effect.
(Note that it is not simply the tilt between the surface of constant
time on which the galaxies live and the surface of constant time at the observer;
rather it is the tilt between the surface of constant time on which the
galaxies live and the \emph{lightcone}.)
From this we understand that for the tilt to be non-unity, and so for there to be an effect at all, it is essential that neither the galaxies nor the messenger
(the photons) are at rest in the observer's frame.
(Of course, the photons are never at rest.)
Indeed, if we boost to a frame in which the galaxies
at $x$ are at rest (so $u^\mu=a^{-1}{\delta^{\sp\mu}}_0$), there is
no lightcone effect.

\subsection{Complete expression of the number density}
\label{sec:number-density}
With the results of the preceding sections we can now give an expression for the redshift-space
density, corrected for the lightcone and lookback time effects. Begin again from
number conservation,
which now reads $n^{(\mrm{s})}_A(\s)\sp\dif^3\s=n_A(\x)(1+\v\cdot\n)\ssp\dif^3\x$.
By similar steps to the ones leading to Eq.~\eqref{eq:ns}, 
we now obtain
\be\label{eq:deltas-all}
\begin{split}
n^{(\mrm{s})}_A(\chi)
&=\frac{1}{\chi^2}\int^\infty_0\chi'^2\dif\chi'
    n_A(\chi') [1+\v(\chi')\cdot\n] \\[-3pt]
&\qquad\qquad\quad
\times\,\delD\big[\chi-\chi'-\delta\chi(\chi')\big]\,,
\end{split}
\ee
where $\delta\chi$ is given by Eq.~\eqref{eq:chi-map}, as before, and we
used the shorthand $n_A^{(\mrm{s})}(\chi)=n_A^{(\mrm{s})}(\tau_0-\chi,\chi\n)$,
$n_A(\chi')=n_A(\tau_0-\chi',\chi'\n)$, etc.
The integration here is along the line of sight, in time and space
(similar to the lensing convergence integral).
Of course Eq.~\eqref{eq:deltas-all} can be expressed in
terms of the overdensity $\delta^{(\mrm{s})}_A(\chi)$, using
$n^{(\mrm{s})}_A(\chi)=\bar{n}^{(\mrm{s})}_A[1+\delta^{(\mrm{s})}_A(\chi)]$.

Equation~\eqref{eq:deltas-all} provides a compact way of accounting for all relevant effects affecting the number density of galaxies. In this expression, wide-angle effects result from the radial measure
$\chi^2\dif\chi$ (number of sources per radial bin varies with
distance for a given solid angle);\footnote{Note there are
additional `wide-angle effects' which arise when computing the
multipoles. In our framework, these are better understood
as projection effects related to the displacement field (as opposed to radial binning).}
fluctuations in the redshift due to large-scale structure enter
through the map in the delta function; and,
as we have discussed in the previous section, the fact that light is
used to probe the positions of tracers yields a velocity correction
to the underlying density (a kind of kinematic selection effect).

\subsubsection{Recovering linear perturbation theory}

It is not difficult to show that the well-known linear 
expression~\eqref{eq:delta-lin} for the overdensity is
contained within Eq.~\eqref{eq:deltas-all}.  First,
linearizing Eq.~\eqref{eq:deltas-all} by expanding
the delta function to first order in $\delta\chi$
(see Ref.~\cite{Dam:2023}, section 4, for details), and
using that
\be
\frac{\dif}{\dif\chi}\sp f(\tau_0-\chi,\chi\n)
=\left(\frac{\partial}{\partial\chi}-\frac{\partial}{\partial\tau}\right)f(\tau,\chi\n)\,,
\ee
we obtain at leading order the compact formula
\bea
\delta_A^{(\mrm{s})}
&=\tilde\delta_A+\frac{1}{\chi^2}
    \left(\frac{\partial}{\partial\tau}-\frac{\partial}{\partial\chi}\right)
    \left(\chi^2\frac{1}{\calH}(\v\cdot\n-\Psi+\Psi_O)\right)\,, \label{eq:del-lin-compact}
\eea
where $\calH=\calH(\tau)$, $\v=\v(\tau,\chi\n)$, $\Psi=\Psi(\tau,\chi\n)$, etc,
and $\tilde\delta_A=b_A\delta+\v\cdot\n$ is the lightcone-corrected overdensity
(as observed in the absence of redshift fluctuations).%
\footnote{With selection effects the right-hand side of
$\tilde\delta_A=\delta_A+\v\cdot\n$ picks up additional terms, also proportional
to $\v\cdot\n$.}
Note that here we have reinstated the $1/\calH(\tau)$ since it is
also differentiated.

The form of Eq.~\eqref{eq:del-lin-compact} shows at once the structure underlying the
various terms, signs, prefactors and derivatives in
Eq.~\eqref{eq:delta-lin}. In particular, the time derivative traces
back to the lookback time where by expanding the full formula~\eqref{eq:deltas-all} we are in effect computing the Jacobian of the spacetime map~\eqref{eq:spacetime-map}.

Now, evaluating the derivatives in Eq.~\eqref{eq:del-lin-compact}
we see that all terms in Eq.~\eqref{eq:delta-lin} are
indeed recovered:
\bea
\delta_A^{(\mrm{s})}
&=\delta_A+\v\cdot\n-\frac1\calH\frac{\partial}{\partial\chi}(\v\cdot\n-\Psi) 
    +\frac1\calH(\dot\v\cdot\n-\dot\Psi) \nonumber\\
&\qquad\qquad-\left(\frac{2}{\calH\chi}+\frac{\dot\calH}{\calH^2}\right)(\v\cdot\n-\Psi+\Psi_O)\,.
\label{eq:lin-recovery}
\eea
In fact, we have also recovered additional terms which do not appear in
Eq.~\eqref{eq:delta-lin} (but which are contained in the relativistic
derivation~\cite{Bonvin:2011,Challinor:2011bk,Yoo:2009au,Jeong:2011as}) such as $\Psi$ and $\dot\Psi$; these are all
of order $(\calH/k)^2$, and so have been neglected in
Eq.~\eqref{eq:delta-lin}.
We could have even gone further and accounted for other
$(\calH/k)^2$ terms. For example,
it is straightforward to add the ISW contribution to the redshift
displacement~\eqref{eq:chi-map}, which would lead to several extra terms in
Eq.~\eqref{eq:lin-recovery}.
% $\delta\chi\to\delta\chi-\calH^{-1}\int^\chi\dif\chi'(\dot\Phi+\dot\Psi)$.
The transverse Doppler effect (a second-order contribution to
redshift) may also be included in a similar way.

That we are able to recover Eq.~\eqref{eq:delta-lin} in very few steps derives from
the fact that the redshift perturbation, taking us from real space to redshift space,
determines most of the terms in the full expression of
$\delta^{(\mrm{s})}$, as was pointed out in Ref.~\cite{DiDio:2020jvo}.
As we can see, the large number of terms is simply due to the
product rule.

\subsection{Complete model of the correlation function}
\label{sec:lightcone_model}
The complete model for the redshift-space correlation function,
including all $\calH/k$ effects, follows from Eq.~\eqref{eq:deltas-all}.
The derivation is similar to the one before, the main difference now being that
the underlying real-space statistics are slightly modified due to the
lightcone effect.
However, Eq.~\eqref{eq:n'} indicates that this simply amounts to
the following replacement throughout the model: $1+\delta_A(\x_1)\to[1+\delta_A(\x_1)][1+\v(\x_1)\cdot\n_1]$
and $1+\delta_B(\x_2)\to[1+\delta_B(\x_2)][1+\v(\x_2)\cdot\n_2]$.
In particular, this leads to the following change
in the normalisation of the density-weighted averages:
\bea
1+\xi_{AB}(r)
&\to\langle[1+\delta_A(\x_1)][1+\v(\x_1)\cdot\n_1] \nonumber\\[-1pt]
&\qquad\times
        [1+\delta_B(\x_2)][1+\v(\x_2)\cdot\n_2]\rangle \label{eq:weighting-replace}\\[1pt]
&=\big[1+\xi_{AB}(r)\big]
\big[1+\llangle v_{\|1}+v_{\|2}+v_{\|1}v_{\|2}\rrangle\big]\,. \nonumber
\eea
Thus the underlying real-space correlation function is
kinematically `modulated'. This means that even before accounting
for RSD we will observe stronger clustering in regions where the
galaxy pair is streaming away from us compared to where it is streaming
towards us.

Finally, by repeating the same steps that led to Eq.~\eqref{eq:gsm}, and
using that the underlying correlation function is replaced with
Eq.~\eqref{eq:weighting-replace}, we have for the complete model
\bea
1+\xi^{(\mrm{s})}_{AB}(\bm\chi)
&=\frac{1}{\chi_1^2\sp\chi_2^2}
\int\dif\chi_1'\sp\chi_1'^2
\int\dif\chi_2'\ssp\chi_2'^2\,
    \big[1+\xi_{AB}(\bm\chi')\big]
     \nonumber\\[2pt]
&\qquad\times
    \big[1+\llangle v_{\|1}+v_{\|2}+v_{\|1}v_{\|2}\rrangle\big]\ssp
    p(\bm\chi\Mid\bm\chi')\,,
\label{eq:xi_final}
\eea
where the time dependence in all quantities is now actively
integrated over since there is a $\chi'$ dependence in $\tau$, e.g.\
$v_{\|i}=\v(\tau_0-\chi'_i,\chi_i'\n_i)\cdot\n_i$.
The form of this expression is essentially the same as
Eq.~\eqref{eq:gsm} (had we absorbed the kinematic
part into the definition of $\xi_{AB}$).
However, the integration is now a line integral directed down the lightcone
instead of on a fixed-time slice.
Apart from the kinematic factor, there is one other difference 
when compared to the previous expression~\eqref{eq:gsm2}. This difference is
found in the cumulants of $p$, which are now weighted by
$\tilde{n}_A(\x)=n_A(\x)(1+\v\cdot\hat\x)$ instead of $n_A(\x)$.
The form of $p$ is otherwise the same.
Unless otherwise stated, double brackets $\llangle\cdots\rrangle$ will
henceforth denote the following \emph{lightcone-corrected}
density-weighted average:
\bea\label{eq:lc-weighting}
\llangle F(\x_1,\x_2)\rrangle
&=\frac{\langle \tilde{n}_A(\x_1)\ssp \tilde{n}_B(\x_2) F(\x_1,\x_2)\rangle}
    {\langle \tilde{n}_A(\x_1)\ssp \tilde{n}_B(\x_2)\rangle}\,,
\eea
replacing the previous one given by Eq.~\eqref{eq:dens-weighting}.
(But note that the distinction between this average and the previous one
is not particularly important for the level of approximation used
in this work, as we will see in Section~\ref{sec:pair}.)

\section{Model inputs}\label{sec:ingredients}
Our focus up to now has been on the formal relation between 
correlations in redshift space and those
in real space, independent of the specific biased tracer used or
how the underlying fields might evolve gravitationally. We will
henceforth specialise to dark matter haloes, as per the
RayGalGroup catalogues (used in the model validation to follow).

Clearly a number of real-space correlation functions are needed
in the model~\eqref{eq:xi_final}. These enter through the components
of $\bm{m}$ and $\mbf{C}$ of the Gaussian distribution function.
Although these components consist of two-point functions,
because of the density weighting they will
generally also contain contributions from higher-order statistics,
namely the (integrated) bispectrum  and the (integrated) trispectrum.
In this work we neglect such contributions and
consider only contributions which are
at most quadratic in fields. We thus approximate,
e.g.\
$\llangle u_\|(\x_1)\rrangle
\simeq\langle u_\|(\x_1)\sp\delta(\x_2)\rangle[1+\xi_{AB}(r)]^{-1}$ and
$\llangle u_\|(\x_1)\sp u_\|(\x_2)\rrangle
\simeq\langle u_\|(\x_1)u_\|(\x_2)\rangle[1+\xi_{AB}(r)]^{-1}$
(where terms such as $\langle u_\|(\x_1)\sp\delta(\x_1)\sp\delta(\x_2)\rangle$
have been neglected). As shown in Appendix~\ref{app:corrfuncs}, with this approximation, $\bm{m}$ and $\mbf{C}$ are composed of seven (scalar) correlation functions between $\delta$, $\u$, and $\psi$: $\xi_{\delta\delta}$, $\xi_{u\delta}$, $\xi_{\psi\delta}$, $\xi_{uu}^\parallel$, $\xi_{uu}^\perp$, $\xi_{u\psi}$ and $\xi_{\psi\psi}$. 
We evaluate these correlations at linear order in perturbation theory and
assume linear bias. As we will see in Section~\ref{sec:results} when comparing with numerical simulations, these approximations work well in the mildly nonlinear regime. 

There is however one exception to this, which is the contribution from $\llangle\psi(\x)-\psi_O\rrangle$ that enters $\bm{m}_\mrm{grav}$ [Eq.~\eqref{eq:mrsd-mgrav}]. For this term, linear perturbation theory is not enough and a good estimate of this inherently nonlinear quantity is needed to reproduce simulation measurements of the dipole (see Section~\ref{sec:results}). 
The reason is that $\llangle\psi(\x)-\psi_O\rrangle$ contains the one-point function $\langle\psi(\x)\sp\delta(\x)\rangle$ (from the density weighting), which is sensitive to small-scale nonlinearities and to shot noise.
In contrast, the analogous one-point function of RSD (entering $\bm{m}_\mrm{RSD}$) vanishes
due to parity, $\langle u_\|(\x)\delta(\x))\rangle=0$. Hence $\langle\psi(\x)\sp\delta(\x))\rangle$ is the only one-point function in $\bm{m}$ for which we need to go beyond linear theory.\footnote{There is also the velocity dispersion, another nonperturbative input. Since this parameter
(which enters $\mbf{C}_\mrm{RSD}$) tends to affect the even multipoles, we
simply estimate it using linear theory. We note that a more principle approach
should include a one-halo correction for the velocity correlations, so that
the velocity dispersion carries a population dependence.}
In practice this nonperturbative input can either be estimated given a prescription of the (highly) nonlinear regime, or it can be measured from numerical simulations, or else it can be treated as a free parameter, fitted together with cosmological parameters to data. In the following we compute $\llangle\psi(\x)-\psi_O\rrangle$
using the halo model~\cite{Cooray:2002dia}. This gives at once
the one-point function contained within.

\subsection{Density-weighted potential: nonlinear treatment}
\label{sec:potential-nl}
Since $\llangle\psi(\x)-\psi_O\rrangle$ is a two-point correlation it decomposes
in the usual way into a one- and two-halo term:
\be\label{eq:psi-halo-split}
\llangle\psi(\x)-\psi_O\rrangle=\llangle\psi(\x)-\psi_O\rrangle_\mrm{1h}+\llangle\psi(\x)-\psi_O\rrangle_\mrm{2h}\,.
\ee
The first term is the contribution from the internal halo structure
(one-halo term) and the second term is the contribution
from clustering as described by perturbation theory
(two-halo term).%
\footnote{Due to the density weighting, the one-halo term
actually arises from an integration over a two-point statistic, while the
two-halo term is due to an integration
over a three-point statistic.  The actual one-halo term vanishes upon integration.
See Appendix~\ref{app:shot} for details.}

The important aspect of this statistic is the non-locality of the
potential $\psi(\x)$,
because of which the one-halo term, nominally the
shot noise contribution, yields a significant correction to the
two-halo term \emph{at all separations}. As we will see in Section~\ref{sec:pair},
the one-point function $\langle\psi(\x)\sp\delta(\x)\rangle$ contained within
is not well modelled using perturbation theory. The nature of density
weighting skews the gravitational potential towards regions of high density
(density contrast of several hundred) where tracers are likely to be found
but where perturbation theory fails.
In effect, the halo model augments the large-scale potential
(described by perturbation theory) by the small-scale
potential of the halo itself. The basic picture is of a smooth potential (sourced by
large-scale structure) punctuated by sharp, local deviations due to dense dark matter haloes.
Hence at the sites of tracers (the haloes), we find a large
enhancement in the size of the one-point function. Since photons are emitted from the halos, they are affected by the small-scale potential. At linear order, this contribution vanishes, since the local potentials of two distant galaxies are uncorrelated and only the large-scale linear potential contributes to the correlation function. At nonlinear order, however, this is not the case any more and the signal is strongly boosted by this contribution.

To compute the one- and two-halo terms in Eq.~\eqref{eq:psi-halo-split} we first
express the number density $n_A(\x_1)$ and $n_B(\x_2)$, which enter through Eq.~\eqref{eq:dens-weighting}, as a superposition of halos of mass $M_A$ and $M_B$
(or range of masses centred on $M_A$ and $M_B$). The gravitational potential
$\psi(\x)$ is sourced by haloes of all masses, distributed
according to some mass function, with each halo
described by the same characteristic density profile. 
Details of the calculation are given in Appendix~\ref{app:shot}; here we summarize
the main results. For the one-halo terms we have
\bea
\begin{split}
\!\!\llangle\psi(\x_1)-\psi_O\rrangle_\mrm{1h}
&=[\phi_A(0)-\phi_A(\chi_1')]-[\phi_B(\chi_2')-\phi_B(r)]\,,\label{eq:1halo}\\
\!\!\llangle\psi(\x_2)-\psi_O\rrangle_\mrm{1h}
&=[\phi_B(0)-\phi_B(\chi_2')]-[\phi_A(\chi_1')-\phi_A(r)]\,,
\end{split}
\eea
where $\chi_1'$ and $\chi_2'$ are the comoving distance, in real space, to $\x_1$ and $\x_2$ and $r$ is the real-space separation between them.
Here $\phi_A$ is the potential of an individual halo
in population $A$;
treated as an isolated body with spherically-symmetric density
profile $\rho_A(r)$, the halo potential is given by
\be
\phi_A(r)=\lambda
    \int^\infty_r\dif r'\sp \frac{GM_{A}(r')}{r'^2}
    \label{eq:phiA}\,,
\ee
with analogous expressions for tracer $B$.
Here $\lambda\equiv\calH^{-1}a^2$, $G$ is Newton's constant of gravitation,
and $M_A(r)=4\pi\int^{r}_0 \rho_A(x)\sp x^2\dif x$ is the
mass profile, normalised so that $M_A(r)=M_A$ for $r>R_\mrm{vir}$,
with $R_\mrm{vir}$ the virial radius.
The tracer dependence is through parameters like the mass
which characterise the density profile.
Note that if $r>R_\mrm{vir}$ we have
$\phi_A(r)=\lambda\sp GM_A/r$, i.e.\ the details of the
density profile are irrelevant and the potential is given
as if it was sourced by a point particle of mass $M_A$. 

The two-halo terms are (see again Appendix~\ref{app:shot} for details)
\bea
\llangle\psi(\x_1)-\psi_O\rrangle_\mrm{2h}
&=\frac{[\xi^\mrm{HM}_{\psi\delta_A}(0)-\xi^\mrm{HM}_{\psi\delta_A}(\chi_1')]-[\xi^\mrm{HM}_{\psi\delta_B}(\chi_2')-\xi^\mrm{HM}_{\psi\delta_B}(r)]}{1+\xi_{AB}(\chi_1',\chi_2',\cos\vartheta)},\label{eq:2halos}\\[-2pt]
\llangle\psi(\x_2)-\psi_O\rrangle_\mrm{2h}
&=\frac{[\xi^\mrm{HM}_{\psi\delta_B}(0)-\xi^\mrm{HM}_{\psi\delta_B}(\chi_2')]-[\xi^\mrm{HM}_{\psi\delta_A}(\chi_1')-\xi^\mrm{HM}_{\psi\delta_A}(r)]}{1+\xi_{AB}(\chi_1',\chi_2',\cos\vartheta)},\nonumber
\eea
where $\xi^\mrm{HM}_{\psi\delta_A}(r)$, corresponding to
$\langle\psi(\x)\sp\delta_A(\x')\rangle$, is given by
\be\label{eq:Delta-psi-2h}
\xi^\mrm{HM}_{\psi\delta_A}(r)
=\int\dif M\frac{\dif n}{\dif M}\int\dif^3\x\,
    \xi_{hh}(\r-\x,M_A,M)\ssp\phi_M(\x)\,,
\ee
and likewise for $B$.
Here $\dif n/\dif M$ is the halo mass function,
$\xi_{hh}$ is the halo correlation function, and
$\phi_M$ is the potential~\eqref{eq:phiA} of a halo of mass $M$
[with $\phi_A=\phi_{M_A}$ and $\phi_B=\phi_{M_B}$].
We assume
$\xi_{hh}(\r,M,M')=b_1(M_A)\sp b_1(M')\sp\xi_{mm}(r)$ with $b_1(M)$
estimated using the peak--background split.
For numerical work we use an NFW profile~\cite{NFW}
(truncated at the virial radius $R_\mrm{vir}$) and a
Sheth--Tormen mass function~\cite{Sheth:1999mn}.

As might be noticed, the normalisation
$1+\xi_{AB}$ from the density weighting is present in the
two-halo term but not in the one-halo term. As shown in
Appendix~\ref{app:shot}, the apparent dependence of
$\llangle\psi(\x)-\psi_O\rrangle_\mrm{1h}$ on the normalisation
drops out upon taking into account the third-order statistic
$\langle\psi\delta\delta\rangle$ contained within (in addition
to the second-order statistic $\langle\psi\delta\rangle$).
This ought to be the case for the one-halo term, which should
not depend on clustering statistics like $\xi_{AB}$.

In Section~\ref{sec:pair} we will revisit these quantities
and show that $\xi^\mrm{HM}_{\psi\delta}(r)$ is
given to a good approximation by linear theory for virtually all separations, 
with corrections being higher derivative (suppressed for $r\gtrsim R_\mrm{vir}$).
More importantly, we will see that the one-halo term represents a sizeable
correction to linear theory.

\section{Numerical results}\label{sec:results}
In this section we validate the model~\eqref{eq:xi_final}
against $N$-body simulations, focussing on the dipole moment.

\subsection{Multipole decomposition}
To extract the multipoles from the model we
will first need to make a change of coordinates. 
In particular, we change from spherical coordinates
$(\chi_1,\chi_2,\cos\vartheta)$, which we have used so far,
to standard coordinates
$(s,d,\mu)$, where $s$ is the separation, $d$ is some
measure of the distance to the pair, and $\mu=\hat\s\cdot\n$
is the angle of the pair with respect to the line of sight
$\n$. 
In the wide-angle regime the latter set of coordinates are not
uniquely specified but depend on the choice of line of sight
(with respect to which the multipoles are defined).
In this work the line of sight is chosen according to the
mid-point parametrisation; this is the most
`symmetric' choice in that it minimises the
wide-angle effects~\cite{Gaztanaga:2015jrs}.\footnote{The relation of the mid-point
parametrisation to the more practical
end-point parametrisation can be found in Ref.~\cite{Dam:2023}.}
Definitions and coordinate relations can be
found in Appendix~\ref{app:formulas}.
After making this change of coordinates, we obtain the
multipoles of $\xi^{(\mrm{s})}$ through numerical integration
of the standard formula:
\be\label{eq:xi_ell}
\xi_\ell(s,d)
=\frac{2\ell+1}{2}\int^1_{-1}\dif\mu\,
    \mathcal{L}_\ell(\mu)\,\xi^{(\mrm{s})}(s,d,\mu)\,,
\ee
where $\mathcal{L}_\ell$ is the Legendre polynomial of degree $\ell$. Note that the multipoles $\xi_\ell$ depend on the separation $s$ \emph{and} the distance $d$ to the
pair.

On the practical matter of
evaluating Eq.~\eqref{eq:xi_ell}, some care is required
at the upper and lower limits, $\mu=\pm1$.
At these points the lines of sight coincide ($\n_1=\n_2$)
meaning that 
Eq.~\eqref{eq:gsm2} is integrated through a 
configuration in which both galaxies `collide' ($\x_1=\x_2$). 
As a result $\mbf{C}$ is singular
and we essentially have a single line of sight, or
one less degree of freedom in our problem.
These edge cases, and those where $\mu$ is close
to but not quite equal to $\pm1$, correspond to extremely
flattened triangles in Figure~\ref{fig:config},
i.e.\ where wide-angle
effects are negligible and one should revert to using
the streaming model in the distant-observer limit given by
\begin{align}
1+\xi^{(\mrm{s})}(s_\|,s_\perp)
&=\int^\infty_{-\infty}\dif r_\|\,\big[1+\xi(r)\big]
\frac{1}{\sqrt{2\pi}\ssp\sigma_{12}(r_\|,s_\perp)}\nonumber\\
&\qquad\times
\exp\left(-
\frac{[s_\|-r_\|-m_\|(r_\|,s_\perp)]^2}{2\sp\sigma_\|^2(r_\|,s_\perp)}\right) \,,
\label{eq:gsm-dol}
\end{align}
where  $r=(r_\|^2+r_\perp^2)^{1/2}$,
$r_\|$ is the (real-space) separation along the line of sight, and
$r_\perp=s_\perp$ is the separation perpendicular to the line of sight.
We also have the first and second moments,
$m_\|(r_\|,r_\perp)=\llangle\Delta\chi\rrangle$
and $\sigma_\|^2(r_\|,r_\perp)=\llangle\Delta\chi^2\rrangle$,
with $\Delta\chi\equiv\delta\chi(\x_1)-\delta\chi(\x_2)$
wherein $\n_1=\n_2$. (In the absence of gravitational redshift
these are the familiar pairwise velocity moments.) As we have shown in Ref.~\cite{Dam:2023} (see Appendix B), 
Eq.~\eqref{eq:gsm-dol} is indeed a limiting case of the full
wide-angle model~\eqref{eq:gsm2}.

\begin{figure*}[t!]
\centering
\begin{subfigure}[b]{0.93\textwidth}
   \includegraphics[width=\linewidth]{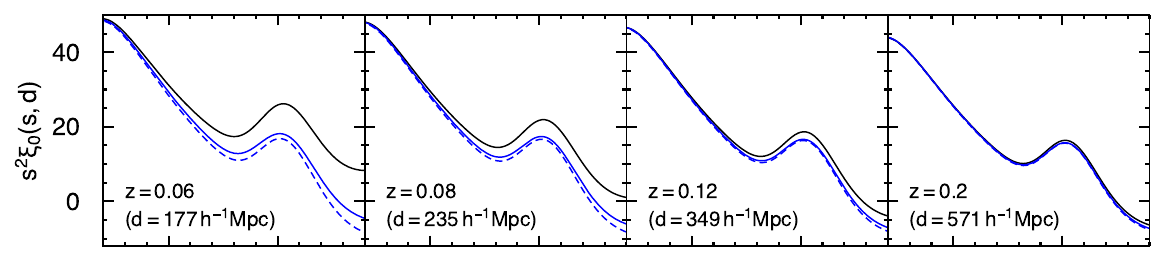}
   % \caption{}
   \label{fig:monopole} 
\end{subfigure}
\vskip -0.6cm
\begin{subfigure}[b]{0.93\textwidth}
   \includegraphics[width=\linewidth]{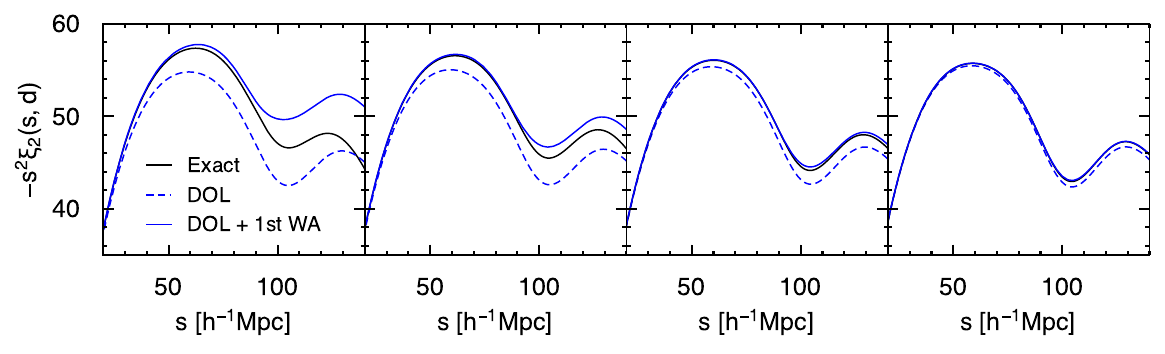}
   % \caption{}
   \label{fig:quadrupole}
\end{subfigure}
\vspace{-20pt}
\caption{Comparison between the Gaussian streaming model in the
        exact wide-angle regime (solid black) and in the usual distant-observer
        limit (DOL; dashed blue). The solid blue curves show how close we can
        get to the exact result by simply adding the leading-order wide-angle
        correction to the DOL curve. Top row shows the monopole, bottom row the
        quadrupole. For this comparison we turn off all effects except RSD and
        associated wide-angle effects (no gravitational redshift, lightcone, etc).
        Here we have used the mid-point parametrisation, adopted a {\it Planck} 2018 cosmology, and assumed no galaxy bias.}
\label{fig:wide-angle-multipoles}
\end{figure*}

\subsection{Impact of wide-angle effects on even multipoles}\label{sec:wa-even}
Before we turn to the complete model, let us first focus exclusively
on the wide-angle aspect of the model, ignoring all effects except
RSD (no gravitational
redshift, lightcone effect, lookback time, etc). The results for the even multipoles
are shown in Figure~\ref{fig:wide-angle-multipoles}, illustrating
the perturbativity of the wide-angle corrections as we go to higher redshift
(or distance).
As a basis of comparison we also show the standard Gaussian
streaming model (in the distant-observer limit). As expected, we see large
wide-angle corrections at large separations. For fixed distance $d$
(or redshift) these corrections grow with the opening angle $\vartheta$,
and in the perturbative regime ($\vartheta\ll1$) we use
$\epsilon\equiv s/d\approx \vartheta$ as a small parameter to organize the
wide-angle corrections.
These large separations are within the linear regime and do not
necessarily imply a need for a nonlinear dynamical treatment;
linear dynamics are sufficient, although perhaps not near the
BAO scale~\cite{Castorina:2018nlb}.
However, one can still have large wide-angle corrections
at small separation if the distance $d$ is small, so that
$\epsilon$ is large (see the lowest redshift bins). 

To get a sense of how nonperturbative the departures are
we also show an approximate model in which we simply add the leading-order
wide-angle correction to the standard streaming model~\eqref{eq:gsm-dol}
(which for even multipoles goes as $\epsilon^2$, not $\epsilon$).
Comparing this model with the exact model
(which contains wide-angle corrections at all orders in $\epsilon$),
it is seen that both models converge by a redshift $z=0.2$, with
quadrupole already well converged at $z=0.12$.
Furthermore, the quadrupole at $z=0.08$ is also reasonably well
accounted for by just the linear correction up to 
$s=100\,h^{-1}\mrm{Mpc}$, while the monopole appears less so.
This is because of the integration of the
correlation function over $\mu$; some wedges are
less susceptible to wide-angle effects than others,
e.g.\ the ones aligned with the line of sight.
The monopole gives equal weight to all wedges ($\mathcal{L}_0=1$),
and in particular those where the galaxy pair is perpendicular
to the line of sight ($\mu\simeq0$); here the opening angle
$s/d$ is largest (for fixed separation $s$) and thus large
wide-angle corrections. By contrast the quadrupole gives
about half weight, $\mathcal{L}_2(\mu)=(3\mu^2-1)/2\simeq-1/2$,
to the same configurations and thus the quadrupole
is driven more by wedges aligned with the line of sight,
where as we mentioned we are less prone to wide-angle effects.

To determine the relevance of wide-angle effects, we compare them with the uncertainties on the measurements of the multipoles. For a survey like the DESI Bright Galaxy Sample, we find that wide-angle corrections for the monopole are of similar size as the 1$\sigma$ uncertainties in the lowest redshift bin ($z=0.15$), see Figure~\ref{fig:BGS} in Appendix~\ref{app:WA-impact}. 
At redshift $z=0.25$, wide-angle corrections are still about half of the 1$\sigma$ uncertainties. Since the goal of surveys like DESI is to keep systematic effects at their minimum, such large biases should be avoided and wide-angle effects should be carefully accounted for in the modelling.

\subsection{Model validation}

\subsubsection{RayGalGroup simulation}
For this comparison we use the publicly available halo catalogues%
\footnote{\url{https://cosmo.obspm.fr/public-datasets/raygalgroupsims-relativistic-halo-catalogs/}}
from the RayGalGroup simulation,
a dark-matter only $N$-body simulation of $4096^3$ particles
in a box of size $(2625\,h^{-1}\mrm{Mpc})^3$~\cite{Breton:2018wzk,Rasera:2021mvk}.
Haloes are identified using a pFoF
halo-finding algorithm with linking length $b=0.2$
and a minimum particle count of 100.
The particle mass-resolution is $1.88\times10^{10}\,h^{-1}M_\odot$.
The simulations were initialised from a power spectrum
using a WMAP7 cosmology with $\Omega_m=0.25733$,
$\Omega_b=0.04356$, $\Omega_r=8.076\times10^{-5}$,
$h=0.72$, $A_s=2.42\times10^{-9}$, and $n_s=0.963$.

The catalogues are constructed from the $4\pi$
lightcone, with the observed position of the halo connected to
its true position by ray tracing backwards from the observer to the
source. The redshift and angular position, obtained by
solving the null geodesic equations in the weak-field approximation,
contain the total effect of Doppler shift
(including the relativistic transverse part),
gravitational redshift, gravitational lensing, time delay,
and integrated Sachs--Wolfe.
In the halo catalogues, each of these contributions to the total redshift
have been separated. In our comparison, following
Ref.~\cite{Saga:2020tqb}, we keep only the contributions due to
Doppler shift and gravitational redshift. 
The lightcone and lookback time effects are not included in this
comparison. 
The redshift range is $z_\mrm{min}=0.05$ to $z_\mrm{max}=0.465$;
the effective redshift is $z=0.341$, corresponding to a distance
$d=950\,h^{-1}\mrm{Mpc}$.

\begin{figure*}[t]
\centering
\begin{subfigure}{.5\textwidth}
  % \centering
  \includegraphics[width=0.9\linewidth]{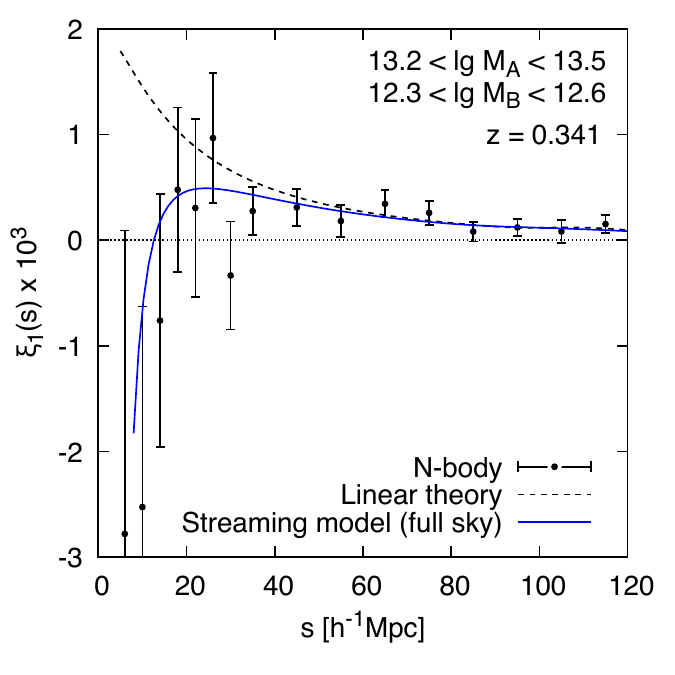}
\end{subfigure}%
\begin{subfigure}{.5\textwidth}
  % \centering
  \includegraphics[width=0.9\linewidth]{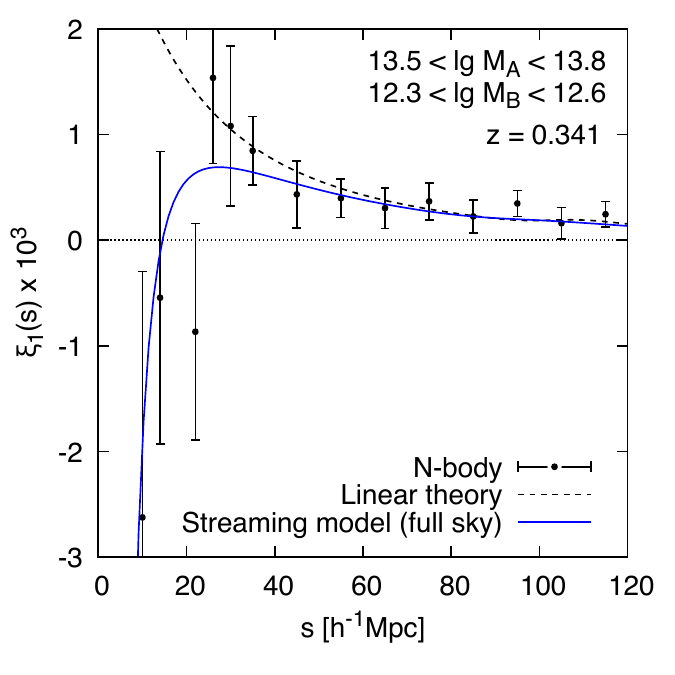}
\end{subfigure}
\caption{Comparison between model prediction~\eqref{eq:xi_final}
    and dipole measurements from the RayGalGroup simulation.
    Left and right panels differ by choice of high-mass haloes:
    on the left we have mean mass $M_A=2.2\times10^{13}\,h^{-1}M_\odot$,
    while on the right $M_A=4.5\times10^{13}\,h^{-1}M_\odot$; in both
    $M_B=0.28\times10^{13}\,h^{-1}M_\odot$.
    The model is evaluated using the mass ranges shown in the plot.
    The statistical uncertainty is estimated using the jack-knife method
    from 32 resamplings.
    Note that measurements at $s\leq32\,h^{-1}\mrm{Mpc}$ have a
    slightly finer binning than those at $s>32\,h^{-1}\mrm{Mpc}$.}
\label{fig:1600x100}
\end{figure*}

\begin{table}[t!]
   \centering
   \caption{Halo catalogues used in this work. Here
   $b_1$ is the estimated linear bias and $c$
   the estimated mean halo concentration.}
   \vspace{1pt}
   \begin{tabular}{@{} lcccr @{}}
     \toprule
      No.\ of haloes & No.\ of particles & Mean mass [$h^{-1}M_\odot$] & $b_1$ & $c$ \\[1pt]
      \hline
      $5.4\times10^6$  & 100--200 & $0.28\times10^{13}$  & $1.08$ & $8.2$ \\
      $1.0\times10^6$  & 800--1600 & $2.2\times10^{13}$  & $1.69$ & $6.6$ \\
      $0.4\times10^6$  & 1600--3200 & $4.5\times10^{13}$ & $2.07$ & $6.1$ \\
      \hline\hline
   \end{tabular}
   \label{tab:raygal}
\end{table}

The dipole is estimated from the cross-correlation of a high-mass
bin with mean mass $2.2\times10^{13}h^{-1}M_\odot$ (tracer $A$) and a
low-mass bin with mean mass $0.28\times10^{12}h^{-1}M_\odot$ (tracer $B$).
We also consider a higher-mass bin with mean mass
$4.5\times10^{13}h^{-1}M_\odot$. See Table~\ref{tab:raygal} for further details.

\subsubsection{Dipole comparison}

To compute the dipole, we evaluate the seven linear two-point correlation functions that enter into $\bm{m}$ and $\mbf{C}$ (using the WMAP7 cosmology, as per the RayGal simulation). Moreover, we compute the nonlinear one-point correlation functions of Eq.~\eqref{eq:1halo} according to the specifications of the halo catalogue. The linear galaxy bias is calculated from
the peak--background split using a Sheth--Tormen mass function~\cite{Sheth:1999mn},
which is integrated over mass ranges given in Table~\ref{tab:raygal}.
Specifically, tracer $A$ ranges from $M_{A1}=10^{13.2}h^{-1}M_\odot$
to $M_{A2}=10^{13.5}h^{-1}M_\odot$, while tracer $B$ ranges from 
$M_{B1}=10^{12.3}h^{-1}M_\odot$ to $M_{B2}=10^{12.6}h^{-1}M_\odot$.
The formula
for $\llangle\psi(\x)-\psi_O\rrangle_\mrm{1h}$, in the case of
mass bins of finite width, has the same form as Eq.~\eqref{eq:1halo} but
with the halo potentials replaced by their mass-weighted average:
\be
\phi_A(\x)\to
\bar\phi_A(\x)
\equiv \frac{1}{\bar{n}_A}
    \int^{M_{A2}}_{M_{A1}}\dif M\ssp\frac{\dif{n}}{\dif M}\,\phi(\x;M)\,,
\ee
where
$\bar{n}_A=\int^{M_{A2}}_{M_{A1}}\dif M\ssp{\dif n}/{\dif M}$ is
the mean density of population $A$ (and likewise for
population $B$).

Figure~\ref{fig:1600x100} compares our theoretical model of the dipole with measurement from the RayGal catalogues, for different galaxy populations. At separations $s\lesssim30\,h^{-1}\mrm{Mpc}$ we see a clear
departure of linear theory (which gives a positive dipole) from the measurements (which cross zero and become negative). Our
model prediction on the other hand provides a better fit to the measurements. 
The point of zero crossing is also broadly consistent
with the simulations in both panels. In the right panel we see however a slight enhancement in the amplitude of the dipole, which is not fully captured by our model. But the large uncertainty in the measurements, due to lower halo numbers, makes it difficult to draw any firm conclusions here.
In Section~\ref{sec:origin} we will discuss in detail the physical origin of the turnover.
In brief, the turnover and eventual zero crossing
is due to a cancellation in the effects of wide-angle RSD and
gravitational redshift, marking a regime
where the gravitational redshift begins to dominate over RSD
and other kinematic effects.

Note that to reproduce the turnover it is essential that we
have a realistic estimate of the one-point functions.
These enter the density-weighted potential~\eqref{eq:psi-halo-split}
by way of $\bm{m}_\mrm{grav}$. As mentioned in Section~\ref{sec:potential-nl},
these one-point functions are estimated using the halo model, for which there is
a contribution coming from the one-halo term and another from the two-halo term.
For the population with mean mass $M_A=2.2\times10^{13}h^{-1}M_\odot$
we find $\phi_A(0)=0.036\,h^{-1}\mrm{Mpc}$, which represents a more than
$100\%$ correction on top of the two-halo contribution
$\xi_{\psi\delta_A}^\mrm{HM}(0)=0.022\,h^{-1}\mrm{Mpc}$.
The linear estimate of the one-point function is
virtually identical to the two-halo estimate (for reasons that will 
become clear in Section~\ref{sec:pot_agnostic}).
By itself, the linear estimate is unable to reproduce the turnover
and needs to be augmented by a small-scale contribution (e.g.\
the one-halo term).

Aside from these small-scale considerations,
we have additionally performed a check on the importance of
$\mbf{C}_\mrm{cross}$ and $\mbf{C}_\mrm{grav}$ on the dipole.
We find almost identical results for the dipole
with and without these contributions (i.e.\ keeping just
$\mbf{C}_\mrm{RSD}$ in the covariance). This tells us that the
important contributions to the dipole come from the mean $\bm{m}$.
We have also checked the impact $\mbf{C}_\mrm{cross}$ and
$\mbf{C}_\mrm{grav}$ have on the monopole and quadrupole.
For $s<150\,h^{-1}\mrm{Mpc}$, we find these contributions also to
be negligible (less than $0.1\%$ at $s=100\,h^{-1}\mrm{Mpc}$,
for example). This is of course not surprising given that in
Fourier space one expects these contributions only to become
important on scales approaching the horizon~\cite{Beutler:2020evf}.

\subsubsection{Impact of lightcone and lookback-time effects}\label{sec:linear-comparison}

\begin{figure}[t!]
  \centering
  \includegraphics[scale=0.75]{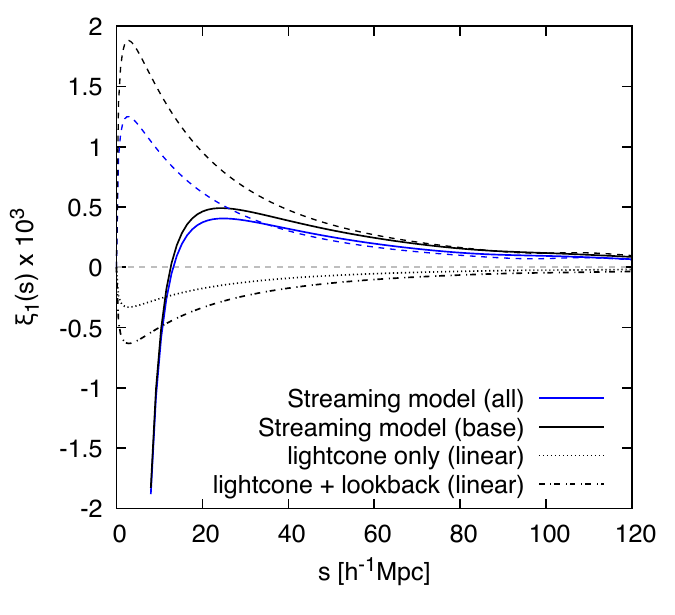}
  \vspace{-5pt}
  \caption{Dipole of the streaming model (solid blue) and from linear
  theory (dashed blue).
  For comparison we show in solid black the dipole from the streaming model
  without lightcone or lookback-time effects (the `base' model), with the
  corresponding linear prediction shown in dashed black. The individual contribution
  from the lightcone effect given by linear theory is shown in dotted black;
  the sum of (linear) lightcone and lookback-time effects is shown in dash-dot black.}
  \label{fig:complete_dipole}
\end{figure}

To assess the impact of the lightcone and lookback time effects
we investigate two versions of the streaming model: one with all the
effects included (`all'), as given by Eq.~\eqref{eq:xi_final}, and one where
we exclude lightcone and lookback time effects, keeping all other effects (`base').
(For the former model we use the lightcone-corrected density weighting~\eqref{eq:lc-weighting},
while for the latter we use Eq.~\eqref{eq:dens-weighting}.)
The result is shown in Figure~\ref{fig:complete_dipole}, in which we also
show the linear predictions given by Eq.~\eqref{eq:delta-lin}.
Comparing the two streaming models we see that the lightcone and
lookback time have a small net negative contribution to the dipole,
consistent with findings elsewhere~\cite{Breton:2018wzk}.
Although the amplitude is lowered slightly,
the overall shape of the dipole remains largely the same.
This rules out either of these effects as the cause of the
turnover. (In Section~\ref{sec:origin} below we will confirm
this.)

An under-appreciated feature of the dipole (and all odd multipoles),
is that it vanishes at zero separation, $\xi_1(s=0,d)=0$.
This is because the overall contribution to the two-point correlation,
$\xi_1(s,d)\sp\mathcal{L}_\ell(\mu)$, for odd $\ell$, 
is anti-symmetric about the axis $s_\|=0$ (the axis orthogonal to
the line of sight) and is therefore required to vanish along this axis.
We see this clearly for the linear theory curves in
Figure~\ref{fig:complete_dipole}, which return to zero at $s=0$. 
In principle, this is also the case for the streaming model,
where we expect a very late turnaround as $s\to0$ (which we are unable
to show due to numerical instabilities in the model at small scales). We thus expect a second turnover in the dipole on very small scales for the
complete model, where by contrast linear theory gives only one.
But an improved treatment of the nonlinear regime (taking into
account the higher-order correlations, halo exclusion, etc) is needed for
a quantitative prediction here.

\section{Anti-symmetries, parity, and pairwise functions}\label{sec:pair}
The way the displacement statistics of Section~\ref{sec:stats} are
currently written obscures their physical ingredients.
In this section we will bring out these ingredients by showing that these
statistics naturally decompose into symmetric and anti-symmetric parts, with
each part able to be reduced to a set of pairwise correlation
functions (as in the distant-observer limit). The advantage of
these pairwise functions is that they each have a distinct physical
meaning, allowing us to isolate the important anti-symmetric effects
(independent of the complications to do with the lines of sight).

\subsection{Decomposition of cumulants into symmetric and anti-symmetric parts}
We consider two galaxies observed at $\s_1$ and $\s_2$, and label
$A$ the galaxy at $\s_1$ and $B$ the galaxy at $\s_2$.
(This label is generic and could describe galaxy bias,
colour, luminosity, etc, although for concreteness we take this
to be the galaxy bias.) In the distant-observer limit, and considering only
the usual RSD effect we have that $\langle AB\rangle=\langle BA\rangle$;
the order in which the two galaxies are paired does not matter.
However, as shown in Ref.~\cite{Bonvin:2013ogt}, when generalising beyond
RSD and the distant-observer limit this is no longer the case.
This motivates us to think of the labels $A$ and $B$ as an
ordered pair given by the 2-tuple $(A,B)$. 
We can then determine which contributions to the two-point function
are invariant under particle exchange, $(A,B)\to(B,A)$, and which contributions are not. 
Ultimately, the properties of
$\langle\delta_A^{(\mrm{s})}(\s_1)\sp\delta_B^{(\mrm{s})}(\s_2)\rangle$
are determined by the cumulants $\bm{m},\mbf{C},\ldots$, and by
decomposing them into symmetric and anti-symmetric parts we can
isolate in a systematic way only those terms which affect
the odd multipoles.

For what follows, it will be convenient to restore labels $A$ and $B$,
writing $\bm{m}\to\bm{m}_{AB}$ and $\mbf{C}\to\mbf{C}_{AB}$.
With this notation we split each term into two parts:
\be\label{eq:decomp}
\bm{m}_{AB}=\bm{m}_{(AB)}+\bm{m}_{[AB]}\,,\quad
\mbf{C}_{AB}=\mbf{C}_{(AB)}+\mbf{C}_{[AB]}\,,
\ee
where each part is given by
\begin{alignat}{2}
\bm{m}_{(AB)}&=\tfrac12(\bm{m}_{AB}+\bm{m}_{BA})\,,&\quad
\bm{m}_{[AB]}&=\tfrac12(\bm{m}_{AB}-\bm{m}_{BA})\,,\nonumber\\
\mbf{C}_{(AB)}&=\tfrac12(\mbf{C}_{AB}+\mbf{C}_{BA})\,,&\quad
\mbf{C}_{[AB]}&=\tfrac12(\mbf{C}_{AB}-\mbf{C}_{BA})\,.\nonumber
\end{alignat}
Under exchange $A\leftrightarrow B$, the first parts are symmetric ($\bm{m}_{(BA)}=\bm{m}_{(AB)}$) 
while the second parts are anti-symmetric ($\bm{m}_{[BA]}=-\bm{m}_{[AB]}$).% 
\footnote{
Note that exchange of label, keeping the positions fixed,
is equivalent to exchanging the positions. The former, which we find
more convenient to use, is a passive relabelling of the particles in
place, whereas the latter actively swaps the particles with their
label attached.
}
Each of these quantities receive contributions from RSD and
gravitational redshift and we write, for instance,
$\bm{m}_{[AB]}=\bm{m}_{[AB]}^\mrm{RSD}+\bm{m}_{[AB]}^\mrm{grav}$.

\subsection{Pairwise correlations}\label{sec:pairwise}
We now apply decomposition~\eqref{eq:decomp} to each term
in $\bm{m}=\bm{m}^\mrm{RSD}+\bm{m}^\mrm{grav}$ [Eq.~\eqref{eq:mrsd-mgrav}].
To avoid inessential complications and introducing too
many novelties at once,
we will present the following decomposition without the lightcone effect,
i.e.\ we assume a symmetric density weighting, as in
Eq.~\eqref{eq:dens-weighting}.
In Section~\ref{sec:pair-discuss} we will discuss why this is justified.

Using that $\langle \psi(\x_1)\sp \delta_A(\x_2)\rangle
=\langle \psi(\x_2)\sp \delta_A(\x_1)\rangle$ and
$\langle\u(\x_2)\sp \delta_A(\x_1)\rangle
=-\langle\u(\x_1)\sp \delta_A(\x_2)\rangle$, and likewise
for $B$, we find that each term decomposes as
\begin{align}
\bm{m}^\mrm{RSD}_{AB}
&=\bm{m}^\mrm{RSD}_{(AB)} + \bm{m}^\mrm{RSD}_{[AB]}
=
\underbrace{\begin{pmatrix}
\phantom{-}\tfrac12\llangle\Delta\u\rrangle\cdot\n_1\, \\[2pt]
-\tfrac12\llangle\Delta\u\rrangle\cdot\n_2
\end{pmatrix}}_{\text{sym.}}
+
\underbrace{\begin{pmatrix}
\llangle\bar\u\rrangle\cdot\n_1 \\[2pt]
\llangle\bar\u\rrangle\cdot\n_2
\end{pmatrix}}_{\text{anti-sym.}}\,,%\label{eq:m-rsd} \\
\nonumber\\
\bm{m}^\mrm{grav}_{AB}
&=\bm{m}^\mrm{grav}_{(AB)} + \bm{m}^\mrm{grav}_{[AB]}
=
\underbrace{\begin{pmatrix}
\llangle\bar\psi\rrangle  \\[2pt]
\llangle\bar\psi\rrangle
\end{pmatrix}}_{\text{sym.}}
+
\underbrace{\begin{pmatrix}
\phantom{-}\tfrac12\llangle\Delta\psi\rrangle\, \\[2pt]
-\tfrac12\llangle\Delta\psi\rrangle\,
\end{pmatrix}}_{\text{anti-sym.}} \,,\label{eq:m-gravz}
\\[-20pt] \nonumber
\end{align}
where the symmetric and anti-symmetric parts are defined
respectively by the first and second terms in the second equality.
Here we have expressed $\u(\x_1)$ and $\u(\x_2)$,
and similarly $\psi(\x_1)-\psi_O$ and $\psi(\x_2)-\psi_O$, in
terms of new variables:
\bea
\Delta\u&=\u(\x_1)-\u(\x_2)\,,
\quad\;
\,\bar\u=\tfrac12[\u(\x_1)+\u(\x_2)]\,,\label{eq:def_DeltaPsi}\\
\Delta\psi&=\psi(\x_1)-\psi(\x_2)\,,
\quad\;
\bar\psi=\tfrac12[\psi(\x_1)-\psi_O+\psi(\x_2)-\psi_O]\,,\nonumber
\eea
so that, e.g.\ $\u(\x_1)=\bar\u+\Delta\u/2$. Note that all quantities on the right-hand side of Eq.~\eqref{eq:m-gravz} depend on the tracers $A$ and $B$ through the density weighting, even though here to simplify the notation we have dropped the subscript $AB$. Hence, e.g.,  the anti-symmetry of $\Delta\psi$ under the exchange of $\x_1$ and $\x_2$ results in an anti-symmetry of the weighted difference $\llangle\Delta\psi\rrangle_{AB}$ under the exchange of $AB$. This is
easily verified upon expanding out the density weighting~\eqref{eq:dens-weighting}:
\bea
\llangle\Delta\psi\rrangle_{BA}
&=(1+\xi_{AB})^{-1}\Big[\langle \psi(\x_1)\sp\delta_A(\x_2)\rangle-\langle \psi(\x_2)\sp\delta_B(\x_1)\rangle \nonumber\\
&\quad\qquad+\langle\psi(\x_1)\sp\delta_A(\x_1) \rangle-\langle\psi(\x_2)\sp\delta_B(\x_2) \rangle \label{eq:assym}\\
&\quad+\langle \psi(\x_1)\sp\delta_A(\x_1)\sp\delta_B(\x_2)\rangle-\langle \psi(\x_2)\sp\delta_A(\x_2)\sp\delta_B(\x_1) \rangle  \Big]\, , \nonumber
\eea
where each line is anti-symmetric under exchange of $A$ and $B$,
so $\llangle\Delta\psi\rrangle_{BA}=-\llangle\Delta\psi\rrangle_{AB}$.

The basis~\eqref{eq:def_DeltaPsi} has the nice property that it is
statistically independent:
$\llangle\bar\u\Delta\u\rrangle
=\llangle\Delta\u\bar\u\rrangle=0$ and
$\llangle\bar\psi\Delta\psi\rrangle=\llangle\Delta\psi\bar\psi\rrangle=0$.%
\footnote{Strictly speaking,
$\llangle\bar\psi\Delta\psi\rrangle=\llangle\Delta\psi\bar\psi\rrangle=0$
when $\chi_1=\chi_2$, e.g.\ when correlating galaxies at the
same redshift. We can also have statistical independence more generally if
we take the view that $\psi_O$ should be fixed and therefore
not averaged over; see Appendix~\ref{app:psi_O}.}
In terms of the covariances, this is just the basis that diagonalizes
$\mbf{C}_\mrm{RSD}$ and $\mbf{C}_\mrm{grav}$ (with the linear
transformations containing all the line-of-sight dependence). The four quantities in Eq.~\eqref{eq:def_DeltaPsi} have a distinct physical meaning and
their density-weighted averages, when reduced to their scalar
parts, provide the basic objects of study (independent of
geometrical factors from the lines of sight). To obtain the
scalar parts of the velocity correlations, we recall that
under statistical isotropy $\llangle\Delta\u\rrangle=\llangle\Delta u\rrangle\hat\r$
or $\llangle\Delta u\rrangle=\hat\r\cdot\llangle\Delta\u\rrangle$.
Here $\llangle\Delta u\rrangle$ is (in our notation) the well-known
pairwise velocity difference~\cite{Peebles_LSS,Scoccimarro:2004},
describing the tendency of infall of galaxy pairs as function of separation.
Similarly, we also have a scalar correlation
$\llangle\bar u\rrangle=\hat\r\cdot\llangle\bar\u\rrangle$, which
we call the pairwise mean velocity (in the frame of the observer).%
\footnote{Often $\llangle\Delta u\rrangle$ is called
the `mean infall velocity' or `mean streaming velocity'.
This should not be confused with $\llangle\bar u\rrangle$, which we call
the `pairwise mean velocity', in keeping with the rest of the terminology.}

The symmetric pairwise correlation functions are thus
\begin{itemize}
\itemsep-1pt
\item $\llangle\Delta u\rrangle(r)\equiv\hat\r\cdot\llangle\Delta\u\rrangle$,
the pairwise velocity difference;%
\item $\llangle\bar\psi\rrangle$, the pairwise mean potential;%
\footnote{Due to the local potentials this quantity depends not just on $r$ but
also on the distances $\chi_1$ and $\chi_2$. It is perhaps more
accurately called the `mean of the potential difference' (with respect to the local
potential); because of the local potentials it remains invariant under
shifts $\psi\to\psi+c$ (as we should expect for any observable of
the potential).}
\end{itemize}
while the anti-symmetric pairwise functions are
\begin{itemize}
\itemsep-1pt
\item $\llangle\bar u\rrangle(r)\equiv
\hat\r\cdot\llangle\bar\u\rrangle$, the pairwise mean velocity;
\item $\llangle\Delta\psi\rrangle(r)$, the pairwise potential difference.
\end{itemize}
Note that $\llangle\Delta u\rrangle$
and $\llangle\Delta\psi\rrangle$ are \emph{not} of the same parity;
the former is symmetric while the latter is anti-symmetric.  As expected, the anti-symmetric functions are identically zero if $A=B$.

The parity of these pairwise functions determines
whether they contribute to the odd or even multipoles (at first order).
Specifically, the symmetric pairwise functions contribute to
the even multipoles, while the anti-symmetric ones contribute to
the odd multipoles. (Note that at nonlinear order we can have
products of these functions contributing to the multipoles,
e.g.\ the product of a symmetric and
anti-symmetric function, which is overall anti-symmetric.)
Based on these properties we can already say that the important
pairwise functions contributing to the dipole are
$\llangle\bar{u}\rrangle$ and $\llangle\Delta\psi\rrangle$.

Another useful way to characterise these pairwise functions is by whether
they are \emph{relative} or \emph{absolute} quantities. This property tells us in which
regime the function enters and therefore whether it is a wide-angle effect, and
if so at what order in $\epsilon=s/d$ it will contribute to the multipoles. To see
this, recall that in the distant-observer limit we have only relative quantities
appearing, $\llangle\Delta u\rrangle$, $\llangle\Delta\psi\rrangle$, etc. 
These generally depend purely on separation $r$ and
are therefore the only types allowed by translation invariance (which exists
in this limit but not in the wide-angle regime).
The appearance in the wide-angle regime of `absolute' quantities, $\llangle\bar{u}\rrangle$ and
$\llangle\bar\psi\rrangle$, breaks translation invariance. For example,
$\llangle\bar\psi\rrangle$ introduces dependence on the distances $\chi_1$ and
$\chi_2$ between the galaxies and the observer (meaning we cannot translate
the galaxy pair
without fundamentally changing the triangle in Figure~\ref{fig:config}
on which the wide-angle correlations depend).
This means that the wide-angle corrections of RSD and gravitational redshift
are given in terms of the `absolute' functions $\llangle\bar{u}\rrangle$ and
$\llangle\bar\psi\rrangle$, respectively, and must enter the multipoles at order
$\epsilon=s/d$ (vanishing in the distant-observer limit when $d\to\infty$).
From these arguments, we can have a term of the form
$\epsilon\sp\dif\llangle\bar{u}\rrangle/\dif s$ entering the dipole
but not $\epsilon\sp\dif\llangle\bar\psi\rrangle/\dif s$ 
(where the derivative is required on dimensional grounds).

\subsubsection{Anti-symmetry of the lightcone effect}
As another illustration, let us apply the decomposition to the kinematic factor in
Eq.~\eqref{eq:xi_final} to show that the lightcone effect
induces an anti-symmetry at leading order. 
We have
\bea
\llangle u_{\|1}+u_{\|2}+u_{\|1}u_{\|2}\rrangle
&\simeq\llangle u_{\|1}+u_{\|2}\rrangle \nonumber\\
&=\llangle \u(\x_1)\rrangle\cdot\n_1+\llangle\u(\x_2)\rrangle\cdot\n_2 \nonumber\\
&=\tfrac12\sp\llangle\Delta u\rrangle\sp(\hat\r\cdot\n_1-\hat\r\cdot\n_2) \nonumber\\
&\quad\;+\llangle\bar{u}\rrangle\sp(\hat\r\cdot\n_1+\hat\r\cdot\n_2)\,,
    \label{eq:lightcone-decomp}
\eea
where we have divided through by $\calH$ and dropped the subdominant
quadratic term (which at any rate is symmetric at leading order so not
important for the odd multipoles).
In the case $\n_1=\n_2$ (distant-observer limit) we see that
only the second term in Eq.~\eqref{eq:lightcone-decomp} remains,
$\llangle\bar{u}\rrangle\propto(b_A-b_B)\langle u\sp\delta\rangle$.
This means that the impact of the lightcone effect at leading order
is anti-symmetric and is present even in the distant-observer limit
(unlike the anti-symmetric contribution from RSD).

Though we can also apply the decomposition to $\mbf{C}_{AB}$ and obtain
its pairwise `dispersion' functions, it will not be necessary.
As we found numerically in
Section~\ref{sec:results}, the impact of the covariance on the dipole
is highly suppressed.
Indeed, evaluated using linear theory, $\mbf{C}_{AB}$ has
no anti-symmetric part, $\mbf{C}_{AB}=\mbf{C}_{(AB)}$, since
any anti-symmetric part can only
come from the density weighting, which for $\mbf{C}_{AB}$
enters at lowest order as a bispectrum contribution
(which we ignore).
In the remainder of this work, we will hence focus only on
$\bm{m}_{AB}$ and its pairwise functions.

\subsubsection{Discussion}\label{sec:pair-discuss}
The decomposition into symmetric and anti-symmetric pairwise functions
underlies the usual way of thinking about parity in terms of the number
of radial derivatives carried by a given term in Eq.~\eqref{eq:delta-lin}.
When correlating with a single tracer (identical particles)
there can be no anti-symmetry due to invariance under pair exchange. When
correlating with distinct tracers (labelled particles) we have the
possibility of anti-symmetry. Typically these parity properties are identified in
the linear correlation function by counting the number of radial
derivatives a given term carries with respect to $\langle\delta\delta\rangle$.
Those with an even (odd) number of derivatives will contain an even (odd)
number of $\mu$'s and therefore contribute to the even (odd) multipoles,
with the dependence on linear bias entering as $b_A+b_B$ ($b_A-b_B$).
We can now see that this originates from the parity of the pairwise functions:
for symmetric functions $\llangle\Delta u\rrangle$ and
$\llangle\bar\psi\rrangle$ the biases enter as $b_A+b_B$, while
for anti-symmetric functions $\llangle\Delta\psi\rrangle$ and
$\llangle\bar{u}\rrangle$ the bias enters as $b_A-b_B$.

Although the parity is easy to see in linear theory, as above,
it should be understood that the parity of these quantities holds at
all orders in perturbation theory. So although we only consider
linear bias in this work ($\delta_A\simeq b_{A}\sp\delta$ and $\delta_B\simeq b_{B}\sp\delta$), the above separation into terms proportional to bias sums
($b_A+b_B$) or
bias differences ($b_A-b_B$) carries to higher-order galaxy bias.
For instance, at second order there is nonlocal tidal bias and this
also enters in the form $b_{K^2A}+b_{K^2B}$ or $b_{K^2A}-b_{K^2B}$.
That this remains true is simply
a consequence of the requirement that the galaxy bias expansion respect the equivalence principle, i.e.\ that galaxies symmetrically cluster with respect to
the underlying matter.\footnote{This means that $\delta_g$ should be composed
of symmetric operators of the tidal field $\partial_i\partial_j\Phi$.}

As mentioned earlier, there is one complication and it is
the effect of the lightcone correction on the pairwise functions.
As we discussed, this changes the density weighting and in effect
modifies the galaxy bias expansion through introduction
of odd parity operators. This complicates the above analysis
because we implicitly assumed that the bias expansion consists
solely of even parity operators (a symmetric density weighting).
As a result, we can no longer say that $\llangle\Delta\psi\rrangle$,
with the lightcone-corrected density weighting~\eqref{eq:lc-weighting},
is totally anti-symmetric;
it can also contain a symmetric part due to the additional factors of
$1+\v\cdot\n$. But this only leads to additional
correlations in the form of $\langle\psi\sp v\rangle$ which is
nominally suppressed relative to the other correlations
$\langle\psi\sp\delta\rangle$ (and which is of the opposite parity anyway).
There are also corrections in the form of a third-order
statistic $\langle\psi\sp v^2\rangle$ which does affect the anti-symmetric part.
But such terms are suppressed relative to $\langle\psi\sp\delta\rangle$.
Altogether, this means that at leading order $\llangle\Delta\psi\rrangle$
remains unchanged and provided we work at this order we can continue to regard
it as anti-symmetric.
We note that similar considerations apply to the
other pairwise functions.%
\footnote{The same considerations also apply to selection effects,
had we included them in our analysis;
see Ref.~\cite{Dam:2023} for further discussion.}

\subsection{The pairwise potential difference}\label{sec:Delta-psi}
The numerical results and arguments above indicate that the key
quantity governing the small-scale behaviour of the dipole is the
pairwise potential difference $\llangle\Delta\psi\rrangle$. As Eq.~\eqref{eq:assym} shows, this quantity contains contributions from the two-point correlation function, but also from the three-point functions that cannot be neglected. 

\subsubsection{Linear contribution}

As a first step, we study the behaviour of $\llangle\Delta\psi\rrangle$ at linear order, and then later in Sections~\ref{sec:phi-HM} and~\ref{sec:pot_agnostic} we compute the nonlinear corrections.

From the first two lines of Eq.~\eqref{eq:assym}, we have at linear order
\bea\label{eq:psi_12-exact}
\llangle\Delta\psi\rrangle
&=
\frac{\big[{\xi_{\psi\delta_A}(0)-\xi_{\psi\delta_A}(r)}\big]
    -\big[{\xi_{\psi\delta_B}(0)-\sp\xi_{\psi\delta_B}(r)}\big]}{1+\xi_{AB}(r)}\,,
\eea
with $\xi_{\psi\delta_A}(r)=\langle\psi(\x_1)\sp\delta_A(\x_2)\rangle$
and $\xi_{\psi\delta_B}(r)=\langle\psi(\x_1)\sp\delta_B(\x_2)\rangle$.
Writing $\xi_{\psi\delta_A}(r)=b_A\sp\xi_{\psi\delta}(r)$ and
$\xi_{\psi\delta_B}(r)=b_B\sp\xi_{\psi\delta}(r)$,
with $b_A$, $b_B$ the linear galaxy biases, and 
$\xi_{\psi\delta}(r)$ given by Eq.~\eqref{eq:xi-psi-delta} in
Appendix~\ref{app:corrfuncs}, and using that at large separation, $\xi_{AB}(r)\ll1$, we have
\bea\label{eq:psi_12}
\llangle\Delta\psi\rrangle
\simeq(b_A-b_B)\,
    \big[{\xi_{\psi\delta}(0)-\xi_{\psi\delta}(r)}\big] \,.
        % {1+b_A\sp b_B\sp\xi(r)} \,,
\eea
This expression, with the galaxy biases made explicit,
makes clear that $\llangle\Delta\psi\rrangle$ is anti-symmetric
under the exchange of $A$ and $B$ and vanishes when $A=B$, as discussed above.

\begin{figure}
  \centering
  \includegraphics[width=1\linewidth]{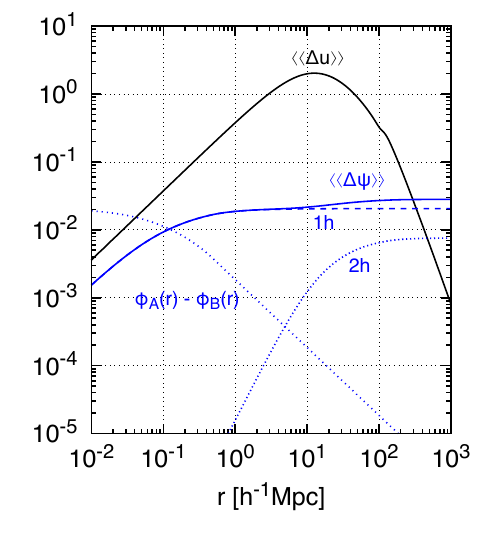}
  \caption{The contributions to the pairwise potential difference
  $\llangle\Delta\psi\rrangle=\llangle\Delta\psi\rrangle_\mrm{1h}
  +\llangle\Delta\psi\rrangle_\mrm{2h}$ (solid blue), where
  $\llangle\Delta\psi\rrangle_\mrm{1h}$
  is given by
  Eq.~\eqref{eq:psi_12_shot} and
  $\llangle\Delta\psi\rrangle_\mrm{2h}$ is
  given by Eq.~\eqref{eq:Delta-psi-2h-2}.
  For comparison, we also show the
  pairwise velocity difference
  $\llangle\Delta u\rrangle$ (which is
  negative valued). The scale dependent part of the one-halo term,
  $\phi_A(r)-\phi_B(r)$ is also shown.
  Note that in linear theory the $\llangle\bar{u}\rrangle$ differs from
  $\llangle\Delta{u}\rrangle$ by an
  order unity proportionality constant,
  $\llangle\bar u\rrangle
    =(b_A-b_B)/(b_A+b_B)\llangle\Delta u\rrangle$.
  All pairwise functions go to zero at $r=0$ and have units
  $h^{-1}\mrm{Mpc}$.
  Here $b_A=2$, $b_B=1$, with $z=0.341$.
  }
  \label{fig:psi_12}
\end{figure}

The key aspect of $\llangle\Delta\psi\rrangle$, setting it apart
from $\llangle\Delta u\rrangle$, is that it does \emph{not}
vanish in the large-scale limit; see Figure~\ref{fig:psi_12}.
This is due to the presence of the one-point
function, $\xi_{\psi\delta}(0)$, which is nonzero because modes
$\psi(\k)$ are in phase with $\delta(\k)$ (or have the same parity).%
\footnote{In this linear analysis we need only relate $\psi$ to
the matter overdensity, since $\delta_g=b\sp\delta$. In this case the
potential is of course in phase with the matter overdensity $\delta$.
One may ask whether this remains true in general---is the
potential in phase with the fully nonlinear galaxy overdensity
$\delta_g$? Yes, because according to the
equivalence principle (on which the galaxy bias expansion rests)
$\delta_g$ can only consist of operators with an even number of
derivatives of the potential~\cite{bias_review}.}
In the large-scale limit
$\llangle\Delta\psi\rrangle\to(b_A-b_B)\,\xi_{\psi\delta}(0)\neq 0$ as 
$r\to\infty$. Therefore, even at large separations $\llangle\Delta\psi\rrangle$ is sensitive to
nonlinearities through $\xi_{\psi\delta}(0)$.%
\footnote{This aspect is separate from the fact that long-wavelength
modes of size $L$ should not affect relative quantities such
as $\llangle\Delta\psi\rrangle$ for pairs separated by $r\ll L$. This
cancellation in the effect of long modes means that in principle
$\llangle\Delta\psi\rrangle$ is sensitive to nonlinearities,
as with the pairwise velocity difference~\cite{Scoccimarro:2004}.
}
By contrast the corresponding one-point function of
$\llangle\Delta u\rrangle$ vanishes, since there are no preferred directions.
So $\llangle\Delta u\rrangle$ goes to zero in the large-scale
limit, with the tendency of pair infall decreasing with separation.

What is the impact of $\llangle\Delta\psi\rrangle$ on the redshift-space
correlations $\xi^{(\mrm{s})}$?
Though the relation between $\xi^{(\mrm{s})}$ and the pairwise
functions is nonlocal, on dimensional grounds and recalling that
$\Delta\psi$ has units length,
we see that at linear order it is the derivative of $\llangle\Delta\psi\rrangle(r)$ with respect to $r$
that contributes to $\xi^{(\mrm{s})}$. Consequently, at this order $\xi_{\psi\delta}(0)$ does not contribute to $\xi^{(\mrm{s})}$. So although $\llangle\Delta\psi\rrangle$
appears to dominate over $\llangle\Delta u\rrangle$ at large $r$,
its contribution is suppressed because of the flatness of
$\llangle\Delta\psi\rrangle$ on these scales. The effect
of the one-point function $\xi_{\psi\delta}(0)$ is however still relevant, since it enters at
second order in $\xi^{(\mrm{s})}$, and leads to a sizeable correction on small scales.
We will show this explicitly in Section~\ref{sec:dipole-expand}.

\subsubsection{Beyond linear theory: halo model approach}\label{sec:phi-HM}

An improved estimate of
$\llangle\Delta\psi\rrangle$, and in particular of the one-point
function contained within, is obtained using the halo model.
Again, since we have compared our model against halo
catalogues, let us identify tracers $A$ and $B$ with
two populations of dark matter haloes, described by mass
$M_A$ and $M_B$ (among other parameters).

We split $\llangle\Delta\psi\rrangle$ into two
contributions, writing
\be\label{eq:Delta-psi-split}
\llangle\Delta\psi\rrangle=\llangle\Delta\psi\rrangle_\mrm{1h}
+\llangle\Delta\psi\rrangle_\mrm{2h}\,.
\ee
By taking the difference of 
$\llangle\psi(\x_1)-\psi_O\rrangle/2$ and
$\llangle\psi(\x_2)-\psi_O\rrangle/2$, using the formulae of Section~\ref{sec:potential-nl}, it is easy to show that we
have for the one- and two-halo terms
\bea
\llangle\Delta\psi\rrangle_\mrm{1h}
&=[\phi_A(0)-\phi_A(r)]-[\phi_B(0)-\phi_B(r)] \,,
\label{eq:psi_12_shot} \\[3pt]
\llangle\Delta\psi\rrangle_\mrm{2h}
&=\frac{[\xi^\mrm{HM}_{\psi\delta_A}(0)-\xi^\mrm{HM}_{\psi\delta_A}(r)]-[\xi^\mrm{HM}_{\psi\delta_B}(0)-\xi^\mrm{HM}_{\psi\delta_B}(r)]}{1+\xi_{AB}(r)}\,,
\label{eq:Delta-psi-2h-2}
\eea
where we recall $\phi_A$ and $\phi_B$ are
given by Eq.~\eqref{eq:phiA}, and
$\xi^\mrm{HM}_{\psi\delta_A}$ and $\xi^\mrm{HM}_{\psi\delta_B}$ 
are given by Eq.~\eqref{eq:Delta-psi-2h}.
Note that the local terms have cancelled out as expected
so that both one- and two-halo terms are functions of $r$ only.

{Let us note that the discreteness of the density field plays a crucial role in Eq.~\eqref{eq:psi_12_shot}. Indeed, in the case where $\delta_A$ is a continuous field, related to the matter density through $\delta_A=b_A\delta$, the contribution from the three-point function to $\llangle\Delta\psi\rrangle$ vanishes due to symmetry. The third line in Eq.~\eqref{eq:assym} becomes
\begin{align}
b_Ab_B \big[\langle \psi(\x_1)\sp\delta(\x_1)\sp\delta(\x_2)\rangle-\langle \psi(\x_2)\sp\delta(\x_2)\sp\delta(\x_1) \rangle\big]=0 \, . 
\end{align}
However, as we show in Appendix~\ref{app:shot}, taking into account the discreteness of the density field, this third line does not vanish and leads to a non-negligible contribution to $\llangle\Delta\psi\rrangle$. This is because when correlating $\delta_A(\x_1)$ with $\psi(\x_1)$ (that is sourced by all masses, as we see from Eq.~\eqref{eq:psi-discrete}), the only contribution that survives is the one due to the potential $\phi_A$. And similarly $\delta_B(\x_2)$ correlated with $\phi(\x_2)$ gives rise to a contribution proportional to $\phi_B$. For $B\neq A$ this leads to a nonzero contribution from the three-point function in Eq.~\eqref{eq:assym}, which enters in the one-halo term $\llangle\Delta\psi\rrangle_\mrm{1h}$. Physically, this term encodes the correlation between the density of the halo and the gravitational potential at the same position. Since this correlation differs for tracers $A$ and $B$, $\phi_A\neq\phi_B$, it contributes to the anti-symmetric part.}

As Figure~\ref{fig:psi_12} shows, the one-halo term~\eqref{eq:psi_12_shot} dominates over the two-halo term
at all scales $r$.
Moreover, the individual contributions from $\phi_A(r)$
and $\phi_B(r)$ are negligible on scales $r\gg1$ (beyond a
few virial radii of the halo).
On the other hand, $\phi_A(0)$ and $\phi_B(0)$
is relevant at all $r$ in that it sets an absolute scale
on the depth of the halo potentials. Since these `one-point
functions' contain no scale dependence they shift $\llangle\Delta\psi\rrangle$ up and down;
the same is also true for the corresponding
$\xi^\mrm{HM}_{\psi\delta_A}(0)$ and
$\xi^\mrm{HM}_{\psi\delta_B}(0)$, although they lead to smaller shifts. In Table~\ref{tab:phi0} we compare the two contributions at zero separation for different halo masses. We see that, except for the lowest mass, the one-halo term dominates over the two-halo term by a factor of $1.5-6$. Note that all of these zero separation terms are
degenerate with each other, however.
The slight scale dependence seen in Figure~\ref{fig:psi_12} at
$r\gtrsim10\,h^{-1}\mrm{Mpc}$ mainly enters through
$\xi^\mrm{HM}_{\psi\delta_A}(r)$ and $\xi^\mrm{HM}_{\psi\delta_B}(r)$.

It is perhaps not surprising that the one-halo term
is dominant: the density weighting biases the effect of
gravitational redshift towards regions of high number density
(galaxy clusters, dark matter haloes, etc), environments
where high mass concentrations drive potential wells deeper
than expectations from linear theory alone.

% Requires the booktabs if the memoir class is not being used
\begin{table}[t!]
   \centering
   %\topcaption{Table captions are better up top} % requires the topcapt package
   \caption{Comparison between
   the one- and two-halo values of the potential depth,
   $\phi_A(0)$ and $\xi_{\psi\delta_A}(0)$, respectively.
   (To get a sense of the size relative to RSD, we have multiplied
   both quantities by $-\calH$; the units are thus $\text{km/s}$.)
   The values are computed at $z=0.2$ using the \emph{Planck} 2018 best-fit cosmology 
   and assuming an NFW profile with concentration $c=9$ and truncated
   at the virial radius.}
   % \begin{ruledtabular}
   \vspace{2pt}
   \begin{tabular}{@{} lcccr @{}}
     \toprule
      $M\ssp[h^{-1}M_\odot]$ & $R_\mrm{vir}\,[h^{-1}\mrm{Mpc}]$ & $b_1$ & $\xi_{\psi\delta_A}(0)\,[\text{km/s}]$ & $\phi_A(0)\,[\text{km/s}]$ \\[1pt]
      \hline
      $10^{12}$  & $0.18$ & $0.87$ & $-0.62$ & $-0.32$ \\
      $10^{13}$  & $0.39$ & $1.39$ & $-0.99$ & $-1.48$ \\
      $10^{14}$  & $0.84$ & $2.80$ & $-2.00$ & $-6.89$ \\
      $10^{15}$  & $1.80$ & $7.68$ & $-5.47$ & $-31.96$ \\
      \hline\hline
   \end{tabular}
   % \end{ruledtabular}
   \label{tab:phi0}
\end{table}

\subsubsection{Beyond linear theory: model-independent approach}\label{sec:pot_agnostic}
Rather than resorting to a specific model of the halo profile,
we will now pursue a more model-independent approach.
In this section (and the next), we will show that the results of the previous
section~\ref{sec:phi-HM}---in particular that there is negligible
contribution to $\llangle\Delta\psi\rrangle$ from $\phi_A(r)$ and
$\phi_B(r)$ for practically all $r$---holds under more general conditions
than was assumed using the halo model.

We are interested in the gravitational potential outside the halo, treated
here as an isolated, spherical but otherwise arbitrary body $\rho_A(\x)$
with compact nonzero support on $|\x|<R_\mrm{vir}$.
(The assumption of spherical symmetry is not a strong one as we will
shortly see.)
In general the halo potential is given by
\be\label{eq:phi-greens}
\phi_A(\x)
=\lambda G\int\dif^3\y\,\frac{\rho_A(\y)}{|\x-\y|}\,.
\ee
Expanding the Green's function $1/|\x-\y|$ in the case
$|\x|>|\y|$ (i.e.\ outside the halo when $|\x|>R_\mrm{vir}$), we have
the following well-known result from electrostatics:
\be
\frac{1}{|\x-\y|}
=\frac{1}{|\x|}-y_i\sp\partial_i\frac{1}{|\x|}
 +\frac12y_iy_j\sp\partial_i\sp\partial_j\frac{1}{|\x|}+\cdots\,,
 \nonumber
\ee
where $\partial_i=\partial/\partial x^i$. Substituting this expansion back into
Eq.~\eqref{eq:phi-greens} and integrating over $\y$
we have, for $|\x|>R_\mrm{vir}$,
\bea\label{eq:phi-series}
\phi_A(\x)
&=\lambda\sp G\left({M}_A - M_A^i\sp\partial_i
+\frac12 M_A^{ij}\ssp\partial_i\sp\partial_j+\cdots\right)\frac{1}{|\x|}\,.
% =\frac{M}{4\pi r} - M_i\sp\partial_i\frac{1}{4\pi r}
% +\frac12 M_{ij}\ssp\partial_i\sp\partial_j\frac{1}{4\pi r}+\cdots\,,
\eea
Here $M_A=\int\rho_A(\x)\sp\dif^3\x$, $M_A^i=\int x^i\rho_A(\x)\sp\dif^3\x$,
$M_A^{ij}=\int x^ix^j\rho_A(\x)\sp\dif^3\x$, etc, are the mass moments
of $\rho_A(\x)$, i.e.\ numbers which completely characterize a given
halo's mass distribution. In general,
the mass moments will differ from halo to halo
(there need not be a universal profile). 

Since we are dealing
with a spherical mass distribution, all but the first term in
Eq.~\eqref{eq:phi-series} vanishes. This implies that the potential at
$|\x|>R_\mrm{vir}$ is given as if it was sourced by a point particle of
mass $M$ and located at the body's centre. This is known as
Newton's shell theorem and means that
outside the halo the potential is completely determined by its mass
with the usual inverse distance fall-off, $\phi\propto M/|\x|$;
no knowledge of the body's internal structure is required.
(At least when dealing with the halo potential, we do not even need to
make the usual halo-model assumption of a universal density profile.)
In Eq.~\eqref{eq:psi_12_shot} the potentials are thus
\begin{subequations}\label{eq:phi-pp}
\bea
\phi_A(r)&=\lambda\sp\frac{GM_A}{r}\,,
\quad 
r>R_{\mrm{vir},A}\,,\\
\phi_B(r)&=\lambda\sp\frac{GM_B}{r}\,,
\quad 
r>R_{\mrm{vir},B}\,,
\eea
\end{subequations}
where the only tracer-dependent parameter is the mass,
$M_A$ and $M_B$. 
Since the largest collapsed objects have $R_\mrm{vir}\simeq 2\,h^{-1}\mrm{Mpc}$,
which is still much smaller than typical separations $r$, these formulae
are broadly applicable.%
\footnote{It is important to keep in mind that in
the streaming model the correlation function is non-locally related to 
pairwise correlations such as $\llangle\Delta\psi\rrangle$. This means
that $r$ can in principle attain values anywhere from infinity down to
small but nonzero separations where Eq.~\eqref{eq:phi-pp} cannot be 
used. However for typical galaxy pair separations this should not be
an issue since these extreme values are associated with very low
probability (and, at any rate, small $r$ are protected due
to halo exclusion).}
Hence, for practically all separations $r$ we indeed satisfy the
condition $r>R_\mrm{vir}$.

These results have of course been obtained under the assumption that haloes are
spherical. But even if we relax this assumption Eq.~\eqref{eq:phi-pp} remains
a good approximation. Since although we would have nonvanishing higher moments,
resulting in additional terms on the right-hand side of Eq.~\eqref{eq:phi-pp},
these yield only small corrections to the basic picture above. On dimensional
grounds, the $n$th moment
is suppressed by a factor of order $(R_\mrm{vir}/r)^n$ with respect to the
leading contribution, $\phi\propto M/r$. In particular, if we make the weaker
assumption that the halo is not spherical but symmetric so $\phi(-\r)=\phi(\r)$,
then the leading correction comes from the quadrupole moment, which is
suppressed by $(R_\mrm{vir}/r)^2$.

Now, although we do not need a full description of the entire profile, we do
require knowledge of the potential at $r=0$ (the depth), where the
point-particle description clearly breaks down.
In keeping with the model-independent approach, one way around
this difficulty is to simply treat
both $\phi_A(0)$ and $\phi_B(0)$ as free parameters (as is often done
in data analysis with the Finger-of-God velocity dispersion parameter).
In fact since these contributions are degenerate with those
from Eq.~\eqref{eq:psi_12}, the total zero-point
$\phi_A(0)+\xi_{\psi\delta_A}(0)-\phi_B(0)-\xi_{\psi\delta_B}(0)$
may be treated as a single unknown quantity.

Any symmetries (or lack thereof) in the
profile carry over to $\llangle\Delta\psi\rrangle_\text{1h}$,
which therefore depends in general on the (signed) quantity $\r=\x_1-\x_2$. 
Here we have assumed spherical symmetry. 
However, for an arbitrary profile Eq.~\eqref{eq:psi_12_shot} is exactly
anti-symmetric under exchange $A\leftrightarrow B$ (or equivalently $\x_1\leftrightarrow\x_2$).
But we do not strictly need to assume spherical symmetry for
Eq.~\eqref{eq:psi_12_shot} to be anti-symmetric; we can still have
anti-symmetry in an approximate sense.
On large scales haloes can be treated as effectively spherical point particles,
regardless of their actual structure, since their aspherical
moments are suppressed, see Eq.~\eqref{eq:phi-kspace} below. At leading order the halo potential
thus falls as $1/r$ so that for sufficiently large separations we can consider
$\llangle\Delta\psi\rrangle_\mrm{1h}$
as a function of $r$ only (and a weak function of $r$ at that).
Whatever complicated structure haloes have when viewed
on small scales is erased on sufficiently large scales, allowing us to
view halos as effectively spherical with
$\llangle\Delta\psi\rrangle_\mrm{1h}$ given by Eq.~\eqref{eq:psi_12_shot}.

\subsubsection{Perturbation theory}\label{sec:pt-effective}
We now show that the two-halo term~\eqref{eq:Delta-psi-2h} is
largely insensitive to the particular details of the halo in that it
differs only marginally from the linear predictions.
This is of course a well known fact of the halo model. But in the
case of the gravitational potential it is interesting to see why this
is from the point of view of effective field theory.

The key observation is that on large scales a separation of scales
emerges between $R_\mrm{vir}$ and $r$.
At larger and larger $r$ the detailed structure of the haloes themselves
should be of diminishing importance in the large-scale
description of clustering, so that to a first approximation one can treat haloes
as point masses with $\rho(\x)=M\delD(\x)$ and $\phi(\x)\propto M/|\x|$.
As is readily checked, the two-halo formula for $\xi_{\psi\delta_A}(r)$
[Eq.~\eqref{eq:Delta-psi-2h}] does indeed recover the linear expression~\eqref{eq:xi-psi-delta}
in the case of point masses and with linear galaxy bias, e.g.\
by writing the convolution in Eq.~\eqref{eq:Delta-psi-2h} in
Fourier space and using that the Fourier transform of
$1/4\pi|\x|$ is $1/k^2$.

If we now allow that haloes are of course extended objects, we can compute the
corrections to the point-particle approximation. On dimensional grounds these
corrections enter as  terms going as the ratio of the size $R_*$ of the
halo to the separation $r$. This is easy to see in Fourier space.
Substituting an arbitrary density written in terms of its mass moments,
$\rho(\k)=M+\im k_iM_i-k_ik_jM_{ij}/2+\cdots$, into the potential
$\phi(\k)=4\pi G\lambda\ssp\rho(\k)/k^2$, we have [cf.~Eq.~\eqref{eq:phi-series}]
\be\label{eq:phi-kspace}
\phi(\k)
=\phi_0(\k)
\left(1+\im k_iM_i/M-\frac12 k_i k_j M_{ij}\sp/\sp M+\cdots\right)\,,
\ee
where $\phi_0(\k)\equiv{4\pi \lambda\sp GM}/{k^2}$ is the potential of a point mass (here we omit tracer labels to ease notation).
Thus the leading correction goes as $kR_*\sim R_*/r$
since $M_{i}\sim MR_*$. In the case of a spherical body the leading correction
goes as $(kR_*)^2\sim (R_*/r)^2$ since $M_{ij}\sim \delta_{ij}MR_*^2$.

These corrections carry over to the two-halo term~\eqref{eq:Delta-psi-2h}.
Writing Eq.~\eqref{eq:Delta-psi-2h} in Fourier form (and setting aside the
unimportant galaxy bias), we find
\bea\label{eq:xi-psi-delta-series}
\xi^\mrm{HM}_{\psi\delta}(\r)
&=\int\frac{\dif^3\k}{(2\pi)^3}\,\rme^{-\im\k\cdot\r}
P_{\delta\delta}(k) \\
&\quad\times\lambda\frac{4\pi G\bar\rho}{k^2}
\left(1+\im k_i A_i-\frac12 k_i k_j A_{ij}+\cdots\right)\,,\nonumber
\eea
where $\bar\rho=\int\dif M(\dif n/\dif M) M$ is the mean mass
density and $A_i,A_{ij}$, etc, are parameters
characterising the spatial extent of haloes as a population average,
i.e.\ integrated over the mass function.
(By contrast $M,M_i,M_{ij}$, etc, characterise only
a particular halo in that population.)
The usual expression of $\xi_{\psi\delta}(r)$
is recovered by dropping all but the first term in Eq.~\eqref{eq:xi-psi-delta-series}.
From the perspective of effective field theory, $A_i,A_{ij},\ldots$ 
are coefficients that absorb the uncertain details of the mass function 
and profile (the two key ingredients of the halo model).

At the population level we can reasonably assume that haloes are
on average spherical so that
$A_i=0$ and $A_{ij}=\delta_{ij}\sp R_*^2$, with $R_*$ a free
parameter corresponding to a characteristic spatial scale 
associated with the second mass moment.
Inserting these back into Eq.~\eqref{eq:xi-psi-delta-series}
% using that $-k^2\rme^{-\im\k\cdot\r}=\nabla^2\rme^{-\im\k\cdot\r}$,
we find $\xi^\mrm{HM}_{\psi\delta}(r)
\simeq\big(1+R_*^2\nabla^2\big)\,\xi_{\psi\delta}(r)$, with
$\xi_{\psi\delta}(r)$ given by the standard expression~\eqref{eq:xi-psi-delta}.
The leading-order correction is thus a higher-derivative term and so is
suppressed on large scales. In fact, for $r\gtrsim2\,h^{-1}\mrm{Mpc}$ we
find very small differences numerically between $\xi_{\psi\delta}(r)$ and 
the halo model $\xi^\mrm{HM}_{\psi\delta}(r)$ given by the full
expression~\eqref{eq:Delta-psi-2h} (ignoring biasing complications).
When evaluated using linear theory and assuming linear bias, we find
that the two-halo term $\llangle\Delta\psi\rrangle_\mrm{2h}$ is 
well approximated by the standard expression.

\section{Origin of the dipole's turnover}\label{sec:origin}
In this section we investigate the origin of the turnover in the dipole, as seen in Figure~\ref{fig:complete_dipole}. 
As mentioned above, this turnover is absent in linear
theory, and is thus a telling signature
of the nonlinear gravitational redshift signal.
Using the results of Section~\ref{sec:pair} and following
the perturbative approach taken in the distant-observer
limit~\cite{Scoccimarro:2004}, we will analytically compute
the dipole
\be\label{eq:xi1-formula}
\xi_{1}(s,d)
=\frac{3}{2}\int^1_{-1}\dif\mu\,
\mathcal{L}_1(\mu)\,\xi^{(\mrm{s})}_{AB}(s,d,\mu)
\ee
by performing an expansion of
$\xi^{(\mrm{s})}_{AB}$~[Eq.~\eqref{eq:gsm2}], identifying the
displacements $\delta\chi$ as the expansion parameter.

We will work to order $\epsilon\equiv s/d$ in the mid-point parametrisation
and consider subleading contributions to Eq.~\eqref{eq:xi1-formula} which
are suppressed (in Fourier space) by a single power of $\calH/k$ with respect to the matter power
spectrum. We will thus ignore terms contributing to Eq.~\eqref{eq:xi1-formula}
at order $\epsilon^2$, $(\calH/k)^2$ and $\epsilon\,\calH/k$, all of
which we will simply call terms $\mathcal{O}(\epsilon^2)$.

\subsection{Small-displacement expansion}\label{sec:expand}
Typical displacements, which we did not require
to be small in our derivation, tend to be small in practice. For
instance, in $\Lambda$CDM 
$\delta\chi\simeq v/H_0\simeq 3\,h^{-1}\mrm{Mpc}$, which
is much smaller than the order $100\,h^{-1}\mrm{Mpc}$
radial length scales ($\chi_1$ and $\chi_2$) associated with typical configurations.
As our notation already suggests,
we can thus think of the radial map~\eqref{eq:chi-map} as
slightly perturbing the sidelengths of the real-space triangle (Figure~\ref{fig:config})
while keeping the opening angle fixed.
On this basis we can expand
`real space around redshift space'~\cite{Scoccimarro:2004}, 
in other words, we expand out the displacement in Eq.~\eqref{eq:gsm}
then perform the resulting integrals analytically. Note that most of the
support of $p$ is tightly centred around $\bm{m}$
(which turns out to be very close to the point at which
$p$ attains its maximum value).

To facilitate the integration, we define the joint six-dimensional
vectors $\bm{X}=(\x_1,\x_2)$, $\bm{S}=(\s_1,\s_2)$, and
$\bm\Delta=\bm{S}-\bm{X}$. 
In these coordinates Eq.~\eqref{eq:gsm} takes the form
\be\label{eq:xis-6d}
1+\xi_{AB}^{(\mrm{s})}(\bm{S})
=\int\dif^6\bm\Delta\,\big[1+\xi_{AB}(\bm{S}-\bm\Delta)\big]\,
    p(\bm\Delta\Mid\bm{S}-\bm\Delta) \,,
\ee
where the integration is now over the displacement $\bm\Delta$ and
we have suppressed the dependence on $A$ and $B$.
Since $\bm\Delta$ is small compared to $\bm{S}$ we expand
$\xi_{AB}(\bm{S}-\bm\Delta)$ and $p(\bm\Delta\Mid\bm{S}-\bm\Delta)$
around $\bm{S}$:
\bea
\xi_{AB}(\bm{S}-\bm\Delta)
&=\big[1-(\bm\Delta\cdot\bm\partial)+\tfrac12(\bm\Delta\cdot\bm\partial)^2+\cdots\big]\,\xi_{AB}(\bm{S})  \,, \nonumber\\
p(\bm\Delta\Mid\bm{S}-\bm\Delta)
% =\rme^{-\Delta_i\sp\partial_i}p(\Delta;S)
&=\big[1-(\bm\Delta\cdot\bm\partial)+\tfrac12(\bm\Delta\cdot\bm\partial)^2+\cdots\big]\sp
    p(\bm\Delta\Mid\bm{S})\,, \nonumber
\eea
where we have defined the joint partial derivative
\be\label{eq:Di}
\partial_a=\partial/\partial\sp S^a\,,
\qquad a=1,2,\ldots,6\,
\ee
acting on six-dimensional Cartesian space.
Substituting these back into
Eq.~\eqref{eq:xis-6d} and using that
\bea
\textstyle\int\dif^6\bm\Delta\sp p(\bm\Delta\Mid\bm{S})\sp \Delta_a&=m^a_{AB}\,, \nonumber\\
\textstyle\int\dif^6\bm\Delta\sp p(\bm\Delta\Mid\bm{S})\sp \Delta_a \Delta_b
&=C^{ab}_{AB}+m^a_{AB}\sp m^b_{AB}\,, \nonumber
\eea
where $\bm{m}_{AB}$ is now a 6-dimensional vector and 
$\mbf{C}_{AB}$ a $6\times6$ matrix, we find up to second order
\bea
&\xi^{(\mrm{s})}_{AB}
=\xi_{AB}-\partial_a\sp m_{AB}^a+\tfrac12 \partial_a\partial_b C_{AB}^{ab} 
-m_{AB}^a\sp \partial_a\sp\xi_{AB} \nonumber\\
&\qquad
+ \tfrac12 \partial_a\partial_b\sp(m_{AB}^a\sp m_{AB}^b)
    - \xi_{AB}\sp \partial_a\sp m_{AB}^a 
    + \tfrac12 C_{AB}^{ab}\sp \partial_a\partial_b\sp\xi_{AB} \nonumber\\
&\qquad
    + \tfrac12 \xi_{AB}\ssp \partial_a\sp \partial_b C_{AB}^{ab}
    + \partial_a\sp\xi_{AB}\sp \partial_b C_{AB}^{ab} 
    + \mathcal{O}(\xi_{AB}^3)\,.
    \label{eq:xis-expand}
\eea
Here repeated indices $a,b$ are summed over (no sum over $A,B$) and all
terms on the right-hand side are
real-space quantities evaluated at the redshift-space position,
i.e.\ at $\bm{X}=\bm{S}=(\s_1,\s_2)$.
The first three terms on the right-hand side
contain the usual leading-order contributions while the
rest of the terms are quadratic in correlations (induced by the nonlinear
part of the redshift map).
In the case of distortions due purely to RSD, we have previously
shown~\cite{Dam:2023} that the first three terms recover the usual Kaiser 
multipoles along with the leading-order wide-angle corrections.\footnote{Note that Eqs.~\eqref{eq:xis-6d} and Eq.~\eqref{eq:xis-expand} are not only valid for the radial displacements considered in this work, but also for angular displacements, like those due to lensing.}

The type of terms appearing in Eq.~\eqref{eq:xis-expand} is revealing.
Although we know from the nonperturbative expression of $\xi^{(\mrm{s})}_{AB}$
[Eq.~\eqref{eq:gsm2}] that the dependence on the underlying real-space
correlations is nonlocal,
the expansion~\eqref{eq:xis-expand} shows that beyond linear order (where $m^a$ or $C_{ab}$ only appear in differentiated form) we also have terms with
local dependence on $m^a$ and $C_{ab}$ (e.g.\ 
$m_{AB}^a\sp \partial_a\sp\xi_{AB}$).
This has a key consequence for the dipole because
$m^a$ (and $C_{ab}$) contain one-point functions (`dispersions')
whose effect on $\xi^{(\mrm{s})}_{AB}$ is not necessarily differentiated
away at second order. In other words, the absolute scale of $m^a$ is important.

\subsection{Perturbative expression of the dipole}\label{sec:dipole-expand}

To capture the turnover occurring on small to intermediate scales,
we now compute the dipole of $\xi^{(\mrm{s})}_{AB}$
using the expansion~\eqref{eq:xis-expand}, keeping terms up to
second order in correlations. Although there are a large number of terms,
fortunately not all of them contribute to the dipole and we can
discard many of them. In particular, since $\mbf{C}_{AB}=\mbf{C}_{(AB)}$
the linear term $\partial_a\partial_bC^{ab}_{AB}$ does not contribute to the
odd multipoles and can be ignored.\footnote{The leading contribution from
$C^{ab}_{AB}$ to the dipole comes from a third-order
statistic, e.g.\ $\langle\delta_A u_\|\psi\rangle$.
Such contributions are ignored throughout this work.}

To further simplify the calculation we will neglect
the lightcone and lookback-time effects. As we saw in
Figure~\ref{fig:complete_dipole}, the turnover is present in the absence of
these effects (the full second-order expression {including these}
effects can be found in Appendix~\ref{app:full-anti}). 
With this simplification $\xi_{AB}$ is just the standard rest-frame
galaxy correlation function and so contains
no anti-symmetric part, $\xi_{AB}=\xi_{(AB)}(s)$. This allows us to further
ignore terms such as $\xi_{AB}\sp \partial_a\sp \partial_b\sp C_{AB}^{ab}
=\xi_{(AB)}\sp \partial_a\sp \partial_b\sp C_{(AB)}^{ab}$,
which is totally symmetric at lowest order.

We are thus left with three anti-symmetric terms,
namely, the second, fourth and sixth terms in Eq.~\eqref{eq:xis-expand}.
These can be written as an overall derivative:
\bea
\xi^{(\mrm{s})}_{[AB]}
&\simeq -(1+\xi_{(AB)})\sp \partial_a\sp m^a_{[AB]}
    -m^a_{[AB]}\sp \partial_a\sp\xi_{(AB)} \label{eq:xis-2nd-2}\\
% &\simeq -\partial_a\sp m^a_{[AB]}-\xi_{(AB)}\sp \partial_a\sp m^a_{[AB]}
%     -m^a_{[AB]}\sp \partial_a\sp\xi_{(AB)} \nonumber\\
&=-\partial_a\big[(1+\xi_{(AB)})\sp m^a_{[AB]}\big] \,.
\label{eq:xis-2nd}
\eea
It is no accident that $\xi^{(\mrm{s})}_{[AB]}$ can be written as an overall 
derivative (or divergence of the `momentum'). Such a term is related to the
conservation of pairs~\cite{Davis_Peebles,Peebles_LSS,Smith:2007gi} which
we have in redshift space as well as in real space, just as we have
conservation of number between real and redshift space (see also Appendix~\ref{eq:rsd-dynamical}).
From a modern field theory perspective, one might expect a term of this
form on the basis that the expansion~\eqref{eq:xis-expand} consists
of all scalars constructed from $1+\xi_{AB}$, $m_{AB}^a$, $C_{AB}^{ab}$ and
derivatives $\partial_a$ thereof.

To extract the dipole from Eq.~\eqref{eq:xis-2nd} we
need to switch to standard coordinates $(s,d,\mu)$.
To do this we first evaluate Eq.~\eqref{eq:xis-2nd}
in coordinates $(\chi_1,\chi_2,\cos\vartheta)$, with which the derivatives
are easy to compute, and then change to $(s,d,\mu)$ using the
coordinate relations in Appendix~\ref{app:formulas}. Since this
coordinate change is nonlinear we work to first order in $\epsilon$.

There are two terms to calculate in Eq.~\eqref{eq:xis-2nd-2},
$\partial_a\sp m^a_{[AB]}$ and $m_{[AB]}^a\sp \partial_a\sp\xi_{(AB)}$.
Using the chain rule and working to leading order in $\epsilon$, we find for
the latter term
\be\label{eq:shift-dipole}
m^a_{[AB]}\sp \partial_a\sp\xi_{(AB)}
=\mu\left(\llangle\Delta\psi\rrangle
    +\epsilon\sp\frac25\sp\llangle\bar{u}\rrangle\right)
        \frac{\dif\sp\xi_{(AB)}}{\dif s}
    +\mathcal{O}(\epsilon^2)\,.
\ee
This is of course a dipole, which we expected by virtue of this term 
being overall anti-symmetric (at order $\epsilon$ there is also
an octupole ($\ell=3$) but which we omit for brevity).
Both pairwise functions appearing here are 
anti-symmetric.

\begin{figure}[t]
  \centering
  \includegraphics[scale=1]{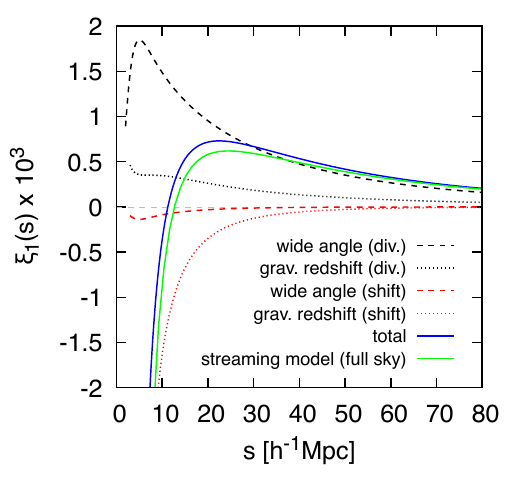}
  \caption{The four contributions to the dipole from the
  perturbative formula~\eqref{eq:xi1-2nd}.
  Here `total' is the sum of the divergence
  contributions `wide angle (div.)' (RSD) and
  `grav.~redshift (div.)' [respectively the fourth and first terms in Eq.~\eqref{eq:xi1-2nd}],
  and 'shift' contributions
  [respectively the third and second terms in Eq.~\eqref{eq:xi1-2nd},
  proportional to $\dif\xi/\dif s$].
  For reference we have also
  plotted the nonperturbative dipole calculated from the streaming model
  (neglecting the connected piece in the covariance~\eqref{eq:mu_cov}). 
  For the gravitational~redshift curve there is a steep rise and fall near $s=0$ due to the NFW cusp.
  (Note that each curve is an odd function so must go to zero at $s=0$, independent of model.)
  Here $b_A=2$, $b_B=1$, and $z=0.341$. }
  \label{fig:xi1_pert}
\end{figure}

\begin{figure}
  \centering
  \includegraphics[scale=1]{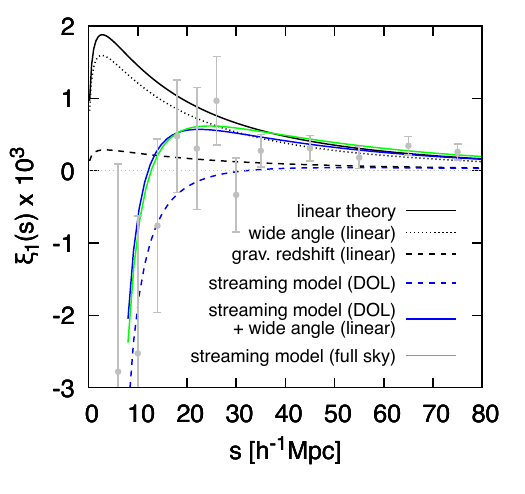}
  \caption{Impact of wide-angle effects on the linear and nonlinear dipoles at $z=0.341$. The linear-theory 
  dipole (solid black) consists of the contributions from gravitational redshift (dashed black)
        and wide-angle effects in RSD (dotted black).
        The other three curves show various nonlinear models of the dipole:
        (i) dashed blue is the
        streaming model in the distant-observer limit (DOL) [Eq.~\eqref{eq:gsm-dol}],
        in which only gravitational redshift contributes to the dipole
        and not RSD (since there are no wide-angle effects);
        (ii) solid blue is similar to model (i) but with the linear wide-angle
        correction added as a post-hoc correction;
        (iii) solid green is the streaming model in the wide-angle regime with
        RSD and gravitational redshift [see Eq.~\eqref{eq:gsm2}].
        Note that: none of these
        models include the effects of lookback time and lightcone;
        each of the linear-theory curves does not by definition contain
        normalisation from the density weighting; and the linear-theory curve
        of gravitational redshift is computed purely
        using linear theory, not the halo model
        (as used in Figure~\ref{fig:xi1_pert}).
        For reference we show the measurements from the RayGal catalogues
        (for $M_A=2.2\times10^{13}\,h^{-1}M_\odot$ and
        $M_B=0.28\times10^{13}\,h^{-1}M_\odot$). }
  \label{fig:800x100}
\end{figure}

Computing $\partial_a\sp m_{[AB]}^a$ is more involved.
For convenience, we split this into a part from RSD and
a part from gravitational redshift:
\be
\partial_a\sp m^{a}_{[AB]}
\equiv \partial\cdot m_{[AB]}
=\partial\cdot m^{\mrm{RSD}}_{[AB]}
+\partial\cdot m^{\mrm{grav}}_{[AB]}\,.
\ee
In spherical coordinates the divergence is given by
\be
\partial\cdot\sp m
=\sum_{J=1,2}\frac{1}{\chi_J^2}\frac{\partial(\chi_J^2\,m^J)}{\partial\chi_J}\,
=\sum_{J=1,2}\left(\frac{\partial}{\partial\chi_J}
     +\frac{2}{\chi_J}\right) m^J \,, \label{eq:div-formula}
\ee
where the sum is over the two radial components of
$\bm{m}_{[AB]}$, see Eq.~\eqref{eq:mrsd-mgrav}.
The absence of the angular gradient on the right-hand side
is because $\bm{m}_{AB}$ is purely radial.

The inverse-distance terms in the divergence
formula~\eqref{eq:div-formula} are wide-angle
terms and so vanish when we take the distant-observer limit ($\epsilon\to0$).
This means that we only need to consider these types of terms for
${m}_{[AB]}^\mrm{RSD}$
since for ${m}_{[AB]}^\mrm{grav}$ these terms lead to corrections of
order $\epsilon\sp\llangle\Delta\psi\rrangle\sim\epsilon^2$.
In other words, we include wide-angle effects for RSD but not for
gravitational redshift.

Substitution of the radial components~\eqref{eq:m-gravz} into
the foregoing equation~\eqref{eq:div-formula} then yields
\bea
\partial\cdot{m}^{\mrm{RSD}}_{[AB]}
&=\left(\frac{\partial}{\partial\chi_1}
     +\frac{2}{\chi_1}\right)
     \mu_1\llangle\bar{u}\rrangle
    +\left(\frac{\partial}{\partial\chi_2}
     +\frac{2}{\chi_2}\right)
     \mu_2\llangle\bar{u}\rrangle\,, \nonumber\\
\partial\cdot{m}^{\mrm{grav}}_{[AB]}
&=\frac{1}{2}\frac{\partial}{\partial\chi_1}
     % +\frac{2}{\chi_1}
     \llangle\Delta\psi\rrangle
    -\frac{1}{2}\frac{\partial}{\partial\chi_2}
     % +\frac{2}{\chi_2}
     \llangle\Delta\psi\rrangle
     +\mathcal{O}(\epsilon^2) \,,\nonumber
    % \label{eq:dmu-anti}
\eea
where $\mu_1\equiv\hat\s\cdot\n_1$ and $\mu_2\equiv\hat\s\cdot\n_2$, and
$\llangle\Delta\psi\rrangle$ and $\llangle\bar{u}\rrangle$ are
evaluated at $r=s$.
To extract the dipole we need to express each
of these terms in coordinates $s,d,\mu$
using the coordinate relations given in Appendix~\ref{app:formulas}.
After a short calculation, we find
\begin{align}
\frac32\int\dif\mu\sp\mathcal{L}_1(\mu)\,
    \partial\cdot m^{\mrm{RSD}}_{[AB]}
&=\epsilon\sp
    \frac{4}{5s^4}\frac{\dif(s^4\llangle\bar{u}\rrangle)}{\dif s}
    +\mathcal{O}(\epsilon^2)\,, \nonumber\\
\frac32\int\dif\mu\sp\mathcal{L}_1(\mu)\,
    \partial\cdot m^{\mrm{grav}}_{[AB]}
&=\frac{\dif\llangle\Delta\psi\rrangle}{\dif s}+\mathcal{O}(\epsilon^2) \,.
\label{eq:xi1-rsd}
\end{align}
These are the linear dipoles due to RSD and gravitational redshift
(although with a correction from the normalisation due to the density weighting).
Note that the RSD dipole, given here in unusually compact form, is in
fact equivalent to the well-known but lengthier expression
(see Appendix~\ref{app:config-rsd-dipole}).

Finally, inserting Eqs.~\eqref{eq:shift-dipole} and
\eqref{eq:xi1-rsd} into Eq.~\eqref{eq:xis-2nd-2} we find for the total dipole
\bea\label{eq:xi1-2nd}
&\xi_{1}(s,d)
=
-(1+\xi_{AB})\frac{\dif\llangle\Delta\psi\rrangle}{\dif s}-\llangle\Delta\psi\rrangle\sp\frac{\dif\xi_{AB}}{\dif s} \\
&\quad
-\epsilon\left(\frac25\llangle\bar{u}\rrangle\frac{\dif\xi_{AB}}{\dif s}
    +(1+\xi_{AB})\frac{4}{5s^4}\frac{\dif(s^4\llangle\bar{u}\rrangle)}{\dif s}\right)
    +\mathcal{O}(\epsilon^2)\,,
    \nonumber
\eea
where we recall the dependence on $d$ enters only through $\epsilon=s/d$.
This formula, expressed in terms of the pairwise functions, contains the 
well-known linear contribution from wide-angle RSD (last term) as well as
the linear gravitational redshift, 
$\frac{\dif}{\dif s}\llangle\Delta\psi\rrangle_{\rm lin}\simeq(b_A-b_B)\langle\frac{\partial\psi}{\partial\chi}\delta\rangle$ (first term).
The other two terms (second and third terms) are
second order in correlators and derive from
$m_{AB}^a\sp \partial_a\sp\xi_{AB}\propto\frac{\dif}{\dif s}\xi_{AB}$.
As a check, note that in the distant-observer limit ($\epsilon\to0$) only
the potential terms remain; the wide-angle RSD contributions, which here
includes a second-order effect, vanish as we expect.%
\footnote{But this does not mean that gravitational redshift is cleanly probed 
in this limit. There are other kinematic contaminants
in this limit which we have not included in the calculation (e.g.\ the
lightcone effect). Had we included this we would find in
Eq.~\eqref{eq:xi1-2nd} an additional kinematic contribution
proportional to $\llangle\bar{u}\rrangle$ but with no dependence on $\epsilon$;
see Eq.~\eqref{eq:lightcone-decomp}.}

\subsection{Discussion}

Figure~\ref{fig:xi1_pert} shows the contributions from each term in
Eq.~\eqref{eq:xi1-2nd}. Both the RSD contribution and the gravitational redshift contributions are split into a linear contribution and a second-order contribution, which we call `shift' since it arises from the shift term $m_{[AB]}^a\sp \partial_a\sp\xi_{(AB)}$ in Eq.~\eqref{eq:xis-2nd-2}. 
Note that the linear contributions are not strictly linear, due to the normalisation $1+\xi_{AB}$, as well as the intrinsic
nonlinear terms within $\Delta\psi$ and $\bar{u}$.

Concerning the turnover in the dipole, we see from Figure~\ref{fig:xi1_pert} that the important term is
due to the shift term in the gravitational redshift:
\be\label{eq:shift}
\xi_1(s,d)
\supset-\llangle\Delta\psi\rrangle(s)\sp\frac{\dif\sp\xi_{AB}}{\dif s}\,.
\ee
The other shift contribution, due to wide-angle RSD, is proportional to
$\llangle\bar{u}\sp\rrangle\frac{\dif}{\dif s}\xi_{AB}$, and, as Figure~\ref{fig:xi1_pert} shows, has negligible
impact on the dipole. In the following, we call the gravitational redshift shift contribution the `shift' for simplicity.

Figure~\ref{fig:xi1_pert} shows that the shift gives a negative contribution to the dipole and rapidly grows in
size towards small scales.
Around $s\simeq 20\,h^{-1}\mrm{Mpc}$ the shift begins to dominate over the
other (net positive) contributions, which are driven mainly by the (linear)
wide-angle effect. Eventually the shift cancels completely with the other
contributions, leading to a sign change in the dipole.

The {scale dependence} of the shift~\eqref{eq:shift} is mainly due
to the slope of $\xi_{AB}$ and not from $\llangle\Delta\psi\rrangle$
itself. As we saw in Figure~\ref{fig:psi_12}, $\llangle\Delta\psi\rrangle$ is 
relatively flat with mild scale dependence, vanishing as $r\to0$.
By contrast, $\xi_{AB}$ contains more small-scale power and so is
increasingly steep towards small $r$, nonvanishing as $r\to0$.

The amplitude of the shift~\eqref{eq:shift} depends on both
$\llangle\Delta\psi\rrangle$ and the slope of $\xi_{AB}$. There
is however more uncertainty in the amplitude compared to the scale dependence
due to the relative uncertainty of $\llangle\Delta\psi\rrangle$.
As discussed in Section~\ref{sec:phi-HM}, $\llangle\Delta\psi\rrangle$
(as with pairwise velocity difference) is sensitive to nonlinearities
in that the large-scale contributions tend to cancel out in the difference,
bringing into relief the small-scale contributions which do not cancel.
The one-point function $\langle\psi(0)\sp\delta_A(0)\rangle$ contained within
is sensitive to small-scale modelling, and corrections to 
the linear estimate of $\langle\psi(0)\sp\delta_A(0)\rangle$ can be large
(from $50\%$ to $200\%$, depending on the mass of the tracers). Uncertainty in this
quantity translates to uncertainty in the scale at which the turnover occurs---or whether it occurs at all. 
While we have assumed linear theory for
$\xi_{AB}$, we have checked that improved modelling
of $\xi_{AB}$ using the resummed Zel'dovich model~\cite{Carlson:2012,White:2014}
leads only to small changes to the shape and size of the shift~\eqref{eq:shift}.
These results suggest that
efforts to improve the modelling of the dipole should focus on $\llangle\Delta\psi\rrangle$
and $\langle\psi(0)\sp\delta_A(0)\rangle$ contained therein.

Although not included in our modelling, we note that
halo exclusion (that haloes cannot overlap)~\cite{Tinker:2006dm,Baldauf:2013hka} 
strongly affects the small-scale behaviour of $\xi_{AB}(r)$  and which we expect
drives another turnover in the dipole around the exclusion scale
$R_\mrm{exc}=R_\mrm{vir,A}+R_\mrm{vir,B}$. We can understand this because $R_\mrm{exc}$
demarcates two distinctly different regimes: for $r\gtrsim R_\mrm{exc}$ we roughly have
the power-law behaviour $\xi_{AB}(r)\propto(r/R_\mrm{exc})^{-\gamma}$ with $\gamma>0$, whereas for $r< R_\mrm{exc}$
we have that $\xi_{AB}(r)=-1$ (as required by the condition that the probability of finding
a halo pair separated by less than $R_\mrm{exc}$ is zero).
This means that we must have a sign change in $\xi_{AB}$ and
$\dif\sp\xi_{AB}/\dif s$, leading to a turnover in the dipole through Eq.~\eqref{eq:shift}.

Since the shift~\eqref{eq:shift} does not vanish in the limit $\epsilon\to0$
it must therefore be present in the distant-observer limit.
This is confirmed by Figure~\ref{fig:800x100}, which shows the predicted dipole
(dashed blue curve) of the standard streaming model in the distant-observer limit but
with the inclusion of gravitational redshift [Eq.~\eqref{eq:gsm-dol}]. 
That
the dipole is negative at small scales and then hovers slightly above zero at
large scales  tells us that it is driven primarily by the shift, as we expect.
However, there is no longer a turnover because the large, positive contribution
from wide-angle RSD is absent from this model.
Interestingly, we find that we can restore the turnover using a hybrid model.
In this model we take the dipole from the aforementioned standard
streaming model
(but generalised to include gravitational redshift), and add to it
a post-hoc correction given by wide-angle dipole [Eq.~\eqref{eq:xi1_rsd-pairwise}]. 
With this simple addition we recover the turnover and find better
agreement with the simulations on scales $s>40\,h^{-1}\mrm{Mpc}$.
But this is not entirely unexpected because we know that at relatively high
redshift ($z=0.341$) the wide-angle effects are in the perturbative
regime and thus able to be accounted for using the first-order correction
(see Figure~\ref{fig:wide-angle-multipoles}).
At lower redshifts such a hybrid approach will not be possible and the full
model in the wide-angle regime will be needed.

\section{Conclusions}\label{sec:conclusions}

{Gravitational redshift is a key observable for future galaxy surveys.
Past works have shown that the mildly nonlinear regime of large-scale structure is well adapted 
to measure this effect, since the signal is boosted on these scales. But to properly isolate the gravitational
redshift signal it is essential to have a robust model of this effect, as well as of all other effects
that contribute to the correlation function at the same order.} 

This work has shown that all such effects---namely, the order
$\calH/k$ relativistic effects---are captured by a simple
formula~\eqref{eq:deltas-all}. The two-point
function is given by Eq.~\eqref{eq:xi_final} and builds on the
wide-angle generalisation of the streaming model developed in
previous work~\cite{Dam:2023}. Compared with that work, here we
have given a complete model by incorporating into the framework
the lightcone effect (Section~\ref{sec:lightcone}) and the
gravitational redshift, our main interest in this work.

This work has focussed on nonlinear modelling of the dipole,
the multipole most sensitive to the gravitational redshift.
With our complete model we found that the gravitational redshift
and the kinematic wide-angle effects are the two most important 
contributions to the dipole, supporting the earlier findings of
Ref.~\cite{Breton:2018wzk}. The contributions from the lightcone and
lookback-time effects are small but not negligible. 

We saw that the dipole's turnover at separation $s\simeq 20\,h^{-1}\mrm{Mpc}$---a key feature
marking a regime where the gravitational redshift begins to dominate over the
kinematic contributions---is well accounted for in our model.
From a perturbative expansion of the model, we showed that this behaviour
is due to an advection-like effect (which
we called the `shift') whereby
pairs of distinct galaxies ($A\neq B$) are asymmetrically transported
along the lines of sight.

The important quantity driving this turnover is the
pairwise potential difference $\llangle\Delta\psi\rrangle$. This quantity
is sensitive to nonlinearities at all scales so is not well modelled by linear theory.
There are several reasons for this.
As with the pairwise velocity difference $\llangle\Delta u\rrangle$,
the long-wavelength contributions to
$\llangle\Delta\psi\rrangle$ tend to cancel out in the difference,
leaving behind the small-scale nonlinearities.
(In fact this cancellation should be more acute for $\llangle\Delta\psi\rrangle$ given
that the potential is longer range, more sensitive to large-scale matter
modes than the velocity field.)

We showed that the dominant contribution to $\llangle\Delta\psi\rrangle$ comes from the one-halo term (see Figure~\ref{fig:psi_12}) and is driven in large part by
the difference between $\langle n_A(\x)\sp\psi(\x)\rangle/\langle n_A(\x)\rangle$
and $\langle n_B(\x)\sp\psi(\x)\rangle/\langle n_B(\x)\rangle$, i.e.\ the
difference between the density-weighted mean potential of each population.
Unlike the velocity, the {weighted} potential does \emph{not} average to zero:
$\langle \delta_A(\x)\v(\x)\rangle=0$
whereas $\langle\delta_A(\x)\sp\psi(\x)\rangle\neq0$ (and likewise for $B$). This follows because, under isotropy,
we are as likely to sample both positive and negative velocities (both
redshift or blueshift) in equal measure. This is true for both the coherent
velocities on large scales and for the random virial motions on small scales:
sampled over sufficiently large volumes velocities have no preferred directions, $\langle \delta_A(\x)\v(\x)\rangle=0$.
By contrast $\langle\delta_A(\x)\sp\psi(\x)\rangle$ tends to have a definite sign.
Since the density weighting biases the potential towards
overdense collapsed regions,
we will tend to see the effect of a gravitational redshift, not a
gravitational blueshift.
Thus, $\llangle\Delta\psi\rrangle$ is sensitive to the immediate
environment through the local potential in which the tracers reside. 

The importance of a nonlinear ingredient to capture the turnover of the dipole at small scales has been pointed out in previous works~\cite{Beutler:2020evf,Saga:2020tqb}. In Ref.~\cite{Beutler:2020evf}, an EFT-like nuisance parameter is added to the modelling of the power spectrum dipole, and fitted from the simulations. In Ref.~\cite{Saga:2020tqb}, a constant nonlinear potential contribution is added to the displacement field of each of the tracers. In this work, we showed how this nonlinear ingredient naturally arises, namely, through the density-weighted pairwise potential difference $\llangle\Delta\psi\rrangle$, which
we estimated using the halo model.

Besides these findings on the gravitational redshift,
we found in the course of our calculations that redshift-space
correlations find their natural expression in terms of pairwise
correlation functions, whether in the distant-observer
limit or the wide-angle regime.
We found, for example, that the dipole generated by RSD
takes on a particularly compact form in terms of the pairwise mean velocity.
These quantities each have distinct physical meaning and
can depend on one-point functions like the velocity dispersion, 
or other nonperturbative inputs
absent in more
straightforward perturbative treatments.
Moreover, these scalar quantities carry parity: either symmetric or anti-symmetric
under pair exchange of distinct tracers (and only symmetric in the case of identical tracers).
We saw that such properties underlie the more conventional way of identifying anti-symmetric
correlations (in terms of derivative counting), and that
these functions provide a convenient way to build up the multipoles perturbatively.
On large scales (terms linear in correlators), the symmetric functions
contribute to the even multipoles, while the anti-symmetric functions
contribute to the odd multipoles. On smaller scales
(terms quadratic or higher in correlators), one counts all
products of the  pairwise functions which are overall symmetric or
anti-symmetric as contributing to the even or odd multipoles, respectively.

Finally, let us mention a few ways in which our model can be improved.
\begin{enumerate}
\item An obvious
starting point is the extension of the gravitational evolution and tracer bias
beyond linear theory. A better approach would be to use,
for example, convolution Lagrangian
perturbation theory~\cite{Carlson:2012,White:2014} to compute all the
density-weighted correlations.
\item
This work used haloes as tracers, as per the RayGal catalogues, but a more realistic application should extend the modelling
to galaxies. This implies an important change in the modelling of
the pairwise potential difference. Since galaxies do not
necessarily lie at the centre of haloes, $\llangle\Delta\psi\rrangle$
needs to be recalculated to account for this offset (e.g.\ through
modelling of the halo occupation statistics).
\item Inclusion of magnification and evolution bias is another obvious
area of improvement.
As we showed in previous work~\cite{Dam:2023}, these selection effects
enter through a change in the density weighting in a similar way to
the lightcone correction so should be straightforward to include
in the model.
\item It would be interesting to 
investigate the impact of non-Gaussian cumulants on the dipole (in analogy
with Ref.~\cite{Uhlemann:2015hqa}). If the first cumulant (the mean) is important for
the anti-symmetric correlations, as we found in this work, then it might reasonably
be expected that the third cumulant (the skew) is the next most relevant. In this context,
numerical simulations of the pairwise statistics of the potential difference and a study
of the scale-dependent distribution should better inform further improvements to
the modelling. We leave these aspects for future work.
\end{enumerate}

\begin{acknowledgments}
We thank Julian Adamek, Anton Chudaykin, Omar Darwish, Enea Di Dio,
Francesca Lepori, Yann Rasera, and Atsushi Taruya for helpful discussions;
Michel-Andr\`es Breton for providing us with the
RayGal dipole measurements;
Jun Koda for correspondence and for sharing with LD his notes on the
$\lambda$ technique;
and especially Sebastian Schulz for insightful discussions on simulations
and the dipole comparison.
This work was supported by the European Research Council (ERC) under the European Union’s
Horizon 2020 research and innovation program (Grant agreement No.~863929;
project title ``Testing the law of gravity with novel large-scale structure observables'').
Use of the Baobab HPC cluster at the University of Geneva
is gratefully acknowledged. 
Code used to obtain the results of this work
is publicly available at
\url{https://github.com/lhd23/relativistic_streaming}.
\end{acknowledgments}

% ~~~~~ end of main text ~~~~~

% \clearpage
\onecolumngrid
% ~~~~~ begin appendix ~~~~~
\appendix
\section{An alternative way to model \texorpdfstring{$p(\bm\chi\Mid\bm\chi')$}{$p(\chi\Mid\chi')$}}\label{app:transitions}
This appendix outlines an alternative strategy for obtaining the
transition probability
$p(\bm\chi\Mid\bm\chi')$ and with it the correlation function $\xi^\mrm{(s)}$.
The basic idea is to use the chain rule to reduce $p(\bm\chi\Mid\bm\chi')$
to the product of two distributions, from RSD and gravitational
redshift, with the assumption
that each effect is independent of the other. This amounts to
ignoring the correlations between velocity and potential (a good approximation
in the case of the dipole; see Section~\ref{sec:results}).

Without making any assumptions about $p(\bm\chi\Mid\bm\chi')$,
we insert Eq.~\eqref{eq:p-chain} into the
correlation function~\eqref{eq:gsm} to get
\bea
1+\xi^{(\mrm{s})}(\bm\chi)
&=\frac{1}{\chi_1^2\,\chi_2^2}
    \int\dif\chi_1'\,\chi_1'^2
    \int\dif\chi_2'\,\chi_2'^2\sp
    [1+\xi(\bm\chi')]\sp
    p(\bm\chi\Mid\bm\chi') \nonumber\\
&=\frac{1}{\chi_1^2\,\chi_2^2}
    \int\dif\chi_1'\,\chi_1'^2
    \int\dif\chi_2'\,\chi_2'^2\sp
    [1+\xi(\bm\chi')]
 \int\dif^2\bm\chi''
        p_\mrm{grav}(\bm\chi\Mid\bm\chi'',\bm\chi')\sp 
        p_\mrm{RSD}(\bm\chi''\Mid\bm\chi')\,,
\label{eq:xi-chain1}
\eea
where $\dif^2\bm\chi''=\dif\chi_1''\dif\chi_2''$.%
\footnote{Note that in pure vector form
$\chi_1^2\sp\chi_2^2=(\bm\chi^\T\mbf{A}\bm\chi)^2$ with
$\mbf{A}=\begin{pmatrix}0&1/2\\1/2&0 \end{pmatrix}$.
}
The crucial assumption that we will now make is that the transitions
are a Markov process so that the current state only depends 
on the immediate previous state, or 
$p_\mrm{grav}(\bm\chi\Mid\bm\chi'',\bm\chi')
=p_\mrm{grav}(\bm\chi\Mid\bm\chi'')$. This allows us to separate
the integral over the initial pair positions $\bm\chi'$
from the integral over the intermediate pair positions $\bm\chi''$.
Indeed, by rearranging the integrand in Eq.~\eqref{eq:xi-chain1} we can write
\bea
1+\xi^{(\mrm{s})}(\bm\chi)
&=\frac{1}{\chi_1^2\,\chi_2^2}
    \int\dif^2\bm\chi''\chi_1''^2\sp\chi_2''^2
\left(\,\frac{1}{\chi_1''^2\sp\chi_2''^2}
 \int\dif^2\bm\chi'\chi_1'^2\sp\chi_2'^2\sp
        [1+\xi(\bm\chi')]\sp
        p_\mrm{RSD}(\bm\chi''\Mid\bm\chi')\right)
        p_\mrm{grav}(\bm\chi\Mid\bm\chi'')\,.
        \label{eq:xi-nest}
\eea
Notice that this expression has a self-similar composition in that
the inner integral has the same form as the outer integral.
Thus defining $1+\xi_\mrm{RSD}(\bm\chi'')$ by the
contents of the parentheses, we can write Eq.~\eqref{eq:xi-nest}
as
\be\label{eq:xis-nested}
1+\xi^{(\mrm{s})}(\bm\chi)
=\frac{1}{\chi_1^2\sp\chi_2^2}
    \int\dif^2\bm\chi''\chi_1''^2\sp\chi_2''^2\sp
    \big[1+\xi_\mrm{RSD}(\bm\chi'')\big]\ssp
    p_\mrm{grav}(\bm\chi\Mid\bm\chi'')\,.
\ee
In fact, this nested form holds for any number
of transitions, provided the Markov property holds for each.
This means that a series of transitions can be modelled by applying successive
integral transformations to the base real-space correlation $1+\xi(\bm\chi')$.
The practical advantage is that 
the correlation function $1+\xi^\mrm{(s)}$ (due to both RSD and 
gravitational redshift) can be obtained by simply applying a
transformation to an existing, well-tested model of
$1+\xi_\mrm{RSD}$
(promoted to the wide-angle regime), i.e.\ by convolving it with
the distribution $p_\mrm{grav}(\bm\chi\Mid\bm\chi'')$.
In this way the problem of modelling RSD is separated from the problem of
modelling the gravitational redshift, which boils down to
modelling the distribution for the pairwise potential
(and which in a further approximation may be modelled
in the distant-observer limit as a one-dimensional probability
distribution; see Section~\ref{sec:origin}).

An obvious assumption one can make in the modelling is to suppose that both
$p_\mrm{RSD}(\bm\chi''\Mid\bm\chi')$ and $p_\mrm{grav}(\bm\chi\Mid\bm\chi'')$
are Gaussian. We will not investigate in any detail the validity of
this assumption, except to note here that the marginal
integral~\eqref{eq:p-chain} for $p(\bm\chi\Mid\bm\chi')$ does not
evaluate to an exact Gaussian, as one might presume, because of
the non-trivial dependence on $\bm\chi''$ through
$\bm{m}_\mrm{grav}(\bm\chi'')$ and $\mbf{C}_\mrm{grav}(\bm\chi'')$
preventing analytic integration. Certainly it does not integrate
to the two-dimensional Gaussian we have used in, e.g.\ Eq.~\eqref{eq:gsm2}.
However, the full numerical result turns out be to close to the 
analytic Gaussian result obtained by simply replacing
$\bm{m}_\mrm{grav}(\bm\chi'')\to\bm{m}_\mrm{grav}(\bm\chi)$ and
$\mbf{C}_\mrm{grav}(\bm\chi'')\to\mbf{C}_\mrm{grav}(\bm\chi)$,
where corrections to this approximation are small because displacements
$\bm\chi-\bm\chi''$ are small compared with the configuration
size $\bm\chi$ itself.

Alternatively, one may wish to simply take
$p_\mrm{grav}(\bm\chi\Mid\bm\chi'')=\delD(\bm\chi-\bm\chi''-\bm{m}_\mrm{grav}(\bm\chi''))$
(obtained by neglecting all but the first cumulant in the cumulant expansion
of Eq.~\eqref{eq:gauss}), on the basis that the covariance is
symmetric at leading order so unimportant for the dipole (see Section~\ref{sec:pair}).

\subsection{Redshift-space distortions as a dynamical equation and the $\lambda$ trick}\label{eq:rsd-dynamical}
In Sections~\ref{sec:p-interpretation} and \ref{sec:origin} we alluded
to a dynamical way of viewing RSD; here we give further support for this picture.

Introduce an auxiliary variable $\lambda$ and consider the map
$\s(\x,\lambda)=\x+\lambda\sp\delta\x(\x)$ which smoothly interpolates between
real space ($\lambda=0$) and redshift space ($\lambda=1$). Define the $\lambda$
density [cf.~Eq.~\eqref{eq:ns}]
\be\label{eq:n-lambda}
n_\lambda(\s,\lambda)
\equiv\int\dif^3\x\, n(\x)\ssp
    \delD\big[\s-\x-\lambda\ssp\delta\x(\x)\big]\,,
\ee
so that we recover the redshift-space density
$n_\lambda(\s,\lambda=1)=n^{\mrm{(s)}}(\s)$ when $\lambda=1$, and
the real-space density $n_\lambda(\s,\lambda=0)=n(\s)=n(\x)$ when
$\lambda=0$. To highlight the novelty in the following
calculation let us restrict to RSD only, so
$\delta\x(\x)=\u_\|(\x)\equiv\n\cdot\u(\x)\sp\n$.
Now differentiate Eq.~\eqref{eq:n-lambda}
with respect to $\lambda$:
\bea
\frac{\partial}{\partial\lambda}\,n_\lambda(\s,\lambda)
=\int\dif^3\x\,n(\x)\,\frac{\partial}{\partial\lambda}
    \int\frac{\dif^3\k}{(2\pi)^3}\,\rme^{-\im\k\cdot[\s-\x-\lambda\sp\u_\|(\x)]}
&=\int\dif^3\x\,n(\x)\ssp\u_\|(\x)\cdot
    \int\frac{\dif^3\k}{(2\pi)^3}\,\im\k\,
        \rme^{-\im\k\cdot[\s-\x-\lambda\sp\u_\|(\x)]} \nonumber\\
&=-\int\dif^3\x\,n(\x)\ssp\u_\|(\x)\cdot
    \frac{\partial}{\partial\s}\,
        \delD\big[\s-\x-\lambda\sp\u_\|(\x)\big]\,,
\label{eq:redshift-continuity1}
\eea
where in the first equality we used the Fourier representation of
the delta function, and in the last equality we exchanged the factor of $\im\k$
for an overall derivative in $\s$.

Equation~\eqref{eq:redshift-continuity1} is an expression of number conservation,
for if we define the $\lambda$ momentum field [cf.~Eq.~\eqref{eq:n-lambda}]
\be\label{eq:p-momentum}
\mbf{p}_\|(\s,\lambda)
\equiv\int\dif^3\x\,n(\x)\ssp\u_\|(\x)\ssp
        \delD\big[\s-\x-\lambda\sp\u_\|(\x)\big]\,,
\ee
(where the redshift-space momentum field is recovered when $\lambda=1$),
then we can write Eq.~\eqref{eq:redshift-continuity1} as
\be\label{eq:redshift-continuity2}
\frac{\partial}{\partial\lambda}\,n_\lambda(\s,\lambda)
+\frac{\partial}{\partial\sp\s}\cdot \mbf{p}_\|(\s,\lambda)=0\,.
\ee
With $\lambda$ acting as a time parameter, this can be viewed as a
continuity equation.
As we let time go from $\lambda=0$ to $\lambda=1$ this equation tells us 
how the real-space density continuously distorts into
the redshift-space density. Note that
$\mbf{p}_\|^\mrm{(s)}$, like $\u_\|$, is a purely radial field.

We can use the continuity equation to obtain the redshift-space analogue
of the pair conservation equation~\cite{Davis_Peebles,Peebles_LSS}.
By evaluating 
Eq.~\eqref{eq:redshift-continuity2} at $\s_1$, multiplying through by
$n_\lambda$ evaluated at $\s_2$, then adding to it the same equation
but with $\s_1\leftrightarrow\s_2$, we obtain
\be\label{eq:continuity-equation-2pt}
\frac{\partial}{\partial\lambda}
\big\langle n_\lambda(\s_1,\lambda)\sp
    n_\lambda(\s_2,\lambda)\big\rangle
+\mu\ssp\frac{\partial}{\partial s_z}\big[
    \big\langle n_\lambda(\s_1,\lambda)\sp
    n_\lambda(\s_2,\lambda)\big\rangle\sp\llangle\Delta u\rrangle_\lambda
    \big]
=0\,,
\ee
where $\partial/\partial s_z=\hat{\mbf{z}}\cdot\partial/\partial\sp\s$,
$\mu=\hat\s\cdot\hat{\mbf{z}}$ (with line of sight along $z$
direction), $\s=\s_1-\s_2$ and 
$\llangle\Delta u\rrangle_\lambda$ is the $\lambda$ pairwise velocity
difference (density weighting given by $n_\lambda$). 
Since the separation of pairs orthogonal to the line of sight
($\mu=0$) do not evolve under this mapping, the corresponding correlations are undistorted, as can be seen from Eq.~\eqref{eq:continuity-equation-2pt}.
Note that here we have assumed the distant-observer limit wherein the
Eq.~\eqref{eq:continuity-equation-2pt} takes a simple form; in the
wide-angle regime there is an additional term with dependence on the
pairwise mean velocity (see Section~\ref{sec:pairwise}).

Since we have touched on the redshift-space momentum field, a prospective cosmological probe~\cite{Okumura:2013zva,Sugiyama:2015dsa,Howlett:2019bky,Dam:2021fff}, let
us note that 
if we have a model or ansatz for the redshift-space density we can
use the continuity
equation~\eqref{eq:redshift-continuity2} to solve for the redshift-space momentum,
that is, without having to use Eq.~\eqref{eq:p-momentum} to model
$\mbf{p}^{\mrm{(s)}}_\|(\s)$ from scratch.
This means that if we have a model of the density
correlations (e.g.~Eq.~\eqref{eq:gsm2}), a model of the momentum
correlations can in principle be `generated' from it by using the
above $\lambda$ trick.
More precisely, by solving Eq.~\eqref{eq:redshift-continuity2}
by Green's method we obtain the redshift-space momentum field
$\mbf{p}^{\mrm{(s)}}_\|(\s)=\mbf{p}_\|(\s,\lambda=1)$ as
\bea
\mbf{p}^\mrm{(s)}_\|(\s)
&=-\frac{\partial}{\partial\lambda}
\int\dif^3\y\ssp
    \frac{(\s-\y)}{4\pi\sp|\sp\s-\y|^3}\ssp n_\lambda(\y,\lambda)\,
    \Big|_{\lambda=1}
=\frac{\partial}{\partial\lambda}
\left(\frac{\partial}{\partial\sp\s}
\int\dif^3\y\,G(\s-\y)\, n_\lambda(\y,\lambda)
    \right)\bigg|_{\lambda=1}\,,
% =
% \int\dif^3\s'\,\frac{\partial G(\s-\s')}{\partial s}\, \frac{\partial n_\lambda(\s',\lambda)}{\partial\lambda}\bigg|_{\lambda=1}
\eea
where $G(\s-\y)=1/4\pi\sp|\sp\s-\y|$ and we used that $\mbf{p}_\|^\mrm{(s)}$ has no
curl component~\cite{Dam:2021fff} so is able to be written as the gradient of a scalar.
The two-point correlation function of the radial momentum
$p_\|^\mrm{(s)}=\mbf{p}_\|^\mrm{(s)}\cdot\n$ is then
\be\label{eq:mom-2pcf}
\big\langle p_\|^\mrm{(s)}(\s_1)\sp p_\|^\mrm{(s)}(\s_2)\big\rangle
=\frac{\partial}{\partial\lambda_1}\frac{\partial}{\partial\lambda_2}
    \int\dif^3\y_1\ssp\frac{\partial}{\partial s_1} G(\s_1-\y_1)
    \int\dif^3\y_2\ssp\frac{\partial}{\partial s_2} G(\s_2-\y_2)\,
    \big\langle n_\lambda(\y_1,\lambda_1)\sp n_\lambda(\y_2,\lambda_2)\big\rangle\,
    \Big|_{\lambda_1=\lambda_2=0}\,,
\ee
where $s_1=|\s_1|=\s_1\cdot\n_1$ and
$s_2=|\s_2|=\s_2\cdot\n_2$.
This formula expresses the redshift-space momentum correlations in
terms of the redshift-space density correlations. It is valid in the full wide-angle regime. Since the relation
between the momentum (or velocity) and the density is nonlocal so too
is the relation between their correlations. This means that the
momentum correlation at a given separation is determined by density
correlations from a range of separations. We thus run into the
problem of needing to integrate down to arbitrarily small separations,
where the model of the density correlations breaks down. 
Furthermore, for wide-angle correlations one needs to contend with
high-dimensional integration. We note however that in the distant-observer limit
one can reduce Eq.~\eqref{eq:mom-2pcf} to a one-dimensional integral
over the line-of-sight separation.

\section{Correlation functions}\label{app:corrfuncs}
Here we list and give explicit formulae for all real-space correlation
functions that constitute the redshift-space correlation function~\eqref{eq:xi_final}.
These enter through the normalisation $1+\xi_{AB}$, the
mean $\bm{m}$ [Eq.~\eqref{eq:mrsd-mgrav}], and the covariance
$\mbf{C}$ [Eq.~\eqref{eq:covs}]. We will work at leading order,
keeping only leading-order statistics such as $\langle u_\|(\x_1)\sp u_\|(\x_2)\rangle$, neglecting third- and fourth-order statistics such as
$\langle u_\|(\x_1)\sp u_\|(\x_2)\sp\delta_A(\x_1)\rangle$ and
$\langle u_\|(\x_1)\sp u_\|(\x_2)\sp\delta_A(\x_1)\sp\delta_B(\x_2)\rangle$.
The components of $\bm{m}_\mrm{RSD}$ and
$\mbf{C}_\mrm{RSD}$ then read
\bea
\llangle u_\|(\x_1)\rrangle
&\simeq \big[1+\xi_{AB}(\x_1,\x_2)\big]^{-1}\hat{n}_1^i\big\langle u_i(\x_1)\sp \tilde\delta_B(\x_2)\big\rangle\,, \\
\llangle u_\|(\x_1)\sp u_\|(\x_2)\rrangle
&\simeq \big[1+\xi_{AB}(\x_1,\x_2)\big]^{-1}
    \hat{n}_1^i\hat{n}_2^j\big\langle u_i(\x_1)\sp u_j(\x_2)\big\rangle\,, 
\eea
while the components of  $\bm{m}_\mrm{grav}$, 
$\mbf{C}_\mrm{grav}$, and $\mbf{C}_\mrm{cross}$ respectively read
\bea
\llangle \psi(\x_1)-\psi_O\rrangle
&\simeq \big[1+\xi_{AB}(\x_1,\x_2)\big]^{-1}
    \big\langle [\psi(\x_1)-\psi_O] [\tilde\delta_A(\x_1)+\tilde\delta_B(\x_2)]\big\rangle\,, \\
\llangle[\psi(\x_1)-\psi_O][\psi(\x_2)-\psi_O]\rrangle
&\simeq \big[1+\xi_{AB}(\x_1,\x_2)\big]^{-1}
    \big\langle[\psi(\x_1)-\psi_O][\psi(\x_2)-\psi_O]\big\rangle\,,\\
\llangle[\psi(\x_1)-\psi_O]\sp u_\|(\x_2)\rrangle
&\simeq \big[1+\xi_{AB}(\x_1,\x_2)\big]^{-1} \hat{n}_2^j
    \big\langle[\psi(\x_1)-\psi_O]\sp u_j(\x_2)\big\rangle\,.
\eea
Here $\tilde\delta_A(\x_1)=\delta_A(\x_1)+\v(\x_1)\cdot\n_1$ and
$\tilde\delta_B(\x_2)=\delta_B(\x_2)+\v(\x_2)\cdot\n_2$
(including the lightcone correction).
Moreover, we will assume
linear theory so $\delta_A=b_A\sp\delta$
and $\delta_B=b_B\sp\delta$, with $\delta$ the linear matter field, and
$b_A$ and $b_B$ the linear biases.
Note that the normalisation $1+\xi_{AB}(\x_1,\x_2)$ [which also includes
the lightcone correction; see Eq.~\eqref{eq:weighting-replace}]
becomes important on small scales and is always retained in numerical
work. In addition to $\xi_{AB}$, there are six other
correlation functions needed.

\textbf{\textit{Velocity correlations.}}
The correlations needed for RSD can be constructed by taking
line-of-sight projections of the basic vector and tensor correlations,
$\langle u_i(\x_1)\sp\delta(\x_2)\rangle$ and
$\langle u_i(\x_1)\sp u_j(\x_2)\rangle$. By statistical isotropy,
these basic correlations have the general forms $\langle u_i(\x_1)\sp\delta(\x_2)\rangle
=\xi_{u\sp\delta}(r)\sp\hat{r}_i$ and
$\langle u_i(\x_1)\sp u_j(\x_2)\rangle=\xi_{u\sp u}^{\sp\|}(r)\hat{r}_i\hat{r}_j
+\xi_{u\sp u}^\perp(r)(\delta_{ij}-\hat{r}_i\hat{r}_j)$, i.e.\ they can
be written in terms of
scalar correlations $\xi_{u\sp \delta}(r)$,
$\xi_{u\sp u}^\|(r)$, and $\xi_{u\sp u}^\perp(r)$.
According to gravitational instability theory, the
growing mode is vorticity free so we can write $u_i$
solely in terms of the (scaled) velocity divergence
$\theta=\nabla_i u^i$ so
$u_i=\im k_i/k^2\theta$ (where $\theta$ has the same units as $\delta$).%
\footnote{Our Fourier convention is
\be
f(\x)=\int\frac{\dif^3\k}{(2\pi)^3}\,\rme^{-\im\k\cdot\x}\tilde{f}(\k)\,,
\quad
\tilde{f}(\k)=\int\dif^3\x\,\rme^{\im\k\cdot\x}{f}(\x)\,.
\ee
}
In terms of the power spectra of $\delta$ and $\theta$, the scalar
correlation functions are obtained through contractions with the unit
separation $\hat\r$:%
\footnote{In keeping with our system of notation, we
use $\xi_{u\sp\delta}$, $\xi_{u\sp u}^{\sp\|}$, and
$\xi_{u\sp u}^\perp$, rather than
$U$, $\Psi_{\|}$, and $\Psi_{\perp}$ more commonly used
in the literature.}
\bea
\xi_{u\sp\delta}(r)
\equiv\hat{r}^i\big\langle u_i(\x_1)\sp\delta(\x_2)\big\rangle
% &\equiv U(r)
&=\frac1\calH\int\frac{k^2\dif k}{2\pi^2}\,\left(\frac{\calH}{k}\right)\ssp j_1(kr)\sp P_{\theta\delta}(k) \,,\label{eq:U} \\
% \xi^{\|,\perp}_{u\:\!u}(r)
% &\equiv\Psi_{\|,\perp}(r)
% =\frac{1}{\calH^2}\int\frac{k^2\dif k}{2\pi^2}\left(\frac{\calH}{k}\right)^2\,K_{\|,\perp}(kr)P_{\theta\theta}(k) \, ,
% \label{eq:Psi}
\xi_{u\sp u}^{\sp\|}(r)
\equiv\hat{r}^i\sp\hat{r}^j\big\langle u_i(\x_1)\sp u_j(\x_2)\big\rangle
&=\frac{1}{\calH^2}\int\frac{k^2\dif k}{2\pi^2}\left(\frac{\calH}{k}\right)^2K_{\|}(kr)\sp P_{\theta\theta}(k)\,, \\
\xi_{u\sp u}^{\perp}(r)
\equiv \frac12\mathcal{P}_\perp^{\sp ki}\,\mathcal{P}_\perp^{\sp kj}\sp\big\langle u_i(\x_1)\sp u_j(\x_2)\big\rangle
&=\frac{1}{\calH^2}\int\frac{k^2\dif k}{2\pi^2}\left(\frac{\calH}{k}\right)^2K_{\perp}(kr)\sp P_{\theta\theta}(k) \label{eq:Psi_para}\,.%\\[3pt]
% \xi_{u\sp u}^\perp(r)
% =\frac{1}{\calH^2}\int\frac{k^2\dif k}{2\pi^2}\left(\frac{\calH}{k}\right)^2\,K_{\perp}(kr)\sp P_{\theta\theta}(k) \, ,
% \label{eq:Psi}
\eea 
Note these depend on $r$ only and the dimensions of these correlations
(as well as those below) are given by the prefactor
$1/\calH^n$ (e.g.\
$\xi_{u\sp\delta}$ has units length and $\xi_{u\sp u}^\perp$
units length squared).
Here  $K_\|(x)=j_0(x)/3 - 2j_2(x)/3$ and $K_\perp(x)=j_0(x)/3 + j_2(x)/3$
are the standard velocity kernels~\cite{1988ApJ...332L...7G}, and
we have defined the projection tensor $\mathcal{P}^\perp_{ij}=\delta_{ij}-\hat{r}_i\hat{r}_j$
which we used to extract the part perpendicular to $\hat{r}$.
The power spectra are defined as usual as
$\langle\delta(\k)\sp\delta(\k')\rangle=(2\pi)^3\delD(\k+\k')P_{\delta\delta}(k)$,
$\langle\theta(\k)\sp\delta(\k')\rangle=(2\pi)^3\delD(\k+\k')P_{\theta\delta}(k)$, and
$\langle\theta(\k)\sp\theta(\k')\rangle=(2\pi)^3\delD(\k+\k')P_{\theta\theta}(k)$.
Note that the odd-parity correlations (those containing an odd
power of $\hat{r}$) have Bessel functions of odd order, while even-parity
correlations have Bessel functions of even order.
This means that, unlike the even-parity correlations, the odd-parity 
correlations vanish at $r=0$.

These formulae are independent of the modelling of $\delta$ and $\theta$.
In this work we use linear theory so $\theta\simeq-f\delta$,
with $f$ the linear growth rate. The above formulae can then be written in terms
of the linear matter power spectrum $P_\mrm{lin}$, with
$P_{\theta\delta}\simeq-f P_\mrm{lin}$ and
$P_{\theta\theta}\simeq f^2P_\mrm{lin}$.

\textbf{\textit{Potential correlations.}}
With the inclusion of gravitational redshift, the mean displacement $\bm{m}$
now contains a correlation between $\psi$ and $\delta$
(due to the density weighting). Working on subhorizon scales, we use 
Poisson's equation $\nabla^2\Psi=4\pi Ga^2\bar\rho\delta=\frac32\ssp\Omega_m\calH^2\delta$ to write
$\xi_{\psi\delta}(r)=\langle\psi(\x_1)\sp\delta(\x_2)\rangle$
in terms of the matter power spectrum as
\bea\label{eq:xi-psi-delta}
\xi_{\psi\delta}(r)
\equiv\big\langle\psi(\x_1)\sp\delta(\x_2)\big\rangle
&=%+\frac1\calH\,A_P
\frac{1}{\calH}\frac{3\Omega_m}{2}
\int\frac{k^2\dif k}{2\pi^2}\left(\frac{\calH}{k}\right)^2 j_0(kr)P_{\delta\delta}(k) \,,
\eea
where $\Omega_m=\Omega_m(z)$ is the fractional matter density.
Note that $\psi$ and $\delta$ are positively correlated due to our
convention $\psi=-\calH^{-1}\Psi$,
so $\xi_{\psi\delta}(r)>0$ for all $r$.
This term is order $(\calH/k)^2$, as with $\langle u^2\rangle$.
There is also a one-point contribution, $\xi_{\psi\delta}(0)$,
which does \emph{not} vanish, unlike $\xi_{u\sp\delta}(0)$ which does
for reasons of symmetry. 

The covariance $\mbf{C}_\mrm{cross}$ [Eq.~\eqref{eq:cov-cross}]
contains correlations between $u$ and $\psi$. Because of the density 
weighting there are also correlations with $\delta$, but these enter
at lowest order in the form of an integrated
bispectrum (proportional to $P_\mrm{lin}^2$).
We will neglect such contributions.
Thus for $\mbf{C}_\mrm{cross}$ we need only
$\langle u_i(\x_1)\sp\psi(\x_2)\rangle
=\xi_{u\sp\psi}(r)\sp\hat{r}_i$, where $\xi_{u\sp\psi}(r)$ is given by
\bea\label{eq:xi-u-psi}
\xi_{u\sp\psi}(r)
\equiv\hat{r}^i\big\langle u_i(\x_1)\sp\psi(\x_2)\big\rangle
&=\frac{1}{\calH^2}\frac{3\Omega_m}{2}
\int\frac{k^2\dif k}{2\pi^2}
    \left(\frac{\calH}{k}\right)^3 j_1(kr) P_{\theta\delta}(k) \,,
% \sim r \int^{k_\mrm{IR}}_0{\dif k}\,k^2
%     \cdot\frac{1}{k^2} \cdot k^{n_s}
% \sim \int^{k_\mrm{IR}}_0 \frac{\dif k}{k^{-n_s}}
\eea
where again we have used that $u_i$ is a gradient flow.
Note that compared with
Eq.~\eqref{eq:U}, $\xi_{u\sp\psi}(r)$
appears to be suppressed by one additional factor of $\calH/k$.
But here $j_1(kr)\sim k$ at low $k$ (whereas $j_0(kr)\sim 1$), so we get an extra factor of $k$ and the overall
suppression on large scales is only $(\calH/k)^2$.
This is important because certain correlations are
infrared divergent if $\calH/k$ is raised to a sufficiently
high power.

The apparent issues with divergences begin with the auto-correlation
of the potential, $\xi_{\psi\psi}(r)=\langle\psi(\x_1)\sp\psi(\x_2)\rangle$:
\be
\xi_{\psi\psi}(r)
\equiv\langle\psi(\x_1)\sp\psi(\x_2)\rangle
=\frac{1}{\calH^2}
\left(\frac{3\Omega_m}{2}\right)^2
\int\frac{k^2\dif k}{2\pi^2}\left(\frac{\calH}{k}\right)^4
    j_0(kr)P_{\delta\delta}(k)\,.
\ee
This integral is formally divergent in the infrared. 
For reasons that we will make clear in
Appendix~\ref{app:pot-corrs}, let us thus define the
renormalised correlation function
\be\label{eq:tilde-xi-spec}
\tilde\xi_{\psi\psi}(r)
=\xi_{\psi\psi}(r)-\sigma_\psi^2
=\frac{1}{\calH^2}
\left(\frac{3\Omega_m}{2}\right)^2
\int\frac{k^2\dif k}{2\pi^2}\left(\frac{\calH}{k}\right)^4
    \big[j_0(kr)-1\big]\sp P_{\delta\delta}(k)\,,
\ee
where $\sigma_\psi^2\equiv\xi_{\psi\psi}(0)$ and the integrand
is well behaved in the infrared.
Correlations of the potential appear in the model always in
this renormalised form and so there is no IR divergence.

As might be noticed, we have written the correlation functions above
in a way that makes explicit their suppression in powers of $\calH/k$,
where typically we work on sub-Hubble scales where $\calH/k\ll1$.
In the case of $\xi_{\psi\psi}$ we see that it is highly suppressed
relative to the RSD auto-correlations $\xi_{u\sp u}^\|$ and $\xi_{u\sp u}^\perp$.
In particular, on
scales of interest (e.g.\ tens of $\mrm{Mpc}$) the contribution
from gravitational redshift to the covariance~\eqref{eq:cov-gravz}
is about two  orders of magnitude smaller than that of RSD, 
see~Eq.~\eqref{eq:Psi_para}; on scales $r\sim100\,h^{-1}\mrm{Mpc}$
it is smaller by about an order of magnitude.
Given that the potential is smooth compared to
the density and so sourced by very large-scale modes, $\tilde\xi_{\psi\psi}$
depends weakly on $r$, varying by a total of $4\%$ from 
$r\simeq 10\,h^{-1}\mrm{Mpc}$ to $r\simeq100\,h^{-1}\mrm{Mpc}$. Since
$\psi(\x)$ is long-range, it is saturated by large-scale, linear density modes, so that we could
have $\psi(\x)\ll1$ even if $\delta(\x)\gg0$. Accordingly, on perturbative scales
we expect $\psi(\x)$ to be a smooth, slowly-varying field, more so
than $\delta(\x)$.

\section{Some aspects of the gravitational potential}\label{app:pot-corrs}
In this appendix we discuss the potential auto-correlations which enter
the covariance $\mbf{C}_\mrm{grav}$ [Eq.~\eqref{eq:cov-gravz}] and
the issue of infrared divergence which apparently affects these correlations.
Although the impact of these correlations on the dipole is suppressed
(so largely irrelevant to the results of the main text), they do
reveal some interesting aspects on the nature of the gravitational 
potential, touching on the equivalence principle.

\subsection{On correlations of the potential}
\begin{figure}
\centering
\begin{subfigure}{.5\textwidth}
  \centering
\includegraphics[width=0.95\linewidth]{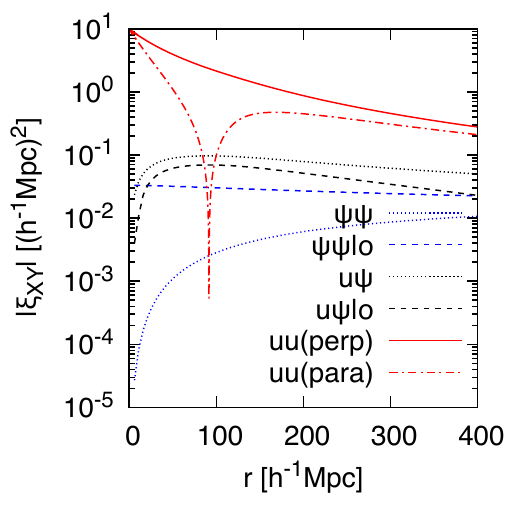}
  % \label{fig:1h}
\end{subfigure}%
\begin{subfigure}{.5\textwidth}
  \centering
  \includegraphics[width=1.02\linewidth]{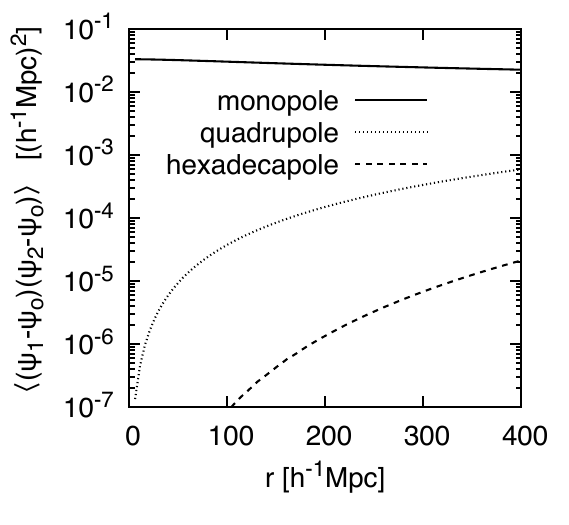}
  % \label{fig:2h}
\end{subfigure}
  \captionsetup{skip=-2pt}
  \caption{\emph{Left panel}: Two-point correlations entering the
        covariance. The RSD contributions are in red,
        gravitational redshift contributions in
        blue, and cross correlations in black.
        Here $uu(\mrm{perp})$ and
        $uu(\mrm{para})$ are given by Eq.~\eqref{eq:Psi_para},
        $\psi\delta$ by Eq.~\eqref{eq:xi-psi-delta}, and
        $\psi\psi$ by Eq.~\eqref{eq:tilde-xi-spec}.
        Note that $\psi\psi$ converges to a positive finite value
        at large separation, although this limit is very slowly
        reached (e.g.\ on the scale of the horizon). Correlations
        involving the local potential $\psi_O$ are indicated by $O$,
        e.g.\ $\psi\psi|O$ corresponds to
        $\langle(\psi_1-\psi_O)\sp(\psi_2-\psi_O)\rangle$.
        These types of correlations are not statistically homogeneous
        (depend not just on $r$), so we show only the angular average
        (monopole).
        Here we have taken $d=834\,h^{-1}\mrm{Mpc}$, corresponding to
        redshift $z=0.3$.
        \emph{Right panel}: Correlation function of the potential
        [Eq.~\eqref{eq:unconstrained2}], broken down into its
        first three non-trivial multipoles, of which the main
        contribution is the monopole (shown also in left panel
        by the dashed blue curve). Here the multipoles are calculated
        for a patch of sky at a distance $\chi_1=\chi_2=834\,h^{-1}\mrm{Mpc}$
        (or $z=0.3$).}
  \label{fig:Psi}
\end{figure}

Without loss of generality consider the off-diagonal component of
$\mbf{C}_\mrm{grav}$ (neglecting the density weighting for simplicity):
\bea
\langle[\psi(\x_1)-\psi_O][\psi(\x_2)-\psi_O]\rangle
&=\langle\psi(\x_1)\sp\psi(\x_2)\rangle
-\langle\psi(\x_1)\sp\psi_O\rangle
-\langle\psi(\x_2)\sp\psi_O\rangle
+ \langle\psi_O^2\rangle \nonumber\\[-1pt]
&=\langle\psi(\x_1)\sp\psi(\x_2)\rangle - \sigma_\psi^2
-\langle\psi(\x_1)\sp\psi_O\rangle + \sigma_\psi^2
-\langle\psi(\x_2)\sp\psi_O\rangle
+ \sigma_\psi^2 \nonumber\\[-1pt]
&=\tilde\xi_{\psi\psi}(r)
-\tilde\xi_{\psi\psi}(\chi_1)-\tilde\xi_{\psi\psi}(\chi_2)
\label{eq:unconstrained2}
\eea
where in the last line we have identified
$\tilde\xi_{\psi\psi}=\xi_{\psi\psi}-\sigma_\psi^2$ 
[Eq.~\eqref{eq:tilde-xi-spec}].
There are several things to note about these correlations.
\begin{enumerate}
\item \textbf{\textit{IR divergence.}}
The subtraction of the one-point function
$\xi_{\psi\psi}(0)=\sigma_\psi^2$ from $\xi_{\psi\psi}(r)$
renormalizes what is otherwise a divergent integral.
Without this subtraction
\be\nonumber
\xi_{\psi\psi}(r)
=\langle\psi(\x_1)\sp\psi(\x_2)\rangle
\propto
\int^\infty_0\frac{k^2\dif k}{2\pi^2}\,\left(\frac{\calH}{k}\right)^4 \!
    j_0(kr)P_{\delta\delta}(k)
% \sim\int^\infty_0{\dif k}\,k^2\cdot\frac{1}{k^4} \cdot 1 \cdot k^{n_s}
\sim\int^{k_{\mrm{IR}}}_0{\dif k}\, k^{n_s-2}
\ee
is infrared divergent for $n_s<1$ (e.g.\ for $\Lambda$CDM
which has $n_s\approx0.96$).

This divergence is related to the fact that $\psi$ and therefore
$\xi_{\psi\psi}(r)=\langle\psi(\x_1)\sp\psi(\x_2)\rangle$ are not in
themselves physical quantities so are not guaranteed to give sensible
answers. Indeed by the equivalence principle we can always
remove gravity in the immediate vicinity by going into a frame of
free fall. This means that at a given point $\psi(\x)$ is not locally
observable for we can always put $\psi=0$ at that point.
What ought to be observable however are differences of
potential, $\Delta\psi=\psi(\x_1)-\psi(\x_2)$. Indeed,
the renormalised correlation function~\eqref{eq:tilde-xi-spec}
can be written in terms of $\Delta\psi$ as
$\tilde\xi_{\psi\psi}(r)=-\tfrac12\langle(\Delta\psi)^2\rangle$, which
tells us that $\tilde\xi_{\psi\psi}(0)=0$ (no gravity at zero lag)
and that the integrand in Eq.~\eqref{eq:tilde-xi-spec} is IR safe.
Castorina and Di Dio~\cite{Castorina:2021xzs} have discussed
these issues in the context of a perturbative model of the power spectrum
(see also Ref.~\cite{Grimm:2020ays}).

Though these manipulations resolve the apparent divergence, at the
end of the day the integration from zero to infinity is mostly a formal 
concern. In practice, the survey's finite size (our `laboratory')
effects an IR cut-off, that is, there exists
a scale $k_f$ below which it does not make
sense to continue integrating modes, as modes with wavelengths
much larger than the survey size leave no discernable imprint on the
gravitational clustering. Another way to put this is that the
long-wavelength modes $k\ll k_f$
affect objects in the survey uniformly, preserving the spacings, angles,
and orientations of galaxy pairs (up to tidal corrections that grow
quadratically with separation).
But this is not to say there is no effect at all. From the point of
view of comoving observers, the long-wavelength
modes do impact the local expansion rate through gravitational time
dilation (clocks run slower in denser regions as perceived by
outside observers). However, this is
in a way that is indistinguishable from the usual
isotropic and homogeneous cosmological expansion, i.e.\ the effects
of these long modes can be absorbed into a new FLRW cosmology,
as in the separate universe description~\cite{Dai:2015jaa}.

Whether or not one is dealing with a finite survey, it seems sensible
to introduce a cut-off anyway, given for instance by the inverse distance
to the surface of last scattering~\cite{Desjacques:2020}. Similar
considerations also apply to correlations of density, velocity, etc
(although their correlations are not IR divergent).

\item {\textbf{\textit{Observer dependence.}}}
The presence of the observer-dependent term $\psi_O$ breaks statistical
homogeneity. As a consequence the two-point function depends 
also on $\chi_1$ and $\chi_2$ (the distance to each tracer).
Due to the second and third terms in Eq.~\eqref{eq:unconstrained2},
$\langle(\psi_1-\psi_O)(\psi_2-\psi_O)\rangle$ acquires a
multipole structure; see Figure~\ref{fig:Psi}.
In the distant-observer limit $\chi_1,\chi_2\to\infty$ ($\vartheta=0$),
Eq.~\eqref{eq:unconstrained2} does not converge to 
$\tilde\xi_{\psi\psi}(r)$, as might be expected, since the
second and third terms are nonzero in this limit.
We will have more to say about this in Section~\ref{app:psi_O} below.

\item \textbf{\textit{{Tidal effects and the equivalence principle.}}}
The physical content of Eq.~\eqref{eq:unconstrained2} and its expression
into multipoles can be better understood by taking the viewpoint of an 
observer locally comoving with the fluid flow (such an observer might
be considered properly inertial if not for cosmic expansion). Since such an
observer experiences no gravitational force we can put to zero the first
derivative of the potential at the observer, $\partial_i\psi|_{\x=0}=0$.
Taylor expanding $\psi(\x)-\psi_O$ about $\x=0$,
we then have at leading order
\be\label{eq:psi-fermi}
\psi(\x)-\psi_O
=\tfrac12 \partial_i\partial_j\sp\psi\sp\big|_{\x=0}x^ix^j+\mathcal{O}(x^3)\,,
\ee
that is, the leading gravitational observable measured by a freely
falling observer is the tidal field, $\partial_i\partial_j\sp\psi$.
(Here the $\mathcal{O}(x^3)$ corrections contain terms which are
higher derivative in the tidal field.)
Observe that the right-hand side of Eq.~\eqref{eq:psi-fermi},
being second derivative in $\psi$, is invariant under arbitrary constant
and pure-gradient shifts,
\be\label{eq:psi-transform}
\psi\to\psi+c+\mbf{b}\cdot\x\,,
\ee
where in general $c$ and $\mbf{b}$ can be time-dependent functions.
In general relativity these shifts correspond to our freedom to set to zero
the metric and its first derivative at any given point by taking up a
reference frame which makes spacetime look locally like Minkowski.%
\footnote{In cosmology one typically seeks not a local inertial frame
but one which looks locally like unperturbed FLRW, up to tidal corrections
which grow quadratically with (comoving) distance.
This is done through constructing `conformal Fermi coordinates'~\cite{Pajer:2013ana,Dai:2015rda,Dai:2015jaa},
a cosmological analogue of the standard Fermi normal coordinates (which realises the strong equivalence principle for a freely
falling observer).
\\
It is worth mentioning here that the construction of this cosmological frame can also
be done in the Newtonian context (without explicit reference to the metric,
curvature, etc). The idea is to change
frame, $\x\to\x+\bm\alpha(\tau)$, i.e.\ a frame accelerated with respect
to a local comoving observer. Here
$\bm\alpha(\tau)$ is chosen so that the reference frame is in free fall,
i.e.\ so that locally the
gravitational force $-\bm\nabla\Psi$ vanishes at (Eulerian) position $\x$.
Noting that the Newtonian potential transforms as $\Psi\to\Psi-(\ddot{\bm\alpha}+\calH\dot{\bm\alpha})\cdot\x$,
we will have that $-\bm\nabla\Psi$ vanishes if
$\ddot{\bm\alpha}+\calH\dot{\bm\alpha}=\bm\nabla\Psi(\x,t)|_{\x=0}$, that is, if
$-\bm\alpha(\tau)$ locally satisfies the equation of motion for a test particle
under gravity \cite{bias_review}.
(Here recall $\Psi=-\calH\psi$.)
We thus see that in Newtonian theory the potential is also frame dependent
and that $-\bm\alpha(\tau)$ is just the Lagrangian displacement, taking us from
the Eulerian frame into the
Lagrangian frame in which gravitational effects are absent.
This transformation, with the particular time dependence of a freely
falling particle, thus realises the equivalence principle and is
behind the `consistency relations of large-scale 
structure'~\cite{Kehagias:2013yd,Peloso:2013zw}.
}
This means that both the potential and the first derivative of the
potential, the gravitational acceleration, are frame-dependent
quantities.
By contrast, the second derivative of the
metric (i.e.\ $\partial_i\partial_j\psi$), with which the Riemann tensor
is built and the field equations formulated, cannot be made to 
vanish locally by a change of frame; indeed it is the second 
derivative that contains objective information about gravity.

Turn now to the correlations.
In terms of the tidal field, Eq.~\eqref{eq:unconstrained2} reads
\be\nonumber
\qquad\langle(\psi_1-\psi_O)(\psi_2-\psi_O)\rangle
\simeq\tfrac{1}{4} x_1^i x_1^j x_2^k x_2^l\,
\big\langle\partial_i\partial_j\psi\ssp\partial_k\partial_l\psi\big\rangle\sp\big|_{r=0}\,.
\ee
Because of isotropy the ensemble average of a rank-4 tensor evaluated at
an arbitrary point has the tensorial structure
$\langle\partial_i\partial_j\psi\ssp\partial_k\partial_l\psi\rangle|_0\propto\delta_{ij}\delta_{kl}+\delta_{ik}\delta_{jl}+\delta_{il}\delta_{jk}$
(where the proportionality constant is fixed by an integral over
the power spectrum),
which, when fully contracted with $x_1^i x_1^j x_2^k x_2^l
=\chi_1^2\ssp\chi_2^2\,\hat{n}_1^i\hat{n}_1^j\hat{n}_2^k\hat{n}_2^l$,
yields at leading order
\be\label{eq:psi-tidal}
\qquad\langle(\psi_1-\psi_O)(\psi_2-\psi_O)\rangle
\propto\chi_1^2\ssp\chi_2^2\sp\big[1+2(\n_1\cdot\n_2)^2\big]\,.
\ee
To obtain the multipoles of this expression we rewrite it in terms of 
coordinates $s,d,\mu$ using the relations in Appendix~\ref{app:formulas}.
The exact forms, while not difficult to calculate, are not particularly
interesting; we simply note here that we have a monopole at order
$\epsilon^0$, a quadrupole at order $\epsilon^2$,
a hexadecapole at order $\epsilon^4$, etc. These are consistent with the orders
of magnitude seen in Figure~\ref{fig:Psi}. In particular, we see that the
slight scale dependence in the monopole is due to the subdominant
wide-angle corrections (order $\epsilon^2$); and that the quadrupole
and hexadecapole does vanish at $r=0$ when $\epsilon=r/d=0$.

Equation~\eqref{eq:psi-tidal} actually finds a more natural basis in spherical 
harmonics. In this basis we have no more than a monopole [the angular average]
and a quadrupole. It is easily understood why.
The monopole is due to the trace, $\nabla^2\psi$, which is associated with isotropic expansion (or contraction), while the quadrupole is
due to the symmetric trace-free part,
$\partial_i\partial_j\sp\psi-1/3\,\delta_{ij}\sp\nabla^2\psi$, which is
associated with (anisotropic) tidal forces.
But the effect of the trace part is indistinguishable from
the cosmological expansion~\cite{Dai:2015jaa}, and because of this
the local scale factor will in general differ from the global one
(cf.~the separate universe). By contrast, the quadrupole tells us something
more unambiguous about gravity.

\end{enumerate}

\subsection{On the local potential}\label{app:psi_O}
\subsubsection{Impact on averages}
The presence of the local potential $\psi_O$ in observables
like $\psi-\psi_O$ requires a slight break from the usual
way of thinking about cosmological averages.
As mentioned earlier, the existence of the observer
spoils statistical homogeneity by singling out one point
for special treatment (the observer's location). This is clear if we
replace the ensemble average
with a volume average, as allowed under ergodicity, for then
averaging the local potential $\psi_O$ is equivalent to averaging
over randomly placed observers, which is not justified;
we make observations from a given point in space---and not any
other point---meaning that when we perform an ensemble average
of $\psi$ (or $\psi-\psi_O$) we should do so by averaging over only
those configurations with $\psi(0)=\psi_O$ for given $\psi_O$~\cite{Desjacques:2020}.

In principle we should replace the unconstrained two-point 
function~\eqref{eq:unconstrained2} with the constrained one.
In practice, however, we can usually get away with using the
unconstrained one. This is because the effect of conditioning
on $\psi_O$ is scale dependent, with impact weakening with distance
from the observer, so that sufficiently far from the observer
the constraint has little
to no effect on correlations calculated in the usual way.
For typical distances encountered in current
surveys, the error we make in using the usual
unconstrained average tends to be small, e.g.\ within the level
of cosmic variance~\cite{Desjacques:2020} (see also
Ref.~\cite{Hall:2019hkt} in the context of weak lensing).

Incidentally, in the case $n_s<1$ and $\psi$ is Gaussian random, the constrained and unconstrained average coincide:
\be\label{eq:equiv}
\big\langle\big(\psi_1-\psi_O\big)\big(\psi_2-\psi_O\big)\Mid O\big\rangle
=\langle(\psi_1-\psi_O)(\psi_2-\psi_O)\big\rangle\,.
\ee
This can be shown by recalling a standard formula of conditional
Gaussian statistics (see, e.g.\ Ref.~\cite{BBKS}, appendix D), namely,
for $A,B,C$ Gaussian variables we have
$\langle AB\Mid C\rangle=\langle AB\rangle
-\langle A\Mid C\rangle\langle B\Mid C\rangle/\langle C^2\rangle$.
Applied to the potential,
\be\label{eq:AB|C}
\begin{split}
\big\langle\big(\psi_1-\psi_O\big)\big(\psi_2-\psi_O\big)\Mid O\big\rangle
=\langle(\psi_1-\psi_O)(\psi_2-\psi_O)\big\rangle 
- \big\langle(\psi_1-\psi_O)\sp\psi_O\big\rangle\sp\big\langle(\psi_2-\psi_O)\sp\psi_O\big\rangle
    \ssp\big/\ssp\big\langle\psi^2_O\big\rangle\,.
\end{split}
\ee
To show that the second term on the right-hand side vanishes, as Eq.~\eqref{eq:equiv}
requires, we introduce an IR cut-off $k_\mrm{IR}$ (which we send to zero
at the end of the calculation)
and replace $\langle\psi_O^2\rangle\equiv\sigma^2_\psi$ with
$\sigma^2_\psi(k_\mrm{IR})$ and
$\tilde\xi_{\psi\psi}(r)$ with $\tilde\xi_{\psi\psi}(r;k_\mrm{IR})$,
i.e.\ we filter out power from scales $k<k_\mrm{IR}$. Then since
$\langle(\psi_1-\psi_O)\sp\psi_O\rangle=\tilde\xi_{\psi\psi}(\chi_1;k_\mrm{IR})$
and $\langle(\psi_2-\psi_O)\sp\psi_O\rangle=\tilde\xi_{\psi\psi}(\chi_2;k_\mrm{IR})$
we find for the second term
\bea
&\big\langle(\psi_1-\psi_O)\sp\psi_O\big\rangle\sp
    \big\langle(\psi_2-\psi_O)\sp\psi_O\big\rangle
        \ssp\big/\ssp\big\langle\psi^2_O\big\rangle
    =\lim_{k_\mrm{IR}\to0}\,
    \tilde\xi_{\psi\psi}(\chi_1;k_\mrm{IR})\ssp
    \tilde\xi_{\psi\psi}(\chi_2;k_\mrm{IR})\,\big/\,\sigma_\psi^2(k_\mrm{IR}) =0\,,\nonumber
\eea
since in the limit $k_\mrm{IR}\to0$ the variance
$\sigma_\psi^2(k_\mrm{IR})\to\infty$,
while $\tilde\xi_{\psi\psi}(r;k_\mrm{IR})$ remains finite. These
formal computations show that when integrating down
to arbitrarily long modes ($k\to0$)
there is no impact from conditioning on the local potential
and the average returns to its unconstrained form.
Of course when dealing with a real survey modes
larger than the survey cannot be measured and so the constrained
and unconstrained averages will generally be different. But as
mentioned, the effect of the constraint tends to be weak for all but
the shallowest of surveys~\cite{Hall:2019hkt}.

These results can be compared with those of Desjacques et al.~\cite{Desjacques:2020}
who have studied the issue of constrained averaging in Fourier space.
Our results however have some differences
with theirs, mainly because the power spectrum is no
longer the Fourier transform of the correlation function so
their results do not straightforwardly carry over to position space.%
\footnote{This is because conditioning on a point necessarily breaks
statistical homogeneity, which is a prerequisite for the Wiener--Khinchin
theorem (that the power spectrum is the Fourier transform of the
correlation function).
This highlights one advantage of using the correlation function over the
power spectrum in that the correlation function is 
unambiguously defined in the absence of any symmetries.}
In particular, estimating the power spectrum by the
usual methods---i.e.\ spatial averaging of the constrained data
assuming ergodicity---only converges to the unconstrained
ensemble average in the limit of an infinite survey volume
(corresponding to our taking $k_\mrm{IR}\to0$ above). Therefore in
this case the constraint's effect completely dies away.
For finite volumes, while the impact of the constraint does
not in general vanish, it is however small. For typical scales
affected, such as those near the fundamental mode of the survey,
the effect is within cosmic variance~\cite{Desjacques:2020}
(see also Ref.~\cite{Hall:2019hkt}).

\subsubsection{Relation to the monopole}\label{sec:psiO}
Since $\psi_O$ is not a local observable it is often neglected
on the basis that, without any angular dependence, it
is absorbed into the mean density, defined as the sky average.
Indeed, $\psi_O$ is itself a monopole so
will indeed drop out in the observed overdensity
$\delta_g-\frac{1}{4\pi}\int\dif^2{\n}\,\delta_g$
(see Appendix~\ref{app:realistic} below). For instance, suppose
$\psi(\chi\n)-\psi_O\subset\delta_g$. Clearly we have
\be\label{eq:psi-no-mono}
\psi-\psi_O-\int\frac{\dif^2\n}{4\pi}\,(\psi-\psi_O)
=\psi-\int\frac{\dif^2\n}{4\pi}\,\psi\,.
\ee
In other words, there is no dependence on the local potential in
the observed overdensity; the dependence on the local potential
gets shuffled into the mean density.
However, this is only the case when $\psi-\psi_O$ (or derivatives 
thereof) enters $\delta_g$ linearly, as in linear perturbation
theory. In general,
averaging and nonlinearity are not commutative operations,
$\langle f(X)\rangle\neq f(\langle X\rangle)$~\cite{Bonvin:2015kea}.
Therefore, in any model of $\delta_g-\frac{1}{4\pi}\int\dif^2{\n}\,\delta_g$
in which $\psi_O$ enters $\delta_g$ nonlinearly, it is important to retain
such local terms, since at nonlinear order we can have products between
terms evaluated at the source and terms evaluated at the 
observer~\cite{Mitsou:2019ocs}.
This means that $\psi_O$ does not in general vanish upon monopole
subtraction and so should not be neglected in, for example,
Eq.~\eqref{eq:mu-map}.
The correct approach is therefore to keep the local terms in the model, compute the monopole associated with the overdensity, and then subtract it from the model, as we discuss below in Appendix~\ref{app:realistic}.

These considerations of course concern full-sky coverage;
for partial-sky coverage, even at linear order one cannot take
for granted the absence of the local potential in the
observed overdensity, as can be seen in Eq.~\eqref{eq:psi-no-mono}
if one replaces $\psi-\psi_O$ with $W(\n)(\psi-\psi_O)$.

\section{Full-sky correlations in the case where the mean density is not known}\label{app:realistic}
Throughout this work we have routinely assumed
that the mean background number density $\bar{n}_g$ is known, thus allowing us to isolate the `theoretical' $\delta_g$ in the usual
way, i.e.\ $\delta_g=(n_g-\bar{n}_g)/\bar{n}_g$. But the mean number
density is not known a priori and this leads to the
`integral constraint'~\cite{1991MNRAS.253..307P,deMattia:2019vdg}, that
since we measure the mean from the survey itself, whether it corresponds
to the true mean or not, the overdensity is required by construction
to average to zero over the survey.
There is also the related issue that perturbations are not guaranteed
to be observable quantities (e.g.\ the local potential) and what one
deems the `background' depends on choice
of gauge (although note that these gauge effects are order $(\calH/k)^2$
and so are suppressed with respect to the level of approximation
used in this work).

These considerations are usually not a concern within the RSD paradigm.
But with the inclusion of the (local) potential in the model
one should take care to operationally define the galaxy
overdensity from observable quantities. Furthermore, one also needs to
be careful about what one means by `average' in the absence of translation
invariance in the wide-angle regime.
Thus, for all-sky coverage over a wide redshift range we define the
mean density as the sky average of the observed density,
$\bar{n}_g\equiv\int_{\n} n_g$, where as a shorthand
$\int_{\n}=\frac{1}{4\pi}\int\dif^2\n$ for the angular average.
Then we extract (in redshift space) the overdensity
$(n_g-\bar{n}_g)/\bar{n}_g=(\delta_g-\int_{\n}\delta_g)/(1+\int_{\n}\delta_g)
\simeq\delta_g-\int_{\n}\delta_g$. That is, the observed
fluctuation is the `theoretical' one, $\delta_g$,
with its monopole part removed~\cite{Desjacques:2020,Yoo:2024tbo}.

A short calculation shows that in going from
$\delta_g\to\delta_g-\int_{\n}\delta_g$
(linear or nonlinear $\delta_g$) the two-point function changes as
\be\label{eq:xi-no-mono}
\xi^{(\mrm{s})}(\chi_1,\chi_2,\cos\vartheta)
\to
\xi^{(\mrm{s})}(\chi_1,\chi_2,\cos\vartheta)
-C_0(\chi_1,\chi_2)/(4\pi)\,,
\ee
where $\xi^{(\mrm{s})}(\chi_1,\chi_2,\cos\vartheta)$ is
any rotationally-invariant correlation
function (e.g.~Eq.~\eqref{eq:gsm}), and we have recognised
that because $\xi^{(\mrm{s})}(\chi_1,\chi_2,\cos\vartheta)$ is
rotationally invariant the average over the lines of sight
$\n_1$ and $\n_2$ is
equivalent to the average over the angular separation:
\bea
\int\frac{\dif^2\n_1}{4\pi}
    \int\frac{\dif^2\n_2}{4\pi}\,
    \xi^{(\mrm{s})}(\chi_1,\chi_2,\cos\vartheta=\n_1\cdot\n_2) 
=\frac12\int^1_{-1}{\dif\cos\vartheta}\,
    \xi^{(\mrm{s})}(\chi_1,\chi_2,\cos\vartheta)
=:C_0(\chi_1,\chi_2)/(4\pi)\,,
\eea
which we have noted is just the definition of the monopole power $C_0$
(up to a factor of $4\pi$).
Thus, the removal of the monopole in the overdensity
translates to the removal of its monopole power $C_0$
in the two-point correlations. In other words, the harmonic
series of the right-hand side of Eq.~\eqref{eq:xi-no-mono} in
angular power $C_\ell$ begins non-trivially at $\ell=1$.

The question arises: does the subtraction of the monopole represent an important correction
to the model~\eqref{eq:gsm2}?
For most surveys, no---the error made in neglecting the monopole
should be small, with relative size $|C_0/(4\pi)\sp /\sp \xi^{(\mrm{s})}|\ll 1$.
For instance, at $z=0.3$ and $s=100\,h^{-1}\mrm{Mpc}$ the correction
is at the few-percent level.
Qualitatively speaking, by changing to coordinates $s,d$, and $\mu$ we find
that the relative correction grows with separation $s$, since the size
of $\xi^{(\mrm{s})}$ goes roughly as the
inverse of $s$, while $C_0$ tends to be insensitive to $s$.
In effect, for sufficiently large redshift (e.g.\ when we
are justified in using the distant-observer approximation), $C_0$
acts as an overall shift and so does not change the
shape of the correlation function (the position of the
BAO peak is unaffected, for example~\cite{Yoo:2024tbo}).
In particular, only the $\ell=0$ multipole is shifted.
On the other hand, at very low redshifts---for instance when
the distant-observer limit breaks down and the curvature of
the shells becomes relevant
(over large $s$)---the monopole $C_0$ can induce in the multipoles
a non-negligible scale-dependent shift.
On the basis of symmetry we expect the leading-order wide-angle
corrections are $\mathcal{O}(\epsilon^2)$ for auto-correlations
(in the mid-point parametrisation) and $\mathcal{O}(\epsilon)$
for cross-correlations.

\begin{figure}[t!]
\centering
\begin{subfigure}{.5\textwidth}
  \centering
  \includegraphics[width=1.02\linewidth]{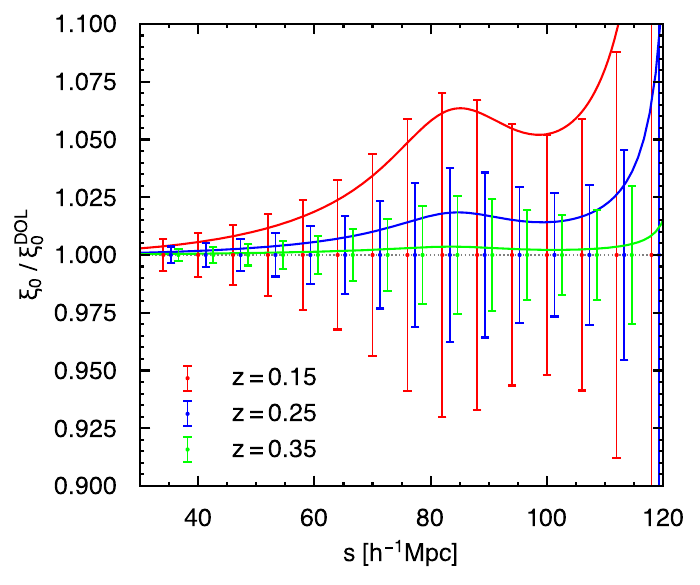}
\end{subfigure}%
\begin{subfigure}{.5\textwidth}
  \centering
  \includegraphics[width=1\linewidth]{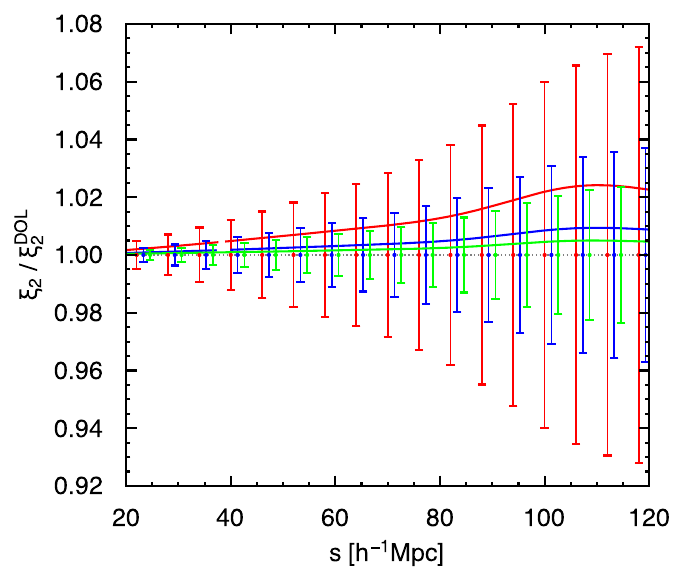}
\end{subfigure}
\caption{Forecasted uncertainties for BGS compared with
        the size of the wide-angle corrections defined
        as $\xi_\ell/\xi_\ell^\mrm{DOL}$
        (solid curves), with $\xi_\ell^\mrm{DOL}$ given by the standard
        GSM [Eq.~\eqref{eq:gsm-dol}] and $\xi_\ell$ the wide-angle generalisation
        [Eq.~\eqref{eq:gsm2}].
        Left panel shows monopole, right panel shows quadrupole.
        We use the fiducial BGS bias $b(z)=1.34/D(z)$, where $D(z)$ is the 
        linear growth factor normalized so that $D(z=0)=1$~\cite{DESI:2016}.
        Here error bars (offset in the plot for clarity) are computed
        assuming DOL binned Gaussian covariances with bins of width
        $6\,h^{-1}\mrm{Mpc}$.
        }
\label{fig:BGS}
\end{figure}

\section{Impact of wide-angle effects on DESI BGS}\label{app:WA-impact}
Given the various novelties in the model~\eqref{eq:xi_final},
it is interesting to see in isolation just how
important wide-angle effects are as a systematic compared to measurement uncertainties from upcoming surveys. Here we focus on the DESI Bright Galaxy
Sample (BGS)~\cite{BGS}, a low-redshift sample covering about
a third of the sky ($f_\mrm{sky}=0.339$) and where we might expect
wide-angle corrections to be important. We show this comparison
in Figure~\ref{fig:BGS}, ignoring all effects
except the wide-angle `corrections' (which we recall are exact to
all orders in $\epsilon=s/d$).
Here it is seen that the monopole tends to be more affected
than the quadrupole. This is not surprising given that the quadrupole is sourced by configurations with smaller opening
angles than those of the monopole (see Section~\ref{sec:wa-even}).
What is also clear is that the monopole at $z=0.15$ is
significantly affected by wide-angle effects. 
However, we should add the caveat that the Gaussian covariances
used to estimate the uncertainties are computed using the
{\sc coffe} code~\cite{Tansella:2018sld},
which is limited to the distant-observer limit.
Nevertheless, Ref.~\cite{Hall:2016bmm} found that
in the case of the dipole the wide-angle corrections to
the variance are negligible (due to cancellations).

\section{Calculation of $\llangle\Delta\psi\rrangle$ using the
halo model}\label{app:shot}
In this appendix we derive Eq.~\eqref{eq:Delta-psi-split}, i.e.\
the one- and two-halo contributions to the pairwise potential difference
\be\label{eq:Delta-psi-weighted}
\llangle\Delta\psi\rrangle
\equiv
\frac{\langle\sp n_A(\x_1)\sp n_B(\x_2)\sp\Delta\psi\rangle}{\langle n_A(\x_1)\sp n_B(\x_2)\rangle}
\ee
where $\Delta\psi=\psi(\x_1)-\psi(\x_2)$, $n_A(\x_1)=\bar{n}_A[1+\delta_A(\x_1)]$ and
$n_B(\x_2)=\bar{n}_B[1+\delta_B(\x_2)]$
(assuming the usual density weighting).
Note that the calculation of
$\llangle\psi(\x_1)-\psi_O\rrangle$ and 
$\llangle\psi(\x_2)-\psi_O\rrangle$, as needed in
$\bm{m}_\mrm{grav}$, is essentially the same as the calculation
of $\llangle\Delta\psi\rrangle$ we give below (with one minor
difference which we point out at the end of this appendix).

Here we recall that the important aspect of this pairwise correlation is that it contains a \emph{nonvanishing}
one-point function, $\langle\psi(\x_1)\sp\delta_A(\x_1)\rangle$. 
As mentioned in the main text, this is because the gravitational redshift
tends to trace overdense regions (i.e.\ potential wells), where
$\langle\psi(\x_1)\sp\delta_A(\x_1)\rangle$ has a definite sign.
By contrast, the analogous velocity statistic,
$\llangle\Delta u\rrangle$, contains a vanishing one-point function
since the mean velocity at a given point is zero (there are no preferred directions).

We carry out the calculation using the formalism of distributions~\cite{1991ApJ...381..349S},
the basic quantity being the discrete distribution function
\be
f_d(\x,M,\tau)=\sum_a\delD(\x-\x_a(\tau))\,\delD(M-M_a)\,,
\ee
where $M_a$ is the $a$th halo's mass and $\x_a(\tau)$ is its
(centre-of-mass) position at time $\tau$.
The idea of the following calculation is to write Eq.~\eqref{eq:Delta-psi-weighted}
in terms of $f_d$ then use the known discrete statistics of $f_d$.
We will not need to know the dynamics of $f_d$; only its statistics
at a given time. We will henceforth suppress the time dependence.

Consider a population of dark matter haloes labelled $A$ in the
mass range $M_{A1}\leq M\leq M_{A2}$. The number density is
\bea\label{eq:n-df}
n_A(\x)
&=\int^{M_{A2}}_{M_{A1}}\dif M\,f_d(\x,M)\,.
% =\int\dif M\,W_A(M)\sp f_d(\x,M)\,,
% \rho_m(\x,\tau)&=\int^{M_2}_{M_1}\dif M\,M\sp f(\x,M,\tau)\,.
\eea
Making the standard assumption that all matter resides within
haloes, each with
characteristic profile $\rho(\x;M)$ (e.g.\ the NFW profile),
the matter density field $\rho_\mrm{tot}(\x)$ is obtained by convolving the point-particle
distribution function with $\rho$, thereby turning the haloes into
extended objects, i.e.\ we have
\bea\label{eq:rho-discrete}
\rho_\mrm{tot}(\x)
&=\int\dif M\int\dif^3\y\sp f_d(\y,M)\ssp
    \rho(\x-\y;M)
=\sum_a\rho(\x-\x_a;M_a)\,,
\eea
Here the integration is over all mass and we require
$\int\dif^3\x\,\rho(\x;M)=M$ to ensure consistency with the predictions
of perturbation theory on large scales~\cite{Schmidt:2015gwz}.
With mass function
$\dif n/\dif M=\langle f_d(\x,M)\rangle$, the
mean density is
$\bar\rho=\langle\rho_\mrm{tot}\rangle=\int\dif M\sp M\sp ({\dif n}/{\dif M})$.

By Poisson's equation the potential sourced by fluctuations to the mean density
$\langle\rho_\mrm{tot}\rangle$ (rather than the density $\rho_\mrm{tot}$
itself) can be written in position space as
\be\nonumber
\psi(\x)=\lambda\sp G\int\dif^3\y\,
    \frac{\rho_\mrm{tot}(\y)-\bar\rho}{|\x-\y|}\,.
\ee
Here $G$ is Newton's constant of gravitation and we have defined
$\lambda\equiv\calH^{-1}a^2$.
By linearity of Poisson's equation, if $\rho_\mrm{tot}(\x)$ is a superposition
of individual densities $\rho$, as it is in Eq.~\eqref{eq:rho-discrete},
then $\psi(\x)$ is a superposition of potentials given by
\bea\label{eq:psi0}
\phi(\x;M)
&=\lambda G\int\dif^3\y\,
    \frac{\rho(\y;M)}{|\x-\y|}\,,
\eea
which individually solve Poisson's equation with boundary
condition $\phi\to0$ as $\x\to\infty$.
(Here we are assuming that haloes can be individually treated as
isolated systems, i.e.\ with negligible tidal forces between them.)
We have also dropped the integral over the mean field $\bar\rho$
since it gives an inconsequential normalisation constant which vanishes
upon taking the difference of potentials (which is the quantity we are
really after). The potential due to the entire distribution is then
[cf.~Eq.~\eqref{eq:rho-discrete}]
\be\label{eq:psi-discrete}
\psi(\x)
=\int\dif M\int\dif^3\y\sp f_d(\y,M)\ssp
    \phi(\x-\y;M)
=\sum_a\phi(\x-\x_a;M_a)\,.
\ee
As with $\rho(\x)$ above, the potential
resulting from the entire distribution is given by the convolution of 
$f_d$ with the individual profiles $\phi$.
But unlike $\rho$ the potential $\phi$, being a nonlocal function
of $\rho$, reaches beyond the halo's material extent (given, for instance,
by the virial radius at which the profile is truncated). 
Therefore at any given point $\psi(\x)$ can in principle receive contributions
from many haloes, while that same point is located in one and only one 
halo (assuming halo exclusion).
Note that in the case of a spherically symmetric profile,
$\rho(\x;M)=\rho(|\x|;M)$,
Eq.~\eqref{eq:psi0} integrates to Eq.~\eqref{eq:phiA} in the main text.

To evaluate Eq.~\eqref{eq:Delta-psi-weighted} we will need the first three
$N$-point functions~\cite{1992LNP...408...65B}:
\bea
\langle f_{d}(\x_1,M_1)\rangle
&\equiv\bar{f}_1
={\dif{n}}(M_1)/{\dif M}\,, \\
\langle f_{d}(\x_1,M_1)\ssp f_{d}(\x_2,M_2)\rangle 
&=\bar{f}_1\sp\delD(\x_1-\x_2)\ssp\delD(M_1-M_2) + \bar{f}_1\bar{f}_2(1+\xi_{12})\,, \\[1pt]
\langle f_{d}(\x_1,M_1)\ssp f_{d}(\x_2,M_2)\ssp f_{d}(\x_3,M_3)\rangle 
&=\bar{f}_1\ssp\delD(\x_1-\x_2)\ssp\delD(M_1-M_2)\ssp\delD(\x_1-\x_3)\ssp\delD(M_1-M_3) \nonumber\\
&\quad+\bar{f}_1\ssp\bar{f}_3\sp(1+\xi_{13})\ssp\delD(\x_1-\x_2)\ssp\delD(M_1-M_2)+\text{2 cyc.} \nonumber\\
&\quad+\bar{f}_1\sp\bar{f}_2\sp\bar{f}_3\sp(1+\xi_{12}+\text{2 cyc.})\,,
\label{eq:f-3pt}
\eea
where $\xi_{ij}\equiv\xi_{hh}(\x_i-\x_j,M_i,M_j)$, $i,j=1,2,3$, is the
cross-correlation function
between haloes of mass $M_i$ and $M_j$. In this work we take 
$\xi(\r,M_A,M)=b_1(M_A)\sp b_1(M)\sp \xi_{mm}(r)$,
with linear bias $b_1(M_A)$ determined by the peak--background split and
$\xi_m(r)$ the matter correlation function.
Here we have assumed Poisson noise and that the clustering is Gaussian
so that we can ignore the connected three-point function that would
otherwise be present in the last line of Eq.~\eqref{eq:f-3pt}. Note
that the structure of $\langle f_{d}(1)\ssp f_{d}(2)\ssp f_{d}(3)\rangle$,
for instance, is that we can have (\emph{i}) all three points in a single halo
(one-halo term; first line in Eq.~\eqref{eq:f-3pt}), (\emph{ii}) two points in
one halo and one point in a separate halo (two-halo term; second-to-last line), 
or (\emph{iii}) each of the three points
in distinct haloes (three-halo term; last line).

\begin{figure}
\centering
\begin{subfigure}{.5\textwidth}
  \centering
  \includegraphics[width=1\linewidth]{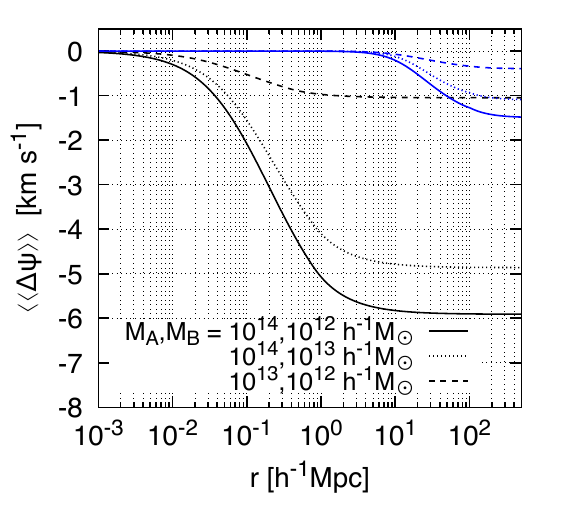}
%  \caption{}
  % \label{fig:1h}
\end{subfigure}%
\begin{subfigure}{.5\textwidth}
  \centering
  \includegraphics[width=1\linewidth]{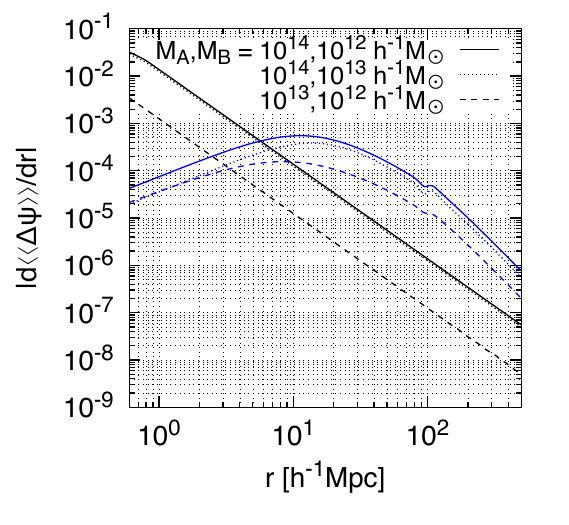} % Delta_Psi_2h_alt
%  \caption{}
  % \label{fig:2h}
\end{subfigure}
\captionsetup{skip=-2pt}
\caption{{\it Left panel:}  Contributions to
$\llangle\Delta\psi\rrangle$ from the one-halo (black curves) and two-halo (blue curves) terms
for three different mass combinations.
(Here $\llangle\Delta\psi\rrangle$ is multiplied
by $-\calH$; the units are thus $\mrm{km}\,\mrm{s}^{-1}$.)
Note that the linear prediction of $\llangle\Delta\psi\rrangle$ is virtually
identical to the two-halo curves.
Note also that this plot is essentially Figure~\ref{fig:psi_12} but
here on a linear scale to show more clearly the relative size.
{\it Right panel:} The corresponding derivatives of each curve in the left panel
[the form in which $\llangle\Delta\psi\rrangle$ enters the multipoles at first order].
Here $z=0.3$ and we assume a truncated NFW profile with concentration $c=9$.
}
\label{fig:1hand2h}
\end{figure}

For simplicity we consider two narrow, non-overlapping mass bins
of width $\delta M_A$ and $\delta M_B$, centred at $M=M_A$ and $M=M_B$,
respectively.
By Eq.~\eqref{eq:n-df} we then have
$n_A(\x_1)\approx \delta M_A\ssp f_d(\x_1,M_A)$ and
$n_B(\x_2)\approx \delta M_B\ssp f_d(\x_2,M_B)$.
The denominator of Eq.~\eqref{eq:Delta-psi-weighted} is given by
the standard result:
\bea\nonumber
\langle n_A(\x_1)\sp n_B(\x_2)\rangle
&=\bar{n}_A\sp\bar{n}_B\left(1+\xi_{AB}(\x_1-\x_2)
    +\frac{\delta_{AB}}{\bar{n}_A}\sp\delD(\x_1-\x_2)\right) \,,
\eea
where $\bar{n}_A\approx\delta M_A\langle f_d(\x,M_A)\rangle=\delta M_A\sp\dif n(M_A)/\dif M$ and
$\bar{n}_B\approx\delta M_B\langle f_d(\x,M_B)\rangle=\delta M_B\sp\dif n(M_B)/\dif M$ are the mean
number densities.

The novel part of the calculation is the evaluation of the numerator of
Eq.~\eqref{eq:Delta-psi-weighted}. In terms of the distribution
function it reads
\bea
&\langle n_A(\x_1)\sp n_B(\x_2)\sp\Delta\psi\rangle 
=\delta M_A\sp\delta M_B\!
\int\dif M\!\int\dif^3\x\,\Big\{
    \big\langle f_d(\x_1,M_A)\sp f_d(\x_2,M_B)\sp f_d(\x,M)\big\rangle
    \big[\phi(\x_1-\x;M)-\phi(\x_2-\x;M)\big]\Big\}\,. \nonumber
\eea
Substituting the three-point function~\eqref{eq:f-3pt} yields several
terms in the integrand. The (unintegrated) one-halo term
[first line in Eq.~\eqref{eq:f-3pt}] is
proportional to $\delD(\x_1-\x_2)$ so vanishes at nonzero separation, while the integrated
three-halo terms yields the clustering contribution.
The remaining terms in Eq.~\eqref{eq:f-3pt},
the integrated two-halo terms, yield the one-halo term which,
after some algebra, can be written as [with $M_1\equiv M_A$, $M_2\equiv M_B$]
\be
\begin{split}
&\langle n_A(\x_1)\sp n_B(\x_2)\sp\Delta\psi\rangle
=\bar{n}_A\ssp\bar{n}_B\sp[1+\xi_{AB}(\x_1-\x_2)] \\
&\qquad\quad\times\int\dif M\int\dif^3\x\,\Big\{\big[
    \delD(\x_1-\x)\ssp\delD(M_1-M)+(1\leftrightarrow2)\big] 
    \big[\phi(\x_1-\x;M)-\phi(\x_2-\x;M)\big]\Big\}
    + \text{2-halo terms}\,.\label{eq:numerator}
\end{split}
\ee
In arriving at this expression we have used that one of the three
two-halo permutations is symmetric in $(\x_1,M_1)$ and $(\x_2,M_2)$ so
therefore vanishes upon integration against the potential difference, which is anti-symmetric in $\x_1$ and $\x_2$.
(In any case it vanishes at nonzero
separation because of the delta function $\delD(\x_1-\x_2)$.)
Integrating the foregoing expression is trivial and gives
\bea\nonumber
\begin{split}
&\langle n_A(\x_1)\sp n_B(\x_2)\sp\Delta\psi\rangle 
=\bar{n}_A\ssp\bar{n}_B\sp\big[1+\xi_{AB}(\x_1-\x_2)\big]
\big[\phi_A(0)-\phi_A(\x_2-\x_1)
+\phi_B(\x_1-\x_2)-\phi_B(0)\big] + \text{2-halo terms}\,,
\end{split}
\eea
where
$\phi_A(\x)\equiv\phi(\x;M_A)$ and 
$\phi_B(\x)\equiv\phi(\x;M_B)$.
Dividing by the normalisation $\langle n_A(\x_1)\sp n_B(\x_2)\rangle=\bar{n}_A\ssp\bar{n}_B\sp[1+\xi_{AB}(\x_1-\x_2)]$,
$\x_1\neq\x_2$, and writing
$\llangle\Delta\psi\rrangle=\llangle\Delta\psi\rrangle_\text{1h}
+\llangle\Delta\psi\rrangle_\text{2h}$, we arrive at the main result:
\be\label{eq:Delta-psi-1halo}
\llangle\Delta\psi\rrangle_\text{1h}
% \equiv\psi_{12}^\mrm{stoch}(r)
=[\phi_A(0)-\phi_A(\x_2-\x_1)]-[\phi_B(0)-\phi_B(\x_1-\x_2)]\,.
\ee
This equation reduces to Eq.~\eqref{eq:psi_12_shot}
in the case of spherical symmetry, 
$\phi_A(r)\equiv\phi_A(\x_2-\x_1)$ and $\phi_B(r)\equiv\phi_B(\x_1-\x_2)$
with $r=|\x_1-\x_2|$.
Notice that Eq.~\eqref{eq:Delta-psi-1halo} is not simply the potential
difference $\phi_B(r)-\phi_A(r)$, but the difference of potentials, normalised to
their respective central values. This ensures that
$\llangle\Delta\psi\rrangle_\mrm{1h}=0$ at $r=0$ (which we should have on account of anti-symmetry).
Note that the one-halo term does not depend on the correlation function
$\xi_{AB}$ (the normalisation completely drops out).

Obtaining the clustering contribution (two-halo term) is 
a more straightforward affair.
From the last line of terms in Eq.~\eqref{eq:f-3pt},
\bea
&\llangle\Delta\psi\rrangle_\text{2h}
=\frac{1}{1+\xi_{AB}(r)}\,\Big\{[\xi_{\psi\delta_A}(0)-\xi_{\psi\delta_A}(r)]
-[\xi_{\psi\delta_B}(0)-\xi_{\psi\delta_B}(r)]\Big\}\,, 
\eea
where $\xi_{\psi\delta_A}(r)$ is given by Eq.~\eqref{eq:Delta-psi-2h}.
A similar expression holds for $\xi_{\psi\delta_B}(r)$
[take $M_A\to M$ and $M\to M_B$].
Note that the integration over $\x$ is a convolution of each halo with its
associated potential $\phi$,%
\footnote{Note that the linear prediction for $\xi_{\psi\delta_A}(r)$
[Eq.~\eqref{eq:xi-psi-delta}] is essentially recovered at large $r$, when
the internal structure of the halo becomes irrelevant to the large-scale
clustering, and haloes can effectively be treated as point masses with
$\rho(\x;M)=M\delD(\x)$ and $\phi\propto1/|\x|$.
Allowing that haloes are extended objects, corrections to the point-mass 
description enter through terms going as the ratio of the typical size of
the halo $R$ to the separation $r$ (or, more precisely, terms of
order $(kR)^2$ when expressed in Fourier space).
}
while the integration
over $M$ is over all mass since the potential is sourced by all matter (not just
the matter belonging to the population of tracer $A$).

Now, as the one-halo term shares the same basic form as the
two-halo term we can therefore write
\bea
\llangle\Delta\psi\rrangle
&=\llangle\Delta\psi\rrangle_\text{1h}
+\llangle\Delta\psi\rrangle_\text{2h} \nonumber\\
&\simeq[\xi_{\psi\delta_A}(0)+\phi_A(0)-\xi_{\psi\delta_A}(r)-\phi_A(r)]
-[\xi_{\psi\delta_B}(0)+\phi_B(0)-\xi_{\psi\delta_B}(r)-\phi_B(r)]\,,
\label{eq:Delta-psi-total}
\eea
where we have dropped the normalisation in the second line.
The last line says that the clustering contributions,
$\xi_{\psi\delta_A}$ and $\xi_{\psi\delta_B}$,
are augmented by a halo correction, $\phi_A$ and $\phi_B$, for both
zero-lag terms and scale-dependent terms.
At the field level we heuristically think of these corrections as coming from
the individual $\phi$'s superposed on the large-scale linear potential,
having the effect of deepening the potential at the sites of tracers (where light is emitted).% 
\footnote{The model of Saga et al.~\cite{Saga:2020tqb} imposes
these corrections at the field level. However, their model misses the scale-dependent
contributions, $\phi_A(r)$ and $\phi_B(r)$, which arises from the nonlocal
nature of the potential.
By contrast, the $z_g(r)$ statistic of Ref.~\cite{Croft:2013taa}
is $\llangle\Delta\psi\rrangle$ minus the zero-lag terms.}

The left panel of Figure~\ref{fig:1hand2h} shows the one- and two-halo contributions
to $\llangle\Delta\psi\rrangle$, while the right panel shows the derivatives thereof.
The first-order contribution to the multipoles comes from the derivative while
at second order we can have products containing $\llangle\Delta\psi\rrangle$ (not differentiated).
The point here is that at first order the two-halo term is more important than the one-halo
term (for $r\gtrsim10\,h^{-1}\mrm{Mpc}$).
While at second order it is the one-halo term which is more important
(on all scales in fact) due to its larger one-point function. 
For two of the mass
combinations (solid and dotted curves) the difference between $\llangle\Delta\psi\rrangle_\mrm{1h}$
and $\llangle\Delta\psi\rrangle_\mrm{2h}$ is significant.
For instance, for $M_{A}=10^{14}h^{-1}M_\odot$ and $M_{B}=10^{12}h^{-1}M_\odot$
at $r=20\,h^{-1}\mrm{Mpc}$ we have that $\llangle\Delta\psi\rrangle_\mrm{1h}$ is about
an order of magnitude larger than the term $\llangle\Delta\psi\rrangle_\mrm{2h}$ (which
we find is very similar to the estimate of linear theory, e.g.\ by writing
$\xi_{\psi\delta_A}(0)=b_1(M_A)\sp\xi_{\psi\delta}(0)$ and evaluating
$\xi_{\psi\delta}(0)$ using Eq.~\eqref{eq:xi-psi-delta}). 
The other mass combination yields a smaller one-halo contribution, albeit one which is still
a factor of two larger that the corresponding two-halo
contribution (the total contribution is a factor of three larger than the two-halo term).

Finally, let us give here the formula for
$\llangle\psi(\x_1)-\psi_O\rrangle_\mrm{1h}$ and $\llangle\psi(\x_2)-\psi_O\rrangle_\mrm{1h}$.
As mentioned at the start of this section, the calculation presented above goes through in much
the same way for these two statistics.
Returning to Eq.~\eqref{eq:numerator} but putting 
$\phi(\x_1-\x;M)-\phi(\x_2-\x;M)\to\phi(\x_1-\x;M)-\phi(-\x;M)$ for the first statistic
and $\phi(\x_1-\x;M)-\phi(\x_2-\x;M)\to\phi(\x_2-\x;M)-\phi(-\x;M)$ for the second,
then repeating many of the same steps above, we find
\bea
\llangle\psi(\x_1)-\psi_O\rrangle_\mrm{1h}
&=[\phi_A(0)-\phi_A(-\x_1)]-[\phi_B(-\x_2)-\phi_B(\x_1-\x_2)]\,,
\label{eq:psi1-psi0}\\
\llangle\psi(\x_2)-\psi_O\rrangle_\mrm{1h}
&=[\phi_B(0)-\phi_B(-\x_2)]-[\phi_A(-\x_1)-\phi_A(\x_2-\x_1)] 
\label{eq:psi2-psi0}\,,
\eea
for $\x_1\neq\x_2$. In the case of spherical symmetry,
$\phi_A(-\x_1)=\phi_A(\chi_1')$ and $\phi_B(-\x_2)=\phi_B(\chi_2')$.
Note that Eq.~\eqref{eq:Delta-psi-1halo} is equal to the difference
between Eq.~\eqref{eq:psi1-psi0} and Eq.~\eqref{eq:psi2-psi0}
(terms depending on radial distances cancel).
Note also that compared with Eq.~\eqref{eq:Delta-psi-1halo}
there is in fact an additional term on the right-hand sides of Eqs.~\eqref{eq:psi1-psi0} and \eqref{eq:psi2-psi0};
this traces back to the first term in the second line of Eq.~\eqref{eq:f-3pt} (which is not identically zero anymore).
However, this term is proportional to
$\delD(\x_1-\x_2)$, i.e.\ a shot noise contribution, so can
be ignored if $\x_1\neq\x_2$, as we have done above.

\section{Configuration-space formula for the dipole induced by RSD in the wide-angle regime}\label{app:config-rsd-dipole}
In the course of the calculations of Section~\ref{sec:origin} we gave a neat
expression for the leading-order RSD dipole in terms of the pairwise mean velocity.
In the mid-point parametrisation it reads
(see Eq.~\eqref{eq:xi1-rsd} in the main text)
\be\label{eq:xi1_rsd-pairwise}
\xi_1^\mrm{RSD}(s,d)
=-\epsilon\sp
    \frac{4}{5s^4}\frac{\dif}{\dif s}\big(s^4\llangle\bar{u}\rrangle\big)
    +\mathcal{O}(\epsilon^2)
=-\epsilon\sp\frac45
    \left(\frac{4}{s}\llangle\bar{u}\rrangle
        +\frac{\dif\llangle\bar{u}\rrangle}{\dif s}
    \right) +\mathcal{O}(\epsilon^2) \,,
\ee
where the first term in the second equality is due to the radial volume effect 
while the second term is due to the projection of distinct lines of sight onto
the mid-point.\footnote{For the dipole given in the end-point parametrisation,
add $-\epsilon\sp (3/5)\sp\xi_2(s)$ to the right-hand side of 
Eq.~\eqref{eq:xi1_rsd-pairwise},
where $\xi_2$ is the usual Kaiser quadrupole;
see Ref.~\cite{Dam:2023}, appendix D.}  Here we confirm that
this is indeed equivalent to the more standard expression of the
RSD dipole
(see, e.g.\ Ref.~\cite{Dam:2023}, equation 61)
\be\label{eq:xi1-rsd-linear}
\begin{split}
\xi_1^\mrm{RSD}(s,d)
&=-\epsilon (b_A-b_B)\sp f 
\int\frac{k^2\dif k}{2\pi^2}
    \left(2\sp\frac{j_1(ks)}{ks}-\frac25 j_2(ks)\right) P_{\delta\delta}(k)
    +\mathcal{O}(\epsilon^2)\,.
\end{split}
\ee
That these two equations are equivalent can be verified by:
substituting into Eq.~\eqref{eq:xi1_rsd-pairwise}
the linear expression
$\llangle\bar{u}\rrangle(s)\simeq-(1/2)(b_A-b_B)\ssp\xi_{u\sp\delta}(s)$, with
$\xi_{u\sp\delta}$ replaced by its Fourier representation~\eqref{eq:U};
evaluating the derivative with respect to $s$;
and using that $\dif j_1(x)/\dif x=j_1(x)/x-j_2(x)$.
Formula~\eqref{eq:xi1_rsd-pairwise} reaffirms that pairwise
functions (and derivatives thereof) form the basic building blocks of
all multipoles, whether in the distant-observer limit or in the
wide-angle regime. Moreover, Eq.~\eqref{eq:xi1_rsd-pairwise},
unlike Eq.~\eqref{eq:xi1-rsd-linear}, has the advantage that it does not explicitly
depend on the modelling details of $\llangle\bar{u}\rrangle$ (dynamics
and bias).

Note that when cross-checking expressions of odd multipoles one may
find an overall sign difference.
This is due to the convention being used for the separation vector:
either $\s=\s_1-\s_2$ (used here) or $\s=\s_2-\s_1$.
(This convention should not be confused with the line-of-sight
convention.)
No such ambiguity exists for the even multipoles since
these are even functions of $\mu=\hat\s\cdot\n$.
For numerical work we multiply our dipole by $-1$, in keeping with
the convention used in the simulation measurements (opposite our own).

\section{Anti-symmetric part of the cross-correlation function up to
second order in Gaussian cumulants}\label{app:full-anti}
For completeness, we give here the expression for the anti-symmetric part of the
cross-correlation function up to second order:
\be\label{eq:xi-anti-full}
\begin{split}
\xi^{(\mrm{s})}_{[AB]}
&=\xi_{[AB]}-\partial_a\sp m^a_{[AB]} + \tfrac12\partial_a\partial_b C^{ab}_{[AB]} 
    + \partial_a\partial_b\sp(m^a_{[AB]}\sp m^b_{(AB)})
    - m^a_{[AB]}\sp \partial_a\sp\xi_{(AB)}
        - m^a_{(AB)}\sp \partial_a\sp\xi_{[AB]} \\
&\quad
    - \xi_{(AB)}\sp \partial_a\sp m^a_{[AB]}
        - \xi_{[AB]}\sp \partial_a\sp m^a_{(AB)}
    + \tfrac12C^{ab}_{(AB)}\sp \partial_a\partial_b\sp\xi_{[AB]}
        + \tfrac12C^{ab}_{[AB]}\sp \partial_a\partial_b\sp\xi_{(AB)} \\
&\quad
    + \tfrac12\xi_{(AB)}\ssp \partial_a\sp \partial_b C^{ab}_{[AB]}
        + \tfrac12\xi_{[AB]}\ssp \partial_a\sp \partial_b C^{ab}_{(AB)}
    + \partial_a\sp\xi_{[AB]}\sp \partial_b C^{ab}_{(AB)}
        + \partial_a\sp\xi_{(AB)}\sp \partial_b C^{ab}_{[AB]}\,,
\end{split}
\ee
where repeated lowercase letters $a,b=1,2,\ldots,6$ are summed over
(no sum over labels $A$, $B$), all quantities are evaluated 
at $\mbf{S}=(\s_1,\s_2)$, and we recall $\partial_a=\partial/\partial\sp S^a$ is
the partial derivative on six-dimensional Cartesian space.
(Note that $A$ and $B$ are not tensorial indices but rather labels
we find convenient to use to distinguish between symmetric and
anti-symmetric parts.)
This expression, valid for both radial and transverse distortions,
follows from Eq.~\eqref{eq:xis-2nd} by inserting into it
$\xi_{AB}=\xi_{(AB)}+\xi_{[AB]}$,
$m^a_{AB}=m^a_{(AB)}+m^a_{[AB]}$, and
$C^{ab}_{AB}=C^{ab}_{(AB)}+C^{ab}_{[AB]}$,
and remembering that the product of a symmetric
and anti-symmetric part is overall anti-symmetric, hence the large
number of cross terms. Except for the first three terms,
all terms in Eq.~\eqref{eq:xi-anti-full} are quadratic in correlators.

Now if we assume no lightcone effect, 
$\xi_{[AB]}=0$ (see Eq.~\eqref{eq:weighting-replace}).
If we keep only the leading contribution to the cumulants,
$C_{[AB]}=0$ (but $m_{[AB]}\neq0$). Under these two approximations
Eq.~\eqref{eq:xi-anti-full} is greatly simplified and we are left with
\bea
\xi^{(\mrm{s})}_{[AB]}
&\simeq
-\partial_a\sp m^a_{[AB]}
    - m^a_{[AB]}\sp \partial_a\sp\xi_{(AB)}
    - \xi_{(AB)}\sp \partial_a\sp m^a_{[AB]}
    + \partial_a\partial_b\sp(m^a_{[AB]}\sp m^b_{(AB)}) \nonumber\\
&=-\partial_a\big[(1+\xi_{(AB)})\, m_{[AB]}^a\big]
    + \partial_a\partial_b\sp(m^a_{[AB]}\sp m^b_{(AB)}) \,.
\eea
Except for the last term, this is Eq.~\eqref{eq:xis-2nd} in the
main text.

The last term is the disconnected piece of
the second moment which is often neglected in streaming models (as we have 
done in Section~\ref{sec:origin}).
Since this is the only term in the
expression which contains a double derivative, developing it requires
some work.
First, distributing the derivatives:
\bea
\partial_a\partial_b\sp(m^{a}_{[AB]}\sp m^{b}_{(AB)})
&=\partial_a \partial_b m^a_{[AB]}\sp m^{b}_{(AB)}
    + m^{a}_{[AB]}\sp \partial_a \partial_b m^{b}_{(AB)}
    + \partial_a  m^{a}_{[AB]}\sp \partial_b m^{b}_{(AB)}
    + \partial_b  m^{a}_{[AB]}\sp \partial_a m^{b}_{(AB)} \nonumber\\
&=m_{(AB)}\cdot \partial\, (\partial\cdot m_{[AB]})
    + m_{[AB]}\cdot \partial\, (\partial\cdot m_{(AB)})
    + (\partial\cdot m_{[AB]})\sp (\partial\cdot m_{(AB)})
    + \partial_b  m^{a}_{[AB]}\sp \partial_a m^{b}_{(AB)} \nonumber
\eea
where in the second line we have used that $\partial_a\partial_b=\partial_b\partial_a$.

In our perturbative scheme recall that we count as the same order terms suppressed by
$\calH/k$ and terms suppressed by $\epsilon$, i.e.\
$\calH/k\sim\epsilon$.
Note that $m^\mrm{grav}_{[AB]}\sim m^\mrm{RSD}_{[AB]}
\sim\epsilon\ssp m^{\mrm{RSD}}_{(AB)}$. 
We have the following divergences:
\bea
\partial\cdot m_{[AB]}^\mrm{grav}
&=\mu\frac{\dif\llangle\Delta\psi\rrangle}{\dif s}+\mathcal{O}(\epsilon\cdot\calH/k)\,,\\
\partial\cdot m_{(AB)}^\mrm{RSD}
&=\mu^2\frac{\dif\llangle\Delta{u}\rrangle}{\dif s}
    +\mathcal{O}(\epsilon^2)\,,\\
\partial\cdot m_{[AB]}^\mrm{RSD}
&=\epsilon\sp\mu(1-\mu^2)\frac{\dif\llangle\bar{u}\rrangle}{\dif s}
    +\mathcal{O}(\epsilon^2)\,,
\eea
where here we only need to compute the wide-angle effect for the anti-symmetric
part of RSD.

\section{Coordinate relations and other useful formulae}\label{app:formulas}
This mathematical appendix gathers together some useful formulae
relating spherical coordinates $\chi_1,\chi_2,\cos\vartheta$
(in which correlations on the full sky are naturally described)
to the standard coordinates $s,d,\mu$ (in which the multipoles
are defined). These formulae are needed in the calculations of 
Section~\ref{sec:origin} but more generally in situations where
one wishes to calculate multipoles in the absence of
translation invariance.

First note $\chi_1=|\s_1|$, $\chi_2=|\s_2|$, 
$s\equiv|\s_1-\s_2|$, while $d=|\mathbf{d}|$ and
$\mu=\hat{\s}\cdot\n$ depend on the choice of $\mbf{d}$ and $\n$;
here they are specified according to the mid-point
prescription, in which $\mathbf{d}\equiv(\s_1+\s_2)/2$ and
$\n\equiv\hat{\mathbf{d}}$. The coordinate relations are
\begin{subequations}\label{eq:sdu-rels}
\bea
s&=\big(\chi_1^2+\chi_2^2-2\chi_1\chi_2\cos\vartheta\big)^{1/2}\,,
\\[5pt]
d&=\tfrac12\big(\chi_1^2+\chi_2^2+2\chi_1\chi_2\cos\vartheta\big)^{1/2}\,, \\
\mu&
% =\frac{\chi_1^2-\chi_2^2}{2sd}
=\frac{\chi_1^2-\chi_2^2}{[(\chi_1^2+\chi_2^2)^2-(2\chi_1\chi_2\cos\vartheta)^2]^{1/2}}\,,
\eea
\end{subequations}
where we recall $\cos\vartheta=\n_1\cdot\n_2$.
These relations all follow from straightforward applications of
the cosine rule, with
analogous relations in real space [replace
$\mu$ by $\mu'$, $s$ by $r$, $\chi_1$ by $\chi_1'$,
$\chi_2$ by $\chi_2'$, etc].
Note that $s$ and $d$ are positive valued;
$\mu<0$ when $\chi_1<\chi_2$;
and $\mu>0$ when $\chi_1>\chi_2$.
Under pair exchange $(\s_1\leftrightarrow\s_2$)
we have $\mu\to-\mu$ and $\chi_1\leftrightarrow\chi_2$.
Note also that the relation between coordinates is unique
since the jacobian is ${2\chi_1^2\chi_2^2}/{s^2d^2}$, which
is nonzero for all physical configurations (i.e.\ excluding
the unphysical case in which one or both points are at the origin,
see Figure~\ref{fig:config}).

An important regime which can be treated perturbatively
concerns small opening angles,
$\epsilon\equiv s/d\ll1$. In this more restricted wide-angle regime
we can linearize about the distant-observer limit
($\epsilon=0$). The radial distances to linear order in $\epsilon$
read
\begin{subequations}
\bea
\chi_1&=d\big(1+\tfrac12{\epsilon}\mu\big) +\mathcal{O}({\epsilon}^2)\,,\\
\chi_2&=d\big(1-\tfrac12{\epsilon}\mu\big)+\mathcal{O}({\epsilon}^2) \,,
% \cos\vartheta&=\n_1\cdot\n_2=1-2\sp\bar\epsilon^2\sp(1-\mu^2)+\mathcal{O}(\epsilon^4)\,.
    % +\bar\epsilon^4(2-8\mu^2+6\mu^4)+\mathcal{O}(\epsilon^6) \,,
\eea
\end{subequations}
where here and below we work in the mid-point parametrisation.
Then since $\s_1=\mathbf{d}+\s/2$ and $\s_2=\mathbf{d}-\s/2$,
the lines of sight $\n_1\equiv\hat\s_1$ and $\n_2\equiv\hat\s_2$
are found to be
\begin{subequations}
\bea
\n_1
&=\n +\tfrac12\epsilon(\hat\s-\mu\n) +\mathcal{O}(\epsilon^2)\,,\\
\n_2
&=\n -\tfrac12\epsilon(\hat\s-\mu\n) +\mathcal{O}(\epsilon^2)\,.
\eea
\end{subequations}
With no transverse displacements we also have
$\n_1=\hat\s_1=\hat\x_1$ and $\n_2=\hat\s_2=\hat\x_2$.
From these it follows that
\begin{subequations}
\bea
\label{eq:chainrule1}
\hat\s\cdot\n_1
&=\mu+\tfrac12\epsilon(1-\mu^2)+\mathcal{O}(\epsilon^2)\,,\\
\label{eq:chainrule2}
\hat\s\cdot\n_2
&=\mu-\tfrac12\epsilon(1-\mu^2)+\mathcal{O}(\epsilon^2)\,.
\eea
\end{subequations}
We also have
$\cos\vartheta=\n_1\cdot\n_2=1+\mathcal{O}(\epsilon^2)$
and $(\hat\s\cdot\n_1)(\hat\s\cdot\n_2)=\mu^2+\mathcal{O}(\epsilon^2)$.

The radial derivatives of these quantities are also needed.
First note $\partial s/\partial\chi_1=\hat\s\cdot\n_1$ and
$\partial s/\partial\chi_2=-\hat\s\cdot\n_2$ (exact).
For an arbitrary function $g(s)$ we have by the chain rule
$\partial g/\partial\chi_1=(\partial s/\partial\chi_1)(\dif g/\dif s)$
and likewise
$\partial g/\partial\chi_2=(\partial s/\partial\chi_2)(\dif g/\dif s)$.
In the distant-observer limit
$\partial\sp g/\partial\chi_1=-\partial\sp g/\partial\chi_2$
since $\partial s/\partial\chi_1=-\partial s/\partial\chi_2$
(or $\hat\s\cdot\n_1=\hat\s\cdot\n_2$).
We also require the following derivatives:
\begin{subequations}
\bea
\partial\sp(\hat\s\cdot\n_1)/\partial\chi_1&=+\big[1-\left(\hat\s\cdot\n_1\right)^2\big]/s\,,\\
\partial\sp(\hat\s\cdot\n_2)/\partial\chi_2&=-\big[1-\left(\hat\s\cdot\n_2\right)^2\big]/s\,.
\eea
\end{subequations}
Alternatively
$\partial\sp(\hat\s\cdot\n_1)/\partial\chi_1=\partial^2s/\partial\chi_1^2$
and $\partial\sp(\hat\s\cdot\n_2)/\partial\chi_2=-\partial^2s/\partial\chi_2^2$.
Notice that the radial derivative of $\hat\s\cdot\n_1$ is given in terms
of a polynomial of $\hat\s\cdot\n_1$
(and likewise for $\hat\s\cdot\n_2$).
In general for every derivative of $\hat\s\cdot\n_1$ we raise
the degree by one so that the $n$th derivative of
$\hat\s\cdot\n_1$ returns a polynomial in $\hat\s\cdot\n_1$
of degree $n+1$ (divided by $s^n$).

% ~~~~~ end of appendix ~~~~~

\bibliography{main}

\end{document}